\documentclass[a4paper,11pt]{article}
\pdfoutput=1 % if your are submitting a pdflatex (i.e. if you have
\usepackage{jcappub} % for details on the use of the package, please
\usepackage[T1]{fontenc} % if needed
\usepackage{booktabs}
\usepackage{graphicx}
\usepackage{amsmath,amssymb}
\graphicspath{{./figures/}}
\usepackage{appendix}
\usepackage{placeins}
\usepackage[normalem]{ulem}
\usepackage{multirow}
\usepackage{bm}
\usepackage{comment}
\usepackage{mathtools}

\usepackage{mathtools}
\usepackage{amsfonts}
\usepackage{upgreek}
\usepackage{latexsym}
\usepackage{stfloats}
\usepackage{afterpage}
\usepackage{caption}
\usepackage{subcaption}
\usepackage{enumitem}
\usepackage{orcidlink}

\newcommand{\be}{\begin{equation}}
\newcommand{\ee}{\end{equation}}
\newcommand{\beq}{\begin{equation}}
\newcommand{\eeq}{\end{equation}}
\newcommand{\bea}{\begin{eqnarray}}
\newcommand{\eea}{\end{eqnarray}}

\newcommand{\bn}{{\mathbf  n}}

\newcommand{\bnabla}{{\bm{\nabla}}}

\newcommand{\HH}{{\cal H}}

\newcommand{\RR}{{\cal R}}

\newcommand{\al}{\alpha}

\newcommand{\de}{\delta}

\newcommand{\ka}{\kappa}

\newcommand{\si}{\sigma}

\newcommand{\Om}{\Omega}

\newcommand{\pa}{\parallel}

\newcommand{\gga}{\gamma}

\newcommand{\bean}{\begin{eqnarray*}}
\newcommand{\eean}{\end{eqnarray*}}

\newcommand{\ii}{\mathrm{i}}

\newcommand{\id}{{\rm 1\kern -2.5pt I}}

\newcommand{\spl}{{\eth^{+}}}
\newcommand{\smin}{{\eth^{-}}}

\title{\boldmath Perturbative and numerical study of nonlinear relativistic effects in weak lensing}

\author[a]{Matteo Magi\,\orcidlink{0000-0001-8062-9447},}
\author[b,c]{Francesca Lepori\,\orcidlink{0009-0000-5061-7138}}
\author[b,c,d]{and Julian Adamek\,\orcidlink{0000-0002-0723-6740}}

\affiliation[a]{Cosmology, Gravity and Astroparticle Physics Group, Center for Theoretical Physics of the Universe, Institute for Basic Science (IBS), Daejeon, 34126, Korea}
\affiliation[b]{Institut f\"ur Astrophysik, Universit\"at Z\"urich, Winterthurerstrasse 190, 8057 Z\"urich, Switzerland}
\affiliation[c]{D\'epartement de Physique Th\'eorique, Universit\'e de Gen\`eve, 24 quai Ernest-Ansermet,\\1211~Gen\`eve~4, Switzerland}
\affiliation[d]{Institut f\"ur Teilchen- und Astrophysik, ETH Z\"urich, Wolfgang-Pauli-Strasse 27, 8093 Z\"urich, Switzerland}

\emailAdd{mmagi@ibs.re.kr}
\emailAdd{francesca.lepori@unige.ch}
\emailAdd{adamekj@ethz.ch}

\abstract{The standard weak lensing formalism assumes that the lensing map relating the observed image of a source to its intrinsic shape depends only on the deflection angle. We show that this description is incomplete beyond linear perturbation theory, even when only scalar perturbations are present at first order. Using the Jacobi map formalism, we derive expressions for the rotation field, shear B-modes, and their angular power spectra at second order in relativistic perturbation theory. In the standard formalism, rotation and shear B-modes share the same spectrum, however, this degeneracy is broken once the parallel transport of the Sachs basis is consistently taken into account. We quantify this correction numerically, finding a difference of about $5\%$ on large angular scales $\ell \sim 5$ for sources at redshift $z_\mathrm{s} = 0.5$.
We also investigate frame-dragging effects, which are usually neglected in weak lensing. We present the first analytical derivation of the corresponding impact on the angular power spectrum of shear B-modes and show that it becomes the dominant contribution on scales $\ell \lesssim 10$. 
While both Sachs-basis rotation and frame dragging significantly affect shear B-modes on large scales, their contribution to the observed ellipticity B-mode angular power spectrum is at the percent level relative to the total ellipticity B-mode signal, making these nonlinear relativistic corrections challenging to detect in practice.

Our results are supported by relativistic simulations of weak lensing observables, including the first numerical study of frame dragging in the power spectra of the lensing convergence and cosmic shear.
}

\begin{document}
\maketitle
\flushbottom

\section{Introduction}
\label{sec:intro}

In general relativity, the trajectory of light rays is deflected by the presence of gravitational fields. As a consequence, the observed images of galaxies do not faithfully represent their intrinsic shapes but are distorted by the gravitational fields of the foreground large-scale structure between the source and our telescope.
At the level of an infinitesimal image, the intrinsic shape of the source can in principle be arbitrary. However, independently of this shape, the effect of lensing on such an infinitesimal source is fully described by the tidal deformation of the light bundle. The corresponding lensing map between transverse separations at the source and observed angular separations can therefore be decomposed into three contributions: an isotropic change in the size of the image, called convergence; a shear that changes the shape anisotropically while preserving the area; and a rotation that preserves both the shape and area of the image.
The `cosmic shear' component has been detected for the first time in 2000 by several groups~\cite{Wittman:2000tc, vanWaerbeke:2000rm, Bacon:2000sy, Kaiser:2000if}, and over the past decades, it has proven to be a powerful cosmological probe~\cite{Dalal:2023olq, DES:2021wwk, Stolzner:2025htz, DES:2026fyc}.

Cosmic shear, as lensing effects in general, directly traces the distribution of matter in the Universe and is not affected by galaxy bias, which is particularly challenging to model in the nonlinear regime. However, several astrophysical effects must be taken into account, including intrinsic alignments and baryonic feedback on small scales. 
Furthermore, the shear field is a spin-2 quantity. Any spin-2 quantity defined on the sphere can be decomposed into spin-weighted spherical harmonics~\cite{Durrer_2020}, leading to two distinct parity components: the even-parity E-modes and the odd-parity B-modes.
Current weak lensing surveys primarily extract cosmological information from the E-modes. B-modes are subdominant in the standard cosmological model and, therefore, their measurements are commonly used as a diagnostic tool to identify the presence of systematics in the measurement~\cite{Asgari:2018hde}. Nevertheless, Stage-IV photometric surveys such as \textit{Euclid}~\cite{Euclid:2024yrr} and the Vera C. Rubin Observatory Legacy Survey of Space and Time (LSST;~\cite{LSST:2008ijt}) will deliver galaxy shape measurements with reduced noise over a large fraction of the sky, enabling significantly cleaner detections of B-modes. Therefore, characterizing all cosmological and astrophysical sources of B-modes is essential for a correct interpretation of these measurements.

In the standard formalism of weak lensing, the lensing map is modelled by the so-called deflection angle, which is defined as the deviation of the photon path between a homogeneous and isotropic background and a perturbed spacetime, projected onto the sphere.
Using this framework, it is shown in~\cite{Hirata:2003ka} (see also~\cite{Pratten:2016dsm, Marozzi:2016uob,Marozzi:2016qxl,Marozzi:2016und,Krause:2009yr}) that the spectra of the different components in the lensing map are not independent and the following relations hold (in the flat-sky limit) at any order in perturbation theory: the convergence and shear E-modes have the same power spectrum; the rotation and the shear B-modes have the same power spectrum.
However, the deflection angle is not a gauge-invariant quantity and, therefore, is not an observable~\cite{Yoo:2018qba, Grimm:2018nto, Fanizza:2022wob}. 
In~\cite{Fanizza:2022wob}, the authors compute the full lensing map at first order in perturbation theory, including vector and tensor modes. This computation shows that the standard relations between convergence, shear, and rotation do not hold when vector and tensor perturbations are included. For example, no image rotation is produced by cosmological perturbations at linear order, while linear vector and tensor modes generate shear B-modes.

In this paper, we show that the standard weak lensing formalism gives an incomplete description of lensing observables beyond linear order, even when only scalar perturbations are present at first order. At second order, these scalar perturbations generate vector contributions, which we include in our analysis, while tensor contributions are neglected. We derive analytical expressions for the resulting corrections to the standard formalism and validate them against relativistic $N$-body simulations~\cite{Adamek:2015eda, Adamek:2016zes, Lepori:2020ifz}, providing the first numerical study of the frame-dragging impact on the weak lensing angular power spectra.

The paper is organized as follows.
In Sec.~\ref{ref:jacobi} we present the nonlinear Jacobi map formalism. In Sec.~\ref{sec:main}, we derive fully relativistic expressions for the lensing observables up to second order in perturbation theory using the Jacobi map formalism in a flat FLRW universe. We provide explicit full-sky expressions for the rotation and the cosmic shear in terms of metric and matter perturbations in the Poisson gauge. In Sec.~\ref{ref:psLin}, we review the linear-order computation of the lensing observables and compute the corresponding angular power spectra in the flat-sky limit. In Sec.~\ref{ref:ps} we compute the angular power spectra of image rotation and shear B-modes at second order in perturbation theory, taking the flat-sky limit of the expressions derived in Sec.~\ref{sec:main}. We focus on the contributions from both scalar and vector perturbations, isolating the corresponding corrections to the shear B-modes that are not captured by the deflection-angle formalism. In Sec.~\ref{ref:num}, we validate our analytical computation against $N$-body simulations. 
In Sec.~\ref{ref:conc}, we discuss the relevance of our results and possible follow-up work.
Additional technical details are provided in the appendices.

\section{Jacobi map formalism for weak lensing observables}
\label{ref:jacobi}

In this section, we introduce the Jacobi map formalism for the description of weak lensing observables. In the standard weak lensing approach, lensing is described by postulating a mapping between the observed angular position of a source and its unlensed position on the sky, with all lensing effects encoded in the two components of the deflection angle. Image distortions and shape deformations are then obtained by taking angular derivatives of the deflection angle. A brief review of this formalism is provided in Appendix~\ref{ap:deflection}.

The Jacobi map formalism provides an alternative relativistic description that does not rely on an assumed lensing ansatz. Instead, it is constructed by solving the geodesic deviation equation for a bundle of null geodesics connecting the source to the observer. In order to define physical image properties, the formalism is supplemented by the parallel transport of an orthonormal screen basis along the light rays, the Sachs basis. The Jacobi map then relates physical separations in the source rest frame to angular separations at the observer in a manifestly covariant manner. By contrast, the standard formalism is formulated in terms of coordinate-dependent quantities and therefore does not provide a strictly observable description of lensing distortions.

In the following, we develop the Jacobi map formalism and derive its exact evolution equation, then simplifying the analysis by means of conformal transformations. Finally, we establish the relation between the Jacobi map and the amplification matrix, thereby connecting this covariant framework to the weak lensing observables.

\subsection{Jacobi map formalism}\label{sub:jacobimap}

We consider a scenario in which an extended luminous source, located within the past light cone of the observer, emits light that is received at the observer location. Geometrically, this setup is described by a congruence of null geodesics~$x(\Lambda, \epsilon)$, where~$\Lambda$ is the affine parameter along each geodesic and~$\epsilon$ is a continuous parameter that labels neighbouring geodesics at fixed~$\Lambda$. We assume that all geodesics in the congruence converge at the observer position~$O$ at $\Lambda = \Lambda_{\rm o}$.

In the weak lensing regime, we are interested in the linearised dynamics of the congruence, which is conveniently described in terms of the displacement field~$\xi$. This vector field, also known as the Jacobi field, connects neighbouring geodesics in the congruence. More concretely, in a given coordinate system, the infinitesimal separation between two such geodesics at fixed affine parameter~$\Lambda$ is expressed as $\xi^\mu(\Lambda) \delta\epsilon$. The displacement field satisfies the geodesic deviation equation, which governs the linear evolution of infinitesimal separations within the light bundle,
    \begin{equation}\label{geodev}
    \frac{D^2}{D\Lambda^2}\xi^\mu = R^\mu{}_{\nu\rho\sigma} k^\nu k^\rho \xi^\sigma~,
    \end{equation}
where $\frac{D}{D\Lambda} = k^\mu \nabla_\mu$ denotes the covariant derivative along a chosen central reference geodesic whose tangent vector is $k^\mu = \frac{dx^\mu}{d\Lambda}$, and~$R^\mu{}_{\nu\rho\sigma}$ is the Riemann tensor constructed from the spacetime metric~$g_{\mu\nu}$. The Riemann tensor encodes the tidal gravitational fields responsible for the relative acceleration of neighbouring geodesics and is evaluated along the central reference geodesic. The affine parameter~$\Lambda$ is physically fixed by requiring that the null tangent vector at the observer position, $k^\mu_{\rm o} = k^\mu(\Lambda_{\rm o})$, coincides with the observed photon wave vector.

The light bundle has a two-dimensional cross section, so the dynamics of the congruence can be described equivalently in terms of its transverse degrees of freedom. However, the displacement field~$\xi$ is, by definition, confined to the three-dimensional subspace orthogonal to the photon wave vector~$k$ at each value of~$\Lambda$. To isolate the two-dimensional transverse components of~$\xi$, we introduce the observer four-velocity~$u$, a unit timelike vector, and impose that at the observer location~$O$, the displacement field is orthogonal to~$u$. Since~$u$ is timelike and~$k$ is null, there exists at~$O$ a two-dimensional spacelike subspace of the tangent space orthogonal to both~$u$ and~$k$. This subspace is known as the screen space. In coordinates,~$\xi$ belongs to the screen space if the following equations are satisfied
    \begin{equation}\label{xi}
    (g_{\mu\nu} \xi^\mu k^\nu)_{\rm p} = 0\,, \qquad\qquad (g_{\mu\nu} \xi^\mu u^\nu)_{\rm o} = 0\,,
    \end{equation}
where the first equation is valid at any point~$P$ on the congruence, whereas the second equation holds only at the observer position~$O$.

We now construct a two-dimensional basis for the screen space. To do so, we introduce a tetrad basis~$e_a$ and its dual~$e^a$. Tetrads are well suited for this purpose because they define an orthonormal frame in which the metric is locally Minkowski,~$\eta_{ab}$\footnote{We adopt the following index conventions:
Greek letters from the middle of the alphabet ($\mu, \nu, \rho, \dots$) denote spacetime coordinate indices running from 0 to 3.
Greek letters from the beginning of the alphabet ($\alpha, \beta, \gamma, \dots$) denote spatial coordinate indices running from 1 to 3.
Latin letters from the beginning of the alphabet ($a, b, c, \dots$) denote tetrad indices running from 0 to 3,
while Latin letters from the middle of the alphabet ($i, j, k, \dots$) are used for spatial tetrad indices running from 1 to 3.}, just like in the laboratory frame. This holds independently of the coordinate system and without altering the global metric. As a result, projecting tensors onto a tetrad basis provides a physically meaningful way to describe local observations in cosmology~\cite{Mitsou:2019nhj}.

A tetrad is a set of four vectors that form a basis of the tangent space at the observer position~$O$. There is, however, a freedom in choosing this basis: different tetrads are related by local Lorentz transformations. To fix this ambiguity, we choose the timelike tetrad vector~$e_0$ to coincide with the observer four-velocity~$u$. This choice fixes the boost degree of freedom. The remaining three tetrad vectors~$e_i$ form a spacelike triad orthogonal to~$e_0\equiv u$, and their orientation is fixed up to a spatial rotation. Once this rotation is specified, the tetrad is fully determined. We will comment in Appendix~\ref{app:paralleltransport} on our choice of the spatial orientation.

To complete the construction of a basis for the screen space, we introduce the screen projector
$\Phi^i_I = (\theta^i, \phi^i)_I$, which projects spatial vectors in the tangent space at the observer onto the two-dimensional subspace orthogonal to the observed photon direction~$n^i=\left(\sin\theta\cos\phi,\sin\theta\sin\phi,\cos\theta \right)^i$ \cite{Grimm:2018nto}. Projecting the spacelike triad yields our choice for the screen basis:
    \bea\label{projtriad}
    e_I = e_i \Phi^i_I \,,
    \eea
and similarly for $e^I$. In our notation, $\Phi^i_1=\theta^i=\tfrac{\partial n^i}{\partial\theta}$ and $\Phi^i_2=\phi^i=\tfrac1{\sin\theta}\tfrac{\partial n^i}{\partial\phi}$.

Note that this construction is only valid at the observer position, however, we need to generalise it to an arbitrary point of the congruence to make consistent comparisons of transverse separations at different points along the light path. To achieve this, we parallel transport the tetrad along the congruence by requiring
    \begin{equation}
    \frac{D}{D\Lambda} e^\mu_I = 0\,,
    \end{equation}
with initial condition at~$O$, ensuring a unique solution. This procedure defines the so-called Sachs basis~$e_I(\Lambda)$~\cite{Sachs:1961zz} as a preferred screen basis. 

A few subtleties arise in connection with the parallel transport of the screen basis. First, one can show that parallel transport along the null geodesics leaves the screen projector~$\Phi^i_I$ invariant. As a result, the only effect of parallel transport concerns the spatial screen triad~$e_i$. When transported to the source position, the spatial triad, together with the parallel-transported observer four-velocity, naturally defines a local reference frame at the source.\footnote{The parallel-transported observer four-velocity defines a family of observers for which the photon frequency is conserved along the ray, and therefore the associated redshift vanishes; see \cite{Synge:1960ueh,Bunn:2008vj}.}
This reference frame does not, in general, coincide with the source rest frame, since the parallel-transported observer four-velocity differs from the source four-velocity. In principle, one would therefore need to perform a local Lorentz boost in order to align the transported frame with the source rest frame. However, it can be shown that this boost has no effect on separations transverse to the photon propagation direction. In practice, this allows us to treat the parallel-transported screen basis as effectively identifying the source rest frame for the purpose of defining weak lensing observables. We refer the reader to Appendix~A of Ref.~\cite{Grimm:2018nto} for a detailed proof and discussion of these points.

We are now in a position to recast the geodesic deviation equation in terms of the displacement field projected onto the screen space $\xi^\mu=\xi^I e^\mu_I$.\footnote{In principle, the displacement vector~$\xi^\mu$ could include a component proportional to the photon wave vector~$k^\mu$. However, this component plays no role when projecting onto the Sachs basis, since by construction the basis vectors~$e^\mu_I$ satisfy~$e^\mu_I k_\mu = 0$. As we are interested only in the projected quantities, we can disregard this component from the beginning.} By direct substitution in Eq.~($\ref{geodev}$), we obtain the Jacobi equation,
    \begin{equation}\label{geodevproj}
    \frac{d^2}{d\Lambda^2} \xi^I = R^\mu{}_{\nu\rho\sigma} k^\nu k^\rho e^\sigma_J e^I_\mu \xi^J \doteq \mathcal{R}^I{}_J \xi^J\,,
    \end{equation}
where the covariant derivative simplifies to a total derivative $\frac{d}{d\Lambda} = k^\mu \partial_\mu$ due to parallel transport, and we defined the projected optical tidal matrix $\RR^{I}{}_J$.

The Jacobi equation is a linear differential equation in the screen-projected displacement field~$\xi^I$, with the initial condition~$\xi^I(\Lambda_{\rm o}) = 0$ reflecting that the observer position~$O$ is the vertex of the light bundle. We therefore seek solutions of the form
    \begin{equation}\label{defJacob}
    \xi^I = \mathcal{D}^I{}_J \left( \frac{d}{d\Lambda} \xi^I \right) \bigg\rvert_{\Lambda_{\rm o}}\,,
    \end{equation}
where we have introduced the Jacobi map~$\mathcal{D}^I{}_J$ as a linear map relating~$\xi^I(\Lambda)$ to a physical initial condition~$( d\xi^I/{d\Lambda} )_{\rm o}$, proportional to the observed angular separation~\cite{Yoo:2016vne,Bernardeau:2009bm}.
Plugging the ansatz above into Eq.~($\ref{geodevproj}$), we derive the exact evolution equation for the Jacobi map,
    \begin{equation}\label{Jacob}
    \frac{d^2}{d\Lambda^2} \mathcal{D}^I{}_J = \mathcal{R}^I{}_K \mathcal{D}^K{}_J\,,
    \end{equation}
with initial conditions
    \begin{equation}\label{init}
    \frac{d}{d\Lambda} \mathcal{D}^I{}_J \bigg\rvert_{\Lambda_{\rm o}} = \delta^I_J~, \qquad\qquad \mathcal{D}^I{}_J(\Lambda_{\rm o}) = 0\,.
    \end{equation}
To recapitulate, Eqs.~($\ref{Jacob}$)-($\ref{init}$) provide a convenient and covariant way to describe how a congruence of null geodesics evolves, specifically how transverse separations between neighbouring rays change along the propagation direction, as projected onto the screen space defined by the Sachs basis at each point along the congruence.

Finally, we note that projected indices $I, J, K, \dots$, which live on the screen space, can be raised and lowered using the Euclidean metric~$\delta_{IJ}$, so the distinction between upper and lower indices is not essential.

\subsection{Conformal Jacobi map}
As we shall see in Sec.~\ref{sec:main}, in the context of
cosmological spacetimes, the Jacobi-map formalism
discussed so far can be significantly simplified by performing a conformal rescaling of the metric such that $g_{\mu\nu} = \Omega^2 \hat{g}_{\mu\nu}$, where the conformal factor~$\Omega$ depends on spacetime position. Although the two conformally related metrics describe different physics, light propagation is not affected because null geodesics are invariant under conformal transformations. However, Levi-Civita connections compatible with $g_{\mu\nu}$ and $\hat{g}_{\mu\nu}$ differ, and as a result, any given null geodesic will have different affine parametrisations in the two manifolds \cite{Wald:1984rg}.

We parametrise the null congruence in the conformal spacetime using the affine parameter~$\lambda$, which is distinct from~$\Lambda$. The relationship between these two parametrisations is proportional to the conformal factor
\begin{equation}\label{lamLam}
\frac{d\Lambda}{d\lambda} = \mathbb{C} \, \Omega^2~,
\end{equation}
with proportionality constant~$\mathbb{C}$ that cannot be determined by the conformal transformation alone. We emphasise that, as we shall see,~$\mathbb{C}$ carries dimensions, since~$\Lambda$ and~$\lambda$ parametrise two physically distinct wave vectors.

We now aim to derive how the governing equation for the Jacobi map, namely Eq.~\eqref{Jacob}, transforms under a conformal rescaling of the metric. Intuitively, since conformal transformations preserve the causal structure of spacetime, we expect that an appropriate rescaling of the Jacobi map exists such that the form of the equation remains invariant.

To proceed, we first determine how the optical tidal matrix transforms under a conformal transformation. This requires several ingredients: the relation between the conformally rescaled wave vector~$\hat{k}^\mu = \frac{d x^\mu}{d\lambda}$ and the original wave vector~$k^\mu = \frac{d x^\mu}{d\Lambda}$, the transformation of the Sachs basis, and the well-known conformal transformation law for the Riemann tensor.

The transformation of the wave vector is determined by Eq.~($\ref{lamLam}$) and reads $k^\mu=\tfrac{\hat k^\mu}{\mathbb{C}\,\Omega^2}$. To compute how the Sachs basis transforms, we use the orthonormality condition for the conformally rescaled metric, $\hat g_{\mu\nu}\hat e^\mu_I\hat e^\nu_J=\delta_{IJ}$, which leads to the relation: $e^\mu_I=\Om^{-1}\hat e^\mu_I$. Substituting these expressions into the definition of the optical tidal matrix and using the conformal transformation of the Riemann tensor, we obtain the conformal rescaling of the optical tidal matrix~\cite{Bernardeau:2009bm}:
    \begin{equation}
    \mathcal{R}_{IJ}=\frac{1}{\mathbb{C}^2\Omega^4}\left[ \hat{\mathcal{R}}_{IJ}+\delta_{IJ}\frac{d^2}{d\lambda^2}\ln\Omega-\delta_{IJ}\left(\frac{d}{d\lambda}\ln\Omega\right)^2\right]\,.
    \end{equation}
To derive the transformation of the Jacobi map, we begin by defining the conformal Jacobi map~$\hat{\mathcal{D}}_{IJ}$ similarly to Eq.~($\ref{defJacob}$),
    \begin{equation}
    \hat\xi^I = \hat{\mathcal{D}}^I{}_J \left( \frac{d}{d\lambda} \hat\xi^I \right) \bigg\rvert_{\lambda_{\rm o}}\,,
    \end{equation}
where~$\hat{\xi}^I$ is the component of the displacement field projected on the screen basis~$\hat{e}^\mu_I$ in the conformal manifold. By expressing the displacement vector~$\xi^\mu$ in the two bases as $\xi^\mu = e^\mu_I\, \xi^I$ and $\xi^\mu = \hat{e}^\mu_I\, \hat{\xi}^I$, we can directly derive the conformal transformation of the projected displacement field. Since the Sachs basis transforms as $e^\mu_I = \Omega^{-1}\, \hat e^\mu_I$ under a conformal rescaling, it follows that $\xi^I = \Omega\, \hat{\xi}^I$. 
Substituting this expression into the definition of the Jacobi map in Eq.~($\ref{defJacob}$), and using the definition of~$\hat{\mathcal{D}}_{IJ}$ above, together with the relation between the affine parameters in Eq.~($\ref{lamLam}$), we find that the Jacobi map transforms as
    \begin{equation}\label{DhatD}
    \mathcal{D}_{IJ}=\mathbb{C}\,\Omega_{\rm o}\,\Omega\,\hat{\mathcal{D}}_{IJ}\,.
    \end{equation}
Indeed, a direct substitution into Eq.~($\ref{Jacob}$) shows that the governing equation for the conformal Jacobi map~$\hat{\mathcal{D}}_{IJ}$ retains the same form as in the original spacetime,
    \begin{equation}\label{JacobConf}
    \frac{d^2}{d\lambda^2} \hat{\mathcal{D}}_{IJ} = \hat{\mathcal{R}}_{IK} \hat{\mathcal{D}}^K{}_J\,, \qquad\quad 
    \frac{d}{d\lambda} \hat{\mathcal{D}}_{IJ} \bigg\rvert_{\lambda_{\rm o}} = \delta_{IJ}\,, \qquad\quad
    \hat{\mathcal{D}}_{IJ}(\lambda_{\rm o}) = 0\,,
    \end{equation}
where $\hat{\mathcal{R}}_{IJ} = \hat{R}_{\mu\nu\rho\sigma} \, \hat{k}^\nu \hat{k}^\rho \, \hat{e}^\mu_I \hat{e}_J^\sigma$ is the conformal optical tidal matrix, and $\hat{R}_{\mu\nu\rho\sigma}$ is the Riemann tensor based on the conformal metric~$\hat g_{\mu\nu}$. We note that~$\hat{\mathcal{R}}_{IJ}$ (just as~$\mathcal{R}_{IJ}$) is symmetric as a consequence of the symmetries of the Riemann tensor. We will return to this important point in the next section.

In practice, Eq.~(\ref{JacobConf}) tells us that we can compute the conformal Jacobi map~$\hat{\mathcal{D}}_{IJ}$ by working in the conformal manifold, where the metric~$\hat{g}_{\mu\nu}$ is simpler than the original metric~$g_{\mu\nu}$. The Jacobi map~$\mathcal{D}_{IJ}$ is then recovered by applying a suitable prefactor that accounts for the conformal rescaling. However, this simplification comes with a trade-off: while the conformal metric is easy to handle, the screen basis~$\hat{e}^\mu_I$ is no longer parallel transported along null geodesics parametrised by~$\lambda$. In contrast to the physical Sachs basis, which satisfies the parallel transport condition $\tfrac{D}{D\Lambda} e^\mu_I = 0$, the conformal basis obeys a different transport law,
    \begin{equation}
    \frac{D}{D\lambda} \hat{e}^\mu_I = - \hat{k}^\mu \, \hat{e}^\nu_I \, \partial_\nu \ln \Omega\,,
    \end{equation}
where $\frac{D}{D\lambda} = \hat{k}^\mu \hat{\nabla}_\mu$ with covariant derivative~$\hat{\nabla}_\mu$ compatible with the conformal metric~$\hat g_{\mu\nu}$. For details on the derivation of this equation, we refer to Appendix~\ref{app:paralleltransport}.

\subsection{Amplification matrix}

The fundamental object in gravitational lensing is the lens map, which associates to each observed direction in the sky the corresponding spacetime emission event on the source worldline. More precisely, for a given observed angle, the lens map identifies the spacetime point reached by tracing the corresponding null geodesic backward from the observer until it intersects the source trajectory. In the weak lensing regime, the lens map is invertible, so that distinct observed directions correspond to distinct emission events and no multiple imaging occurs \cite{Perlick:2010zh}.

Although the lens map fully encodes the effects of gravitational lensing, it is not the most convenient object for practical calculations. Instead, one typically works with its linearisation around a chosen reference direction on the observer's sky. This linearised mapping is described by the amplification matrix\footnote{The object we refer to as the amplification matrix is also commonly called the distortion matrix in the literature. Some authors define the amplification matrix as the inverse of the convention adopted here.}~$\mathcal{A}$, which relates infinitesimal angular separations between neighbouring light rays at the observer to the intrinsic transverse separations~$\xi^I_{\rm s}=\xi^I(\Lambda_{\rm s})$ of the corresponding emission events in the source rest frame.
In our notation, we define the amplification matrix as follows
    \begin{equation}\label{amplif}
    \xi^I_{\rm s} = \mathcal{A}^I{}_J \, \bar{\xi}^J_z\,.
    \end{equation}
Here~$\bar{\xi}^I_z$ is the observed transverse separation inferred from the observed redshift and angular position.

The amplification matrix is typically decomposed into its trace, symmetric trace-free, and antisymmetric parts:
\bea\label{A}
    \mathcal{A}\doteq\begin{pmatrix}
    1-\kappa & 0\\
    0 & 1-\kappa
    \end{pmatrix}
    -
    \begin{pmatrix}
    \gamma_1 & \gamma_2\\
    \gamma_2 & -\gamma_1
    \end{pmatrix}
    -
    \begin{pmatrix}
    0 & -\omega\\
    \omega & 0
    \end{pmatrix}\,.
    \eea
Although this decomposition is purely algebraic and applies to any $2 {\scriptstyle \times} 2$ matrix, the specific definition of the amplification matrix in Eq.~($\ref{amplif}$) allows a direct physical interpretation of the resulting components. These are the lensing convergence~$\kappa$, the two components of cosmic shear~$\gamma_1$ and~$\gamma_2$, and the rotation~$\omega$, which together characterise the evolution of an infinitesimal light bundle.

The convergence~$\kappa$ describes an isotropic magnification (or demagnification for negative~$\kappa$) of the intrinsic source shape and is related to fluctuations in the luminosity distance~\cite{Yoo:2018qba}. The shear components~$\gamma_1$ and~$\gamma_2$ encode anisotropic distortions, corresponding to the stretching and compression of the source shape along orthogonal directions. In a coordinate system where the photon propagation direction is along the $z$-axis,~$\gamma_1$ induces stretching along the $x$-axis and compression along the $y$-axis, while~$\gamma_2$ produces stretching along the $y=x$ direction and compression along the $y=-x$ direction. Finally, the antisymmetric component~$\omega$ describes a rotation of the source image, corresponding to a counter-clockwise rotation for positive~$\omega$.\footnote{Strictly speaking, this simple geometric interpretation of the lensing observables applies only at linear order in perturbation theory. In Sec.~\ref{ellipt} we discuss the relation between ($\kappa,\gamma_1,\gamma_2,\omega$) and fully non-perturbative lensing quantities.}

Since the lens map is constructed from the backward propagation of null geodesics, its linearisation is expressed in terms of the backward evolution of an infinitesimally thin bundle of light rays. The dynamics of such a bundle is governed by the geodesic deviation equation, making the Jacobi map formalism particularly well suited for the computation of the amplification matrix. Indeed, combining the definition of the Jacobi map in Eq.~(\ref{defJacob}) with that of the amplification matrix in Eq.~(\ref{amplif}), we find that the two quantities are linearly related:
    \begin{equation}\label{ampliJacob}
    \mathcal{A}_{IJ}=-\frac{\nu_{\rm o}}{\bar D_\mathrm{A}}\,\mathcal{D}_{IJ}\,.
    \end{equation}
The ratio between the observed photon frequency~$\nu_{\rm o}$ and the background angular diameter distance~$\bar D_\mathrm{A}$ to the source enters when rewriting the initial condition of the Jacobi map~$(d\xi^I/d\Lambda)_{\rm o}$ in terms of the transverse separation~$\bar{\xi}^I_z$ inferred from the observed redshift and angular position. In particular, the initial condition can be shown to be proportional to the observed angular separation, with a proportionality factor given by~$-\nu_{\rm o}$~\cite{Yoo:2016vne}.

Several important remarks follow from the expression of the amplification matrix in terms of the Jacobi map. First, the resulting amplification matrix is manifestly coordinate independent. Indeed, the right-hand side of Eq.~($\ref{ampliJacob}$) does not depend on the choice of spacetime coordinates. The Jacobi map is a quantity projected onto the Sachs basis, as indicated by its screen-space indices, thus it is unaffected by coordinate transformations. Moreover, the observed photon frequency is, by definition, an observable quantity, and the angular diameter distance is constructed from the observed redshift and angular separations, which are likewise coordinate-independent.

As a consequence, the amplification matrix and therefore the associated lensing observables derived within the Jacobi map formalism are defined in a fully coordinate-independent manner. This property should be contrasted with the standard weak lensing formalism, in which the amplification matrix is obtained from angular derivatives of the deflection angle. Since the deflection angle is defined in terms of spacetime coordinates, that construction is manifestly coordinate-dependent.

The origin of this difference lies in the fact that the standard formalism relies exclusively on global coordinates and does not incorporate a local rest-frame construction. Instead, the Jacobi map formalism explicitly makes use of tetrads to project spacetime separations onto the rest frame of the source. This additional geometric structure ensures that lensing observables are defined intrinsically, rather than as coordinate-dependent quantities.

A more technical way to highlight the shortcoming of the standard weak lensing formalism is that it assumes the simplifying condition~$e_I^\mu \propto\delta^\mu_i\Phi^i_I$ at the source position, effectively ignoring parallel transport altogether \cite{Grimm:2018nto}. Without parallel transport, there is no meaningful way to compare directions or shapes defined at different spacetime points. As a result, the standard formalism leads to expressions that are not truly observable and suffer from gauge dependence, making them unphysical.

As anticipated in the previous subsection, and as will become even more evident in what follows, it is computationally convenient to work with the conformal Jacobi map. For this reason, we now express the relation to the amplification matrix in Eq.~($\ref{ampliJacob}$) directly in terms of~$\hat{\mathcal{D}}_{IJ}$,
    \begin{equation}\label{JacobitoDistortion}
    \mathcal{A}_{IJ}=-\Omega_{\rm s}\,\frac{\mathbb{C}\left(\Omega\nu\right)_{\rm o}}{\bar D_\mathrm{A}}\,\hat{\mathcal{D}}_{IJ}  \,,
    \end{equation}
where Eq.~($\ref{DhatD}$) has been used to replace~$\mathcal{D}_{IJ}$ in terms of~$\hat{\mathcal{D}}_{IJ}$.

Combining Eqs.~($\ref{JacobitoDistortion}$) and ($\ref{A}$), we obtain the expression for the lensing convergence,
    \begin{equation}\label{lensingobs0}
    \kappa=1+\Omega_{\rm s}\,\frac{\mathbb{C}\left(\Omega\nu\right)_{\rm o}}{\bar D_\mathrm{A}}\,\left(\frac{\hat{\mathcal{D}}_{11}+\hat{\mathcal{D}}_{22}}2\right)\,,
    \end{equation}
the two independent components of the cosmic shear,
    \begin{equation}\label{lensingobs1}
    \gamma_1=\Omega_{\rm s}\,\frac{\mathbb{C}\left(\Omega\nu\right)_{\rm o}}{\bar D_\mathrm{A}}\,\left(\frac{\hat{\mathcal{D}}_{11}-\hat{\mathcal{D}}_{22}}2\right)\,,\qquad\qquad
    \gamma_2=\Omega_{\rm s}\,\frac{\mathbb{C}\left(\Omega\nu\right)_{\rm o}}{\bar D_\mathrm{A}}\,\left(\frac{\hat{\mathcal{D}}_{12}+\hat{\mathcal{D}}_{21}}2\right)\,,
    \end{equation}
and finally the image rotation,
  \begin{equation}\label{lensingobs2}
  \omega=\Omega_{\rm s}\,\frac{\mathbb{C}\left(\Omega\nu\right)_{\rm o}}{\bar D_\mathrm{A}}\,\left(\frac{\hat{\mathcal{D}}_{21}-\hat{\mathcal{D}}_{12}}2\right)\,.
    \end{equation}
    
Note that while the decomposition of the amplification matrix into convergence, shear, and rotation is entirely general, it is not necessarily the most useful or physically meaningful one when it comes to describing actual observables. Although the quantities extracted from the Jacobi map are coordinate-independent and thus theoretically well defined, they are not directly measurable. This distinction becomes especially important in perturbation theory, and we will return to this point in the next section when we discuss the distinction between cosmic shear and ellipticity.

\section{Weak lensing observables beyond linear theory}\label{sec:main}

The derivations presented in the previous section are exact and apply to general spacetimes. However, our goal is to compute the weak lensing observables in Eqs.~($\ref{lensingobs0}$)-($\ref{lensingobs2}$) at second order in cosmological perturbation theory. To this end, we now specify the spacetime metric~$g_{\mu\nu}$ by considering perturbations around a flat Friedmann–Lemaître–Robertson–Walker (FLRW) background. Given that lensing observables derived within the Jacobi map formalism are coordinate independent, the physical results are insensitive to the choice of gauge. We therefore adopt the Poisson gauge, in which the spacetime line element is given by
    \begin{equation}
    ds^2 = a^2(\eta)\left[ -e^{2\psi}d\eta^2-2B_\alpha dx^\alpha d\eta +\left(e^{2\varphi}\delta_{\alpha\beta}+h_{\alpha\beta}\right)dx^\alpha dx^\beta \right]
    \,,
    \end{equation}
where~$a(\eta)$ is the scale factor as a function of conformal time, and~$x^\al$ are comoving coordinates. Throughout the analytical part of this work we use units in which $c=1$.

We can rewrite the metric in a more convenient way by highlighting the conformal factor $\Omega=a\,e^{\varphi}$ and defining the Weyl potential $\Psi=(\psi-\varphi)/2$
    \begin{equation}\label{metric}
    ds^2 = \left(a\,e^{\varphi}\right)^2\left[ -e^{4\Psi}d\eta^2-2B_\alpha dx^\alpha d\eta +\left(\delta_{\alpha\beta}+h_{\alpha\beta}\right)dx^\alpha dx^\beta \right]\,.
    \end{equation}
It is important to note that this rewriting of the metric is no longer exact. In our perturbative framework, we assume that only scalar perturbations are present at linear order, since scalar, vector, and tensor decouple at that level. Scalar perturbations are therefore included at both first and second order, while vector and tensor perturbations are treated as scalar-induced second-order quantities, sourced by first-order scalars,
   \begin{equation}\label{SVT}
    e^{2\varphi}=1+2\varepsilon \,\varphi_{(1)}+2\varepsilon^2 \,(\varphi_{(2)}+\varphi_{(1)}^2)+\mathcal{O}(\varepsilon^3)\,,
    \qquad
    B_\alpha=\varepsilon^2 \,B_\alpha^{(2)}+\mathcal{O}(\varepsilon^3)\,,
    \qquad
    h_{\alpha\beta}=\varepsilon^2 \,h_{\alpha\beta}^{(2)}+\mathcal{O}(\varepsilon^3)\,,
    \end{equation}
where the parameter~$\varepsilon$ tracks the perturbative order and the labels (1) and (2) indicate individual contributions. From this naive perturbative counting, vector and tensor perturbations appear to enter at the same order. However, in the context of the $\Lambda$CDM model, tensor perturbations are typically much smaller in amplitude than vector perturbations. We return to this point in the next section.

To solve the Jacobi map equation perturbatively, we now see the practical advantage of working in the conformal manifold. The key ingredient controlling the evolution of the Jacobi map is the optical tidal matrix, which is determined by the Riemann tensor along the light path.
Computing the Riemann tensor~$\hat R_{\mu\nu\rho\sigma}$ based on the conformal metric, which in our setup corresponds to the expression inside the square brackets in Eq.~($\ref{metric}$), we find that~$\hat R_{\mu\nu\rho\sigma}$ starts at first order in perturbations, and so does the optical tidal matrix~$\hat{\mathcal{R}}_{IJ}$.
This has a useful consequence: solving the conformal Jacobi map in Eq.~($\ref{JacobConf}$) at second order requires only the first-order solution~$\hat{\mathcal{D}}^{(1)}_{IJ}$, which in turn depends on the zeroth-order solution. The latter corresponds to unperturbed light propagation in flat spacetime and yields~$\hat{\mathcal{D}}^{(0)}_{IJ}=\lambda\,\delta_{IJ}$. Had we worked instead in the physical spacetime, the optical tidal matrix~$\mathcal{R}_{IJ}$ would contain a non-vanishing background contribution, making the perturbative expansion more cumbersome.

Finally, we specify the link between the conformal Jacobi map and the amplification matrix.
The conformal constant~$\mathbb{C}$, introduced in Eq.~($\ref{lamLam}$) to relate the different affine parametrisations of null geodesics, is not fixed by the conformal transformation itself and thus remains a free parameter. We are free to choose its value, and it is convenient to set the combination~$\mathbb{C}(\Omega \nu)_{\rm o} \equiv 1$ to unity at the position of the observer \cite{Yoo:2018qba}. In our setup, this choice corresponds to~$\mathbb{C}=(a\,\nu)^{-1}_{\rm o} e^{-\varphi_{\rm o}}$ .

We now introduce~$\lambda_z$, defined in terms of the observed redshift as~$\lambda_z=-(1+z)\bar{D}_\mathrm{A}$, which corresponds to the (negative) comoving distance~$\bar r_z$ to the source inferred from the observed redshift~$z$. We also define the fluctuation in the observed redshift via $1+\delta z=a_{\rm s}(1+z)$, where~$a_{\rm s}$ is the scale factor at the source position.
With these definitions and our choice of conformal constant, Eq.~($\ref{JacobitoDistortion}$) becomes
    \begin{equation}\label{JacobitoDistortion2}
    \mathcal{A}_{IJ}=\frac{e^\varphi}{\lambda_z}\left( 1+\delta z\right)\hat{\mathcal{D}}_{IJ}\,,
    \end{equation}
and the lensing observables in Eqs.~($\ref{lensingobs0}$)-($\ref{lensingobs2}$):
\begin{align}\label{lensingobs}
\kappa   &= 1 - \frac{e^\varphi}{\lambda_z}(1+\delta z)\left( \frac{\hat{\mathcal{D}}_{11} + \hat{\mathcal{D}}_{22}}{2} \right)\,,
&
\omega &= \frac{e^\varphi}{\lambda_z}(1+\delta z)\left( \frac{\hat{\mathcal{D}}_{12} - \hat{\mathcal{D}}_{21}}{2} \right)\,,
\nonumber\\
\gamma_1 &= \frac{e^\varphi}{\lambda_z}(1+\delta z)\left( \frac{\hat{\mathcal{D}}_{22} - \hat{\mathcal{D}}_{11}}{2} \right)\,,
&
\gamma_2 &= -\frac{e^\varphi}{\lambda_z}(1+\delta z)\left( \frac{\hat{\mathcal{D}}_{12} + \hat{\mathcal{D}}_{21}}{2} \right)\,.
\end{align}
Although these equations are still exact, they now apply to the cosmological setting of our interest.
In this section, we compute these observables in perturbation theory for the spacetime defined in Eq.~(\ref{metric}), with particular emphasis on the second-order expressions for the cosmic shear and image rotation.

\subsection{First-order Jacobi map}

At first order, the governing equation for the conformal Jacobi map in Eq.~($\ref{JacobConf}$) reads
    \begin{equation}
    \frac{d^2}{d\lambda^2}{}\hat{\mathcal{D}}^{(1)}_{IJ}=\lambda\,{}\hat{\mathcal{R}}_{IJ}^{(1)}\,,
    \end{equation}
where we used the background solution~$\hat{\mathcal{D}}^{(0)}_{IJ}=\lambda\,\delta_{IJ}$.
We integrate the equation above from the observer position, parametrised by~$\lambda_{\rm o}=0$, up to a generic affine parameter~$\lambda$ on the congruence to derive
    \begin{equation}\label{J1}
    \frac{d}{d\lambda}\hat{\mathcal{D}}_{IJ}-\de_{IJ}=\int_0^\lambda d\lambda'\,\lambda'\,\hat{\mathcal{R}}_{IJ}(\lambda')\,,
    \end{equation}
where we suppressed the explicit counting of perturbative order. Upon further integration up to the source position parametrised by the value of the affine parameter~$\lambda_{\rm s}\equiv\lambda_z+\delta\lambda_{\rm s}$ we obtain
    \begin{equation}
    \hat{\mathcal{D}}_{IJ}(\lambda_{\rm s})=(\lambda_z+\delta\lambda_{\rm s})\delta_{IJ}+\int_0^{\lambda_z} d\lambda\, (\lambda_z-\lambda)\lambda\,\hat{\mathcal{R}}_{IJ}(\lambda)\,,
    \end{equation}
where~$\delta\lambda_{\rm s}$ captures the distortion in the source position compared to the comoving distance $\bar r_z=-\lambda_z$ derived from the observed redshift in a FLRW universe.

As we have already remarked, the optical tidal matrix~$\hat{\mathcal{R}}_{IJ}$ is symmetric $\hat{\mathcal{R}}_{IJ}=\hat{\mathcal{R}}_{JI}$ at any order in perturbation theory.  Consequently, the Jacobi map~$\hat{\mathcal{D}}_{IJ}$ is symmetric at first order, and so is the amplification matrix~$\mathcal{A}_{IJ}$. Therefore, like in the standard formalism, also in the Jacobi map approach the image rotation~$\omega$ vanishes at first order.

The non-vanishing lensing observables at linear order are thus the convergence~$\kappa$ and the shear components~$\gamma_1$ and~$\gamma_2$. Starting from their definitions in Eq.~($\ref{lensingobs}$), and using the expressions for the optical tidal matrix~$\hat{\mathcal{R}}_{IJ}$ provided in Appendix~\ref{app:sources}, we derive the following results:
    \bea\label{lin}
    \kappa&=&-\left(\varphi+\delta z-\frac{\delta\lambda_{\rm s}}{\bar r_z}\right)+\int_0^{\bar r_z}d\bar r\,\left(\frac{\bar r_z-\bar r}{\bar r_z\bar r}\right)\left(\partial^2_\theta+\cot{\theta}\,\partial_\theta+\frac{1}{\sin^2\theta}\partial^2_\phi\right)\Psi\,,
    \nonumber\\
    \gamma_1&=&\int_0^{\bar r_z}d\bar r\,\left(\frac{\bar r_z-\bar r}{\bar r_z\bar r}\right)\left(\partial^2_\theta-\cot{\theta}\,\partial_\theta-\frac{1}{\sin^2\theta}\partial^2_\phi\right)\Psi\,,
    \nonumber\\
    \gamma_2&=&2\int_0^{\bar r_z}d\bar r\,\left(\frac{\bar r_z-\bar r}{\bar r_z\bar r}\right)\left(\frac{\partial_{\theta}\partial_{\phi}-\cot\theta\,\partial_\phi}{\sin\theta}\right)\Psi
    \,,
    \eea
where we have replaced the affine parameter with the background comoving distance via~$\bar{r} = -\lambda$, and we left implicit the expressions for the redshift fluctuation~$\delta z$ and the shift in the source affine parameter~$\delta\lambda_{\rm s}$, which are provided in Appendix~\ref{app:sources}. For later notational convenience, we define the lensing kernel $W(a,b)=\tfrac{a-b}{ab}\theta(a-b)$ where~$\theta$ is the Heaviside step function, such that $\tfrac{\bar r_z-\bar r}{\bar r_z\bar r}\equiv W(\bar r_z,\bar r)$.

\subsection{Second-order Jacobi map}

Having derived the Jacobi map at first order, we now compute the second-order solution. We begin by expanding Eq.~($\ref{JacobConf}$) to second order:
    \begin{equation}
    \frac{d^2}{d\lambda^2}{}\hat{\mathcal{D}}_{IJ}^{(2)} = \hat{\mathcal{R}}^{(2)}_{IK}\hat{{\mathcal{D}}}^K{}_J^{(0)}+{}\hat{\mathcal{R}}^{(1)}_{IK}\hat{{\mathcal{D}}}^K{}_J^{(1)}\,.
    \end{equation}
Substituting the background and first-order solutions for the Jacobi map in Eq.~($\ref{J1}$), we obtain the following governing equation, where perturbative order superscripts have been omitted for clarity
    \begin{equation}
    \frac{d^2}{d\lambda^2}{}\hat{\mathcal{D}}_{IJ}=\lambda\,\hat{\mathcal{R}}_{IJ}+\hat{\mathcal{R}}_{IK}\left(\int_0^{\lambda} d\lambda'\, (\lambda-\lambda')\lambda'\,\hat{\mathcal{R}}^{K}{}_J(\lambda')\right)
    \,.
    \end{equation}
    Integration from $0$ to $\lambda$ leads to
    \begin{equation}
    \frac{d}{d\lambda}\hat{\mathcal{D}}_{IJ}-\de_{IJ}=\int_0^\lambda d\lambda'\,\left( \lambda'\hat{\mathcal{R}}_{IJ}+\hat{\mathcal{R}}_{IK}\int_0^{\lambda'}d\tilde{\lambda}\,(\lambda'-\tilde{\lambda})\,\tilde{\lambda}\,\hat{\mathcal{R}}^K{}_J\right)\,,
    \end{equation}
and the second integration from $0$ to $\lambda_{\rm s}$ yields the second-order Jacobi map
    \begin{equation}\label{jacob2interm}
    \hat{\mathcal{D}}_{IJ}=\lambda_{\rm s}\,\de_{IJ}+\int_0^{\lambda_{\rm s}}d\lambda\,(\lambda_{\rm s}-\lambda)\lambda \hat{\mathcal{R}}_{IJ}+\int_0^{\lambda_z}d\lambda\,(\lambda_z-\lambda)\hat{\mathcal{R}}_{IK}\int_0^\lambda d\lambda'(\lambda-\lambda')\lambda'\hat{\mathcal{R}}^K{}_J\,.
    \end{equation}
The first integral is evaluated along the perturbed photon trajectory and therefore contains post-Born corrections once it is expanded around the background geodesic. In contrast, the nested integrals are performed directly along the background trajectory and do not generate additional contributions beyond those already included.

Including the post-Born contribution by expanding the conformal Riemann tensor around the background geodesic, and using the background comoving distance~$\bar{r}$
as the integration variable, we reach the final expression for Eq.~($\ref{jacob2interm}$),
   \begin{multline}\label{jacob2}
    \hat{\mathcal{D}}_{IJ} = \lambda_{\rm s}\,\de_{IJ}+\lambda_z\int_0^{\bar r_z}d\bar r\,\bar r^2W(\bar r_z,\bar r)\Phi^i_I\Phi^j_J\left(1+\delta x_{\bar r}^\mu \partial_\mu\right)\hat\RR_{ij}(\bar r)-\lambda_z\frac{\delta\lambda_{\rm s}}{\bar r_z}\int_0^{\bar r_z}\frac{d\bar r}{\bar r}\,\bar r^2\hat\RR_{IJ}(\bar r)
    \\
    +\lambda_z\int_0^{\bar r_z}d\bar r\,\bar r^2 W(\bar r_z,\bar r)\hat \RR_{IK}(\bar r)\int_0^{\bar r} d\bar r'\,\bar r'^2 W(\bar r,\bar r')\hat \RR^K{}_J(\bar r')\,.
    \end{multline}
Here~$\delta x_{\bar r}^\mu$ represents the coordinate deviation from the background light path, whose explicit form can be found in Appendix~\ref{app:sources}.
    
It is important to emphasise that although the optical tidal matrix~$\hat\RR_{IJ}$ remains symmetric at all orders, the Jacobi map~$\hat{\mathcal{D}}_{IJ}$ is \textit{not} symmetric at second order, precisely due to the presence of the nested integrals in the last line of the equation above. The image rotation at second order is therefore non-vanishing and receives its leading contribution from Eq.~(\ref{lensingobs}),
    \begin{multline}\label{rotation}
    2\omega = \int_0^{\bar r_z}d\bar r\,\bar r^2W(\bar r_z,\bar r)\left(\hat \RR_{11}-\hat \RR_{22}\right)\int_0^{\bar r}d\bar r'\,\bar r'^2W(\bar r,\bar r')\hat \RR_{12}
    \\
    +\int_0^{\bar r_z}d\bar r\,\bar r^2 W(\bar r_z,\bar r)\hat \RR_{12}\int_0^{\bar r}d\bar r'\,\bar r'^2W(\bar r,\bar r')\left(\hat \RR_{22}-\hat \RR_{11}\right)\,.
    \end{multline}
Substituting the expression for the optical tidal matrix from Appendix~\ref{app:sources} we obtain the following explicit result
    \begin{multline}\label{rot}
    \omega = 2\int_0^{\bar r_z}d\bar r\,W(\bar r_z,\bar r)
    \left[\Psi_{,\theta\theta}-\cot\theta \Psi_{,\theta}-\frac{\Psi_{,\phi\phi}}{\sin^2\theta} \right]\int_0^{\bar r}d\bar r'\,W(\bar r,\bar r')\left[\frac{\Psi_{,\theta\phi}-\cot\theta\Psi_{,\phi}}{\sin\theta}\right]
    \\
    -2\int_0^{\bar r_z}d\bar r\,W(\bar r_z,\bar r)\left[\frac{\Psi_{,\theta\phi}-\cot\theta\Psi_{,\phi}}{\sin\theta}\right]
    \int_0^{\bar r}d\bar r'\,W(\bar r,\bar r')\left[\Psi_{,\theta\theta}-\cot\theta \Psi_{,\theta}-\frac{\Psi_{,\phi\phi}}{\sin^2\theta} \right]
    \,.
    \end{multline}
We note that the expression above coincides with the dominant contribution to the image rotation obtained in the standard weak lensing formalism based on the deflection angle~\cite{Pratten:2016dsm,Carron:2025qxj}. In the Jacobi map framework, image rotation is defined physically by comparing image distortions with respect to a local tetrad basis that is parallel transported along the light path, whereas in the standard formalism the rotation is defined with respect to global coordinates and is therefore not strictly a physical quantity. However, at linear order, the image rotation vanishes in the Jacobi map formalism, and it also vanishes in the standard approach when only scalar perturbations are considered, as is commonly assumed. Accounting for the parallel transport of the screen basis does not alter the leading non-vanishing contribution to the image rotation.

The computation of cosmic shear is more involved than that of lensing rotation, as it requires evaluating the optical tidal matrix~$\hat{\mathcal{R}}_{IJ}$ at second order. Combining Eq.~($\ref{jacob2}$) and Eq.~($\ref{lensingobs}$), we derive the two components of the cosmic shear,
    \begin{multline}\label{shear1}
    2\gamma_1 = \left(1+\varphi+\de z\right)\int_0^{\bar r_z}d\bar r\,\bar r^2W(\bar r_z,\bar r)\left(\phi^i\phi^j-\theta^i\theta^j\right)\left(1+\delta x_{\bar r}^\mu \partial_\mu\right)\hat\RR_{ij}-\frac{\delta\lambda_{\rm s}}{\bar r_z}\int_0^{\bar r_z}\frac{d\bar r}{\bar r}\ \bar r^2\left( \hat\RR_{22}-\hat\RR_{11}\right)
    \\
    +\int_0^{\bar r_z}d\bar r\,\bar r^2W(\bar r_z,\bar r)\hat \RR_{22}\int_0^{\bar r}d\bar r'\,\bar r'^2W(\bar r,\bar r')\hat \RR_{22}
    -\int_0^{\bar r_z}d\bar r\,\bar r^2 W(\bar r_z,\bar r)\hat \RR_{11}\int_0^{\bar r}d\bar r'\,\bar r'^2 W(\bar r,\bar r')\hat \RR_{11}\,,
    \end{multline}
    \begin{multline}\label{shear2}
    2\gamma_2 = -2\left(1+\varphi+\de z\right)\int_0^{\bar r_z}d\bar r\,\bar r^2W(\bar r_z,\bar r)\theta^i\phi^j\left(1+\delta x_{\bar r}^\mu \partial_\mu\right)\hat\RR_{ij}-\frac{\delta\lambda_{\rm s}}{\bar r_z}\int_0^{\bar r_z}\frac{d\bar r}{\bar r}\ \bar r^2\hat\RR_{12}
    \\
    -\int_0^{\bar r_z}d\bar r\,\bar r^2 W(\bar r_z,\bar r)\hat \RR_{12}\int_0^{\bar r}d\bar r'\,\bar r'^2W(\bar r,\bar r')\left(\hat \RR_{22}+\hat \RR_{11}\right)
    -\int_0^{\bar r_z}d\bar r\,\bar r^2W(\bar r_z,\bar r)
    \left(\hat \RR_{22}+\hat \RR_{11}\right)
    \\
    {\scriptstyle \times}\int_0^{\bar r}d\bar r'\,\bar r'^2W(\bar r,\bar r')\hat \RR_{12}\,.
    \end{multline}
Substituting the results of Appendix~\ref{app:sources} for the optical tidal matrix~$\hat\RR_{IJ}$, and combining the equation for~$\gamma_2$ with the rotation in Eq.~($\ref{rot}$), we obtain
    \begin{multline}\label{shear1}
    2\gamma_1 = \left(1+\varphi+\de z\right)\int_0^{\bar r_z}d\bar r\,W(\bar r_z,\bar r)\left\{
    2 \Psi_{,\theta\theta}-\frac{2}{\sin^2\theta}\Psi_{,\phi\phi}-2\cot\theta\Psi_{,\theta}
    +2\bar r^2(\theta^i\theta^j-\phi^i\phi^j)\right.
    \\
    \left.
    {\scriptstyle \times}\delta x_{\bar r}^\mu \partial_\mu\Psi_{,ij} 
    +\frac{B_{\parallel,\phi\phi}}{\sin^2\theta}+\cot\theta B_{\parallel,\theta}-B_{\parallel,\theta\theta}-\frac{1}{\sin\theta}\frac{d}{d\bar r} \left[\bar r(B_{\phi,\phi}+B_{\theta}\cos\theta-B_{\theta,\theta}\sin\theta)\right]
    \right.
    \\
    \left.
    +\frac{\bar r^2}{2}\left[\frac{d^2}{d\bar r^2}\left(h_{\phi\phi}-h_{\theta\theta}\right)- n^\al \left(\phi^i\phi^j-\theta^i\theta^j\right)\frac{d}{d\bar r}\left( \partial_{(j}h_{i)\al} \right)+n^\al n^\beta \left(\phi^i\phi^j-\theta^i\theta^j\right)\partial_i\partial_j h_{\alpha\beta}\right]
    \right.
    \\
    \left.+4 (\Psi_{,\theta})^2-4 \left(\frac{\Psi_{,\phi}}{\sin\theta}\right)^2+4 \left(\frac{\Psi_{,\phi\phi}}{\sin^2\theta}-\Psi_{,\theta\theta}+\cot\theta\Psi_{,\theta}\right)\left[ 2\Psi-2\Psi_{\rm o}+(\mathcal{U}_\parallel)_{\rm o}+2 \int_0^{\bar r}dr\,\Psi'\right]
    \right.
    \\
    \left.
    +\frac4{\sin\theta}(\Psi_{,\phi}-\bar r\Psi_{,\bar r\phi})\left[ (\mathcal{U}_\phi )_{\rm o} +2\int_0^{\bar r}\frac{dr}{r\sin\theta}\,\Psi_{,\phi}\right]-4(\Psi_{,\theta}-\bar r\Psi_{,\bar r\theta})\left[ (\mathcal{U}_\theta )_{\rm o} +2\int_0^{\bar r}\frac{dr}r\,\Psi_{,\theta}\right]
    \right\}
    \\
    -2\frac{\delta\lambda_{\rm s}}{\bar r_z}\int_0^{\bar r_z}\frac{d\bar r}{\bar r}\left( \Psi_{,\theta\theta}-\frac{\Psi_{,\phi\phi}}{\sin^2\theta}-\cot\theta\Psi_{,\theta}\right)
    \\
    +4\int_0^{\bar r_z}d\bar r\,W(\bar r_z,\bar r)\left[ \frac{\Psi_{,\phi\phi}}{\sin^2\theta}+\cot\theta\Psi_{,\theta}+\bar r\Psi_{,\bar r}\right] \int_0^{\bar r}d r\,W(\bar r,r)\left[ \frac{\Psi_{,\phi\phi}}{\sin^2\theta}+\cot\theta\Psi_{,\theta}+r\Psi_{,r}\right]
    \\
    -4\int_0^{\bar r_z}d\bar r\,W(\bar r_z,\bar r)\left( \Psi_{,\theta\theta}+\bar r\Psi_{,\bar r} \right)\int_0^{\bar r}dr\, W(\bar r,r)\left( \Psi_{,\theta\theta}+r\Psi_{,r} \right)\,,
    \end{multline}

    \begin{multline}\label{Gamma2}
    \gamma_2 = \omega
    +\left(1+\varphi+\de z\right)\int_0^{\bar r_z}d\bar r\,W(\bar r_z,\bar r)\left\{
    \frac{2}{\sin\theta}\Psi_{,\theta\phi}-2\frac{\cot\theta}{\sin\theta}\Psi_{,\phi}
    +2\bar r^2\theta^i\phi^j\delta x_{\bar r}^\mu \partial_\mu\Psi_{,ij}\right.
    \\\left.- \frac1{\sin\theta}(B_{\parallel,\theta\phi}-\cot\theta B_{\parallel,\phi})+  \frac{1}{2\sin\theta}\frac{d}{d\bar r}\left[\bar r(B_{\theta,\phi}-\cos\theta B_\phi+\sin\theta B_{\phi,\theta})\right]
    \right.
    \\
    \left.
    -\frac{\bar r^2}{2}\left[\frac{d^2}{d\bar r^2}h_{\theta\phi} - n^\al \theta^i\phi^j\frac{d}{d\bar r}\left( \partial_{(j}h_{i)\al} \right)+n^\al n^\beta \theta^i\phi^j\partial_i\partial_j h_{\alpha\beta}\right]+\frac4{\sin\theta} \Psi_{,\theta}\Psi_{,\phi}
    -\frac4{\sin\theta}\left( \Psi_{,\theta\phi}-\cot\theta\Psi_{,\phi} \right)
    \right.
    \\
    \left.
    {\scriptstyle \times}\left[ 2\Psi-2\Psi_{\rm o}+(\mathcal{U}_\parallel)_{\rm o}+2 \int_0^{\bar r}dr\,\Psi'\right]    -\frac2{\sin\theta}(\Psi_{,\phi}-\bar r\Psi_{,\bar r\phi})\left[ (\mathcal{U}_\theta )_{\rm o} +2\int_0^{\bar r}\frac{dr}{r}\,\Psi_{,\theta}\right]
    \right.
    \\
    \left.
    -2(\Psi_{,\theta}-\bar r\Psi_{,\bar r\theta})\left[ (\mathcal{U}_\phi )_{\rm o} +2\int_0^{\bar r}\frac{dr}{r\sin\theta}\,\Psi_{,\phi}\right]
    \right\} -2\frac{\delta\lambda_{\rm s}}{\bar r_z}\int_0^{\bar r_z}\frac{d\bar r}{\bar r}\left( \frac{\Psi_{,\theta\phi}}{\sin\theta}-\frac{\cot\theta}{\sin\theta}\Psi_{,\phi}\right)
    \\
    -4\int_0^{\bar r_z}d\bar r\,W(\bar r_z,\bar r)\left( \Psi_{,\theta\theta}+\bar r\Psi_{,\bar r} \right)\int_0^{\bar r}dr\, W(\bar r,r)\left[\frac{\Psi_{,\theta\phi}-\cot\theta\Psi_{,\phi}}{\sin\theta}\right]
    \\
    -4\int_0^{\bar r_z}d\bar r\,W(\bar r_z,\bar r)\left[\frac{\Psi_{,\theta\phi}-\cot\theta\Psi_{,\phi}}{\sin\theta}\right]\int_0^{\bar r}dr\, W(\bar r,r)\left[ \frac{\Psi_{,\phi\phi}}{\sin^2\theta}+\cot\theta\Psi_{,\theta}+r\Psi_{,r}\right]\,.
    \end{multline}
We note that, unlike the image rotation, the shear components receive contributions not only from scalar perturbations but also from vector and tensor perturbations. In our specific setup we assume that vector and tensor perturbations are sourced by linear-order scalars. However, this assumption has not been used to derive the expressions above, which are thus fully general.
Another important difference is that the image rotation is sourced purely by integrated effects along the unperturbed light path, whereas the shear also includes contributions evaluated explicitly at the observer and source positions. The presence of these boundary terms is essential to ensure coordinate independence of the shear and to prevent spurious infrared divergences \cite{Magi:2023jnl}.

We do not present the second-order computation of the convergence in this work, as it lies outside the scope of our analysis. The convergence is directly related to the angular diameter distance, which has already been computed at second order in the literature \cite{Ben-Dayan:2012lcv,Umeh:2012pn,Magi:2022nfy}. In the following subsection, we clarify more explicitly the specific objectives of the present paper.

\subsection{E/B-mode decomposition of the cosmic shear}

The lensing observables introduced so far are functions of the observed redshift and angular position on the sky. Scalar observables defined on the observer's sphere, such as the convergence and the image rotation, can be expanded in spherical harmonics, defining the multipole coefficients~$\kappa_{\ell m}$ and~$\omega_{\ell m}$. However, cosmic shear has different transformation properties under rotations. In particular, the shear components are invariant under a rotation by an angle~$\pi$ around the line-of-sight direction~$n^i$. This behaviour reflects the spin-2 nature of the cosmic shear and motivates the construction of the corresponding spin-2 fields as follows:
    \begin{equation}\label{spin2}
    \prescript{}{\pm}{\gamma}\doteq \gamma_1\pm \ii\gamma_2\,, \qquad\qquad\quad \prescript{}{\pm}{\gamma}(\bm n)=\sum_{\ell m} \prescript{}{\pm}{\gamma}_{\ell m} \,\prescript{}{\pm2}{Y}_{\ell m}(\bm n)\,.
    \end{equation}
The spin-weighted spherical harmonics~$\prescript{}{\pm2}{Y}_{\ell m}$ fully encode the spin-2 nature of cosmic shear. 

Owing to the parity properties of the spin-weighted spherical harmonics, the cosmic shear can be decomposed into a parity-even (electric) component and a parity-odd (magnetic) component, conventionally referred to as the E- and B-modes:
    \begin{equation}\label{EB}
    \gamma_{\ell m}^\mathrm{E} \doteq -\frac{1}{2}( \prescript{}{+}{\gamma}_{\ell m} +\prescript{}{-}{\gamma}_{\ell m}) \,, \qquad \qquad \quad                                       \gamma _{\ell m}^\mathrm{B} \doteq -\frac{1}{2\ii}(  \prescript{}{+}{\gamma}_{\ell m} -\prescript{}{-}{\gamma}_{\ell m})\,,
    \end{equation}
with $\prescript{}{\pm}{\gamma}_{\ell m}\equiv \gamma _{\ell m}^\mathrm{E}\pm\ii\gamma _{\ell m}^\mathrm{B}$ obtained by inverting the second equation in Eq.~($\ref{spin2}$).
   
In harmonic space, weak lensing is therefore fully characterised by the multipole coefficients $\kappa_{\ell m}$, $\omega_{\ell m}$, $\gamma^\mathrm{E}_{\ell m}$, and $\gamma^\mathrm{B}_{\ell m}$.
However, in the standard weak lensing formalism based on the deflection angle, only two independent degrees of freedom are present. As a consequence, among these four quantities only two are independent: the convergence is related to the E-mode of the shear, while the image rotation is related to the B-mode. The exact relations between these pairs of observables can be derived, and they take their simplest form in the flat-sky approximation:
\begin{equation}\label{std}
\kappa^{\rm std}_{\bm\ell}=\gamma^{\mathrm{E, std}}_{\bm\ell}\,,\qquad\qquad\qquad \omega^\mathrm{std}_{\bm\ell}=\gamma^{\mathrm{B, std}}_{\bm\ell}\,.
\end{equation}
In the flat-sky approximation, the observer's sky is treated as a locally Cartesian patch, and the expansion in spherical harmonics is replaced by a two-dimensional Fourier expansion on the plane orthogonal to a reference line of sight. In this limit, the multipole~$\bm \ell$ is equivalent to the dimensionless transverse Fourier wave vector. Further details on the flat-sky approximation and its implementation are discussed in Sec.~\ref{sec:wick}.
We emphasise that the relations written in Eq.~($\ref{std}$) hold only within the standard weak lensing formalism, as indicated by the label ``std''. In the Jacobi map formalism, the lensing observables are independent quantities, and there is no a priori reason for these relations to hold.

To assess whether the standard relations are satisfied, we expand the perturbations in Fourier space and perform the E/B decomposition in the flat-sky limit, where Eq.~($\ref{EB}$) reads
\begin{equation}\label{EBflat}
\gamma^\mathrm{E}_{\bm\ell}=\gamma_{1,\bm\ell}\cos({2\alpha_\ell})+\gamma_{2,\bm\ell}\sin({2\alpha_\ell})\,,\qquad\qquad \gamma^\mathrm{B}_{\bm\ell}=-\gamma_{1,\bm\ell}\sin({2\alpha_\ell})+\gamma_{2,\bm\ell}\cos({2\alpha_\ell})\,,
\end{equation}
with $\alpha_\ell$ the polar angle of~$\bm \ell$ in Fourier space.
At linear order, combining Eqs.~($\ref{lin}$) and ($\ref{EBflat}$), we find that the convergence coincides with the E-mode of the shear when the contributions in round brackets are neglected, while both the image rotation and the shear~B-mode vanish and are therefore equal. Beyond linear order, the convergence and the shear E-mode receive subleading corrections, so their approximate equality remains valid $\kappa_{\bm\ell}\approx\gamma^\mathrm{E}_{\bm\ell}$. The situation is different for the image rotation and the shear~B-mode, which receive their leading contributions at second order. As a result, the equality that holds at linear order is, in principle, violated. The main purpose of this paper is to quantify this difference.

\subsection{Observed galaxy ellipticity}\label{ellipt}

Studying the Jacobi map formalism, we have highlighted several key differences from the standard approach to weak lensing. To summarise, the standard formalism defines lensing observables purely in terms of the deflection angle and leads to incomplete and gauge-dependent expressions for the lensing observables.
In contrast, the Jacobi map formalism provides a fully relativistic and coordinate-independent description of lensing observables by tracking the distortion of the screen-projected cross section of an infinitesimal bundle of light rays. The convergence, shear, and rotation derived from this approach are therefore physically meaningful in a way that their counterparts in the standard formalism are not.

Nevertheless, the components of the amplification matrix provide only an intermediate description of image distortions. Beyond linear order, they do not correspond directly to fully observable properties of lensed images, independently of the specific lensing formalism adopted. For example, in actual observations, one does not measure directly the symmetric trace-free part of the amplification matrix $(\gamma_1,\gamma_2)$, but rather the ellipticity~$\epsilon$ of galaxy images, defined in terms of the ratio of the principal axes of the lensed image. Only at linear order are these quantities proportional to each other. Similar considerations apply to~$\kappa$ and~$\omega$. Beyond linear order, the ratio of the physical area in the rest frame of the source to the area inferred from the observed angular size is given by~$\tilde\kappa$, which is a nonlinear combination of the components of the amplification matrix, rather than by the convergence alone.

An analogous remark applies to the physical rotation of images $\tilde\omega$. However, since the rotation vanishes at linear order and first appears at second order, our expression for~$\omega$ in Eq.~($\ref{rot}$) correctly captures the physical rotation up to second order, while nonlinear corrections arise only at third order in perturbation theory and therefore $\omega\approx\tilde\omega$ up to third-order corrections. The situation is different for the convergence and the shear, which do not vanish at linear order and hence do not provide a faithful description of observable quantities at second order.

At fully nonlinear order, the relation between the cosmic shear and the ellipticity reads~\cite{Lepori:2020ifz}:
    \begin{align}
    \gamma_1 &= - \frac{1-\tilde\kappa}{2 \sqrt{2 + \sqrt{4 + \epsilon_1^2 + \epsilon_2^2}}}
    \left(\epsilon_1 \cos \tilde\omega - \epsilon_2 \sin \tilde\omega \right)\,, 
    \label{eq:ellip1}
    \\
    \gamma_2 &= - \frac{1-\tilde\kappa}{2 \sqrt{2 + \sqrt{4 + \epsilon_1^2 + \epsilon_2^2}}}
    \left(\epsilon_1 \sin \tilde\omega + \epsilon_2 \cos \tilde\omega \right)\, ,
    \label{eq:ellip2}
    \end{align}
which reduces at second order to
    \begin{equation}\label{epspm}
    \prescript{}{\pm}{\epsilon}\approx-4\prescript{}{\pm}{\gamma}(1+\kappa)\,.
    \end{equation}
At leading order, the ellipticity is proportional to the cosmic shear and therefore also transforms as a spin-2 quantity, admitting an $\rm E/ \rm B$-mode decomposition. The nonlinear contributions arise from products of the linear-order cosmic shear with the lensing convergence. These nonlinear corrections have been studied in previous work~\cite{Shapiro:2008yk, Krause:2009yr} and it has been shown that it is necessary to include them in analyses of \textit{Euclid} data to avoid biases in the inferred cosmological parameters~\cite{Deshpande:2019sdl}.

\section{First-order angular power spectra}\label{ref:psLin}
In this section, we compute the angular power spectra of first-order lensing observables within the Jacobi map formalism. Since the Jacobi map is symmetric at linear order, the only non-vanishing observables in harmonic space are the convergence and the shear, in particular the shear~E-modes.
In the standard weak lensing formalism, the angular power spectra of these two quantities are proportional to each other, whereas this relation does not hold in the Jacobi map formalism. This discrepancy at linear order was already noted, e.g., in~\cite{Fanizza:2022wob}, where the angular power spectra of the convergence and shear~E-modes were computed including both scalar and tensor contributions. In our setup, first-order lensing observables are sourced exclusively by scalar perturbations. In addition, unlike~\cite{Fanizza:2022wob}, we adopt a number of simplifying assumptions, such as the flat-sky and Limber approximations, which will also be employed in the next section devoted to second-order angular power spectra.

\subsection{Flat-sky limit and Limber approximation}\label{subsec:flat}

Let us start by considering a statistically homogeneous and isotropic Gaussian field $X(\bm r)$, with dimensionful power spectrum~$P_{X}(k)$ at a given redshift
    \begin{equation}\label{ps3d}
    \langle X(\bm k)X(\bm k') \rangle=(2\pi)^3\delta_D(\bm k+\bm k')P_{X}(k)\,.
    \end{equation}
Here, $X$ denotes a generic scalar perturbation entering the expressions for the lensing observables, such as the Weyl potential~$\Psi$. Ultimately, all such perturbations can be related to the primordial curvature perturbation~$\zeta$.

In order to compute angular power spectra of lensing observables, we first expand the angular dependence of~$X(\bm r)$ in harmonic space. Working in the flat-sky approximation, the expansion in spherical harmonics is replaced by an expansion in plane waves~\cite{Hu:2000ee,Matthewson:2020rdt}, corresponding to a two-dimensional Fourier transform
    \begin{equation}\label{flatfour}
    X(\bm r)=\int\frac{d^2 \bm\ell}{(2\pi)^2}\,X_{\bm\ell}(\bar r)\,e^{\ii \bm\ell\cdot\bm\vartheta}\,.
    \end{equation}
Here~$\bm\ell$ is the dimensionless wave vector conjugate to the angular coordinate~$\bm\vartheta$, defined on the flat-sky plane. This plane is taken to be orthogonal to a reference direction~$\bm e_z$, obtained by expanding the line-of-sight direction~$\bm n(\theta,\phi)$ for small values of~$\theta$ around $\theta=\phi=0$, such that $\bm n=\bm e_z+\bm\vartheta$. Note that we implicitly treat~$X$ as a perturbation evaluated along the unperturbed light path, with the spatial position given by $\bm r=\bar r\,\bm n$.

By inverting Eq.~($\ref{flatfour}$), the two-point correlation function in harmonic space can be expressed in terms of the three-dimensional power spectrum~$P_X(k)$ at a given redshift. In general, we consider correlations evaluated at two different redshifts,
    \begin{equation}\label{nolimber}
    \langle X_{\bm\ell}(\bar r)X_{\bm\ell'}(\bar r')\rangle=(2\pi)^{2}\frac{1}{\bar r^{2}} \delta_D\left(\bm\ell +\bm\ell '\frac{\bar r}{\bar r '}\right)\int_{-\infty}^{\infty} \frac{d k_{\parallel }}{ 2\pi } e^{\ii(\bar r -\bar r ') k_{\parallel }} P_{X}\left(\sqrt{k^2_\parallel+\ell^2/\bar r^2}\right)\,,
    \end{equation}
where we decomposed the Fourier wave number $\bm k$ as $\bm k=k_\parallel\bm e_z+\bm k_\perp$, with the Dirac delta enforcing $\bm k_\perp=\bm\ell/\bar r$. Adopting the Limber approximation, the equation above becomes
    \begin{equation}\label{limber}
    \langle X_{\bm\ell}(\bar r)X_{\bm\ell'}(\bar r')\rangle=(2\pi)^{2}\delta_D\left(\bm\ell +\bm\ell '\right)\frac{\delta_D\left(\bar r-\bar r '\right)}{\bar r^{2}} P_{X}\left(\frac{\ell}{\bar r}\right)\,.
    \end{equation}

Having established how to compute angular correlations in the flat-sky approximation, we now turn to the lensing observables themselves, which are initially expressed in full-sky (see Sec.~\ref{sec:main}). We therefore derive a systematic procedure to take their flat-sky limit.

In our construction, the flat sky is defined as the tangent plane at the north pole, parametrised by the angular coordinates~$\bm\vartheta$. Expanding the line-of-sight direction~$\bm n$ for small polar angle~$\theta$, we obtain
\begin{equation}
\bm\vartheta = (\theta\cos\phi,\theta\sin\phi,0) \doteq (x,y,0)\,.
\end{equation}
The origin of the flat-sky plane, $x=y=0$, corresponds to a line of sight parallel to the reference direction $\bm e_z=(0,0,1)$.

The flat-sky limit can be regarded as a change of coordinates from the sphere to the tangent plane. In this limit, angular derivatives on the sphere are replaced by derivatives with respect to the flat-sky coordinates,
\begin{equation}
\partial_\theta = \cos\phi\,\partial_x + \sin\phi\,\partial_y\,,\qquad
\partial_\phi = -\theta\sin\phi\,\partial_x + \theta\cos\phi\,\partial_y\,,
\end{equation}
and trigonometric functions of~$\phi$ are expressed in terms of~$x$, $y$, and $\theta=\sqrt{x^2+y^2}$.

To obtain flat-sky expressions for the lensing observables, we proceed as follows: starting from the full-sky result, we expand all trigonometric functions of~$\theta$ to leading order in the small-angle limit, replace angular derivatives using the relations above, and finally evaluate the expression at the origin of the flat-sky plane. We then move to harmonic space by taking the Fourier transform. From Eq.~(\ref{flatfour}), derivatives with respect to~$x$ and~$y$ correspond to multiplication by~$\ii\ell_x$ and~$\ii\ell_y$, respectively.

\subsection{Shear E-modes spectrum}

In the flat-sky approximation, the shear components in Eq.~(\ref{lin}) reduce to
    \begin{equation}
    \gamma_1=\int_{0}^{\bar r_z}d\bar r\,W(\bar r_z,\bar r)\left(\Psi_{,xx}-\Psi_{,yy}\right)
    \,,\qquad\qquad \gamma_2=2\int_{0}^{\bar r_z}d\bar r\,W(\bar r_z,\bar r)\Psi_{,xy}\,.
    \end{equation}
Taking the two-dimensional Fourier transform on the flat sky, we derive
    \begin{equation}
    \gamma_{1,\bm\ell}=\int_{0}^{\bar r_z}d\bar r\,W(\bar r_z,\bar r)\left(\ell^2_y-\ell^2_x\right)\Psi_{\bm\ell}
    \,,\qquad\qquad \gamma_{2,\bm\ell}=-2\int_{0}^{\bar r_z}d\bar r\,W(\bar r_z,\bar r)\ell_x\ell_y\Psi_{\bm\ell}\,.
    \end{equation}
To extract the E/B-modes, we make use of the flat-sky definition in Eq.~(\ref{EBflat}). 
Decomposing the Fourier mode~$\bm\ell$ into Cartesian components through the polar angle~$\alpha_{\ell}$, i.e., $\bm\ell=\ell(\,\cos\al_\ell,\sin\al_\ell\,)$, we find that the shear B-modes vanish, while the E-modes are given by
    \begin{equation}
    \gamma^\mathrm{E}_{\bm\ell}=-\ell^2\int_{0}^{\bar r_z}d\bar r\,W(\bar r_z,\bar r)\Psi_{\bm\ell}(\bar r)\,.
    \end{equation}

The shear E-modes angular power spectrum $C^{\gamma_\mathrm{E}\gamma_\mathrm{E}}_\ell$ is defined as
    \begin{equation}
    \langle \gamma^\mathrm{E}_{\bm\ell}\gamma^\mathrm{E}_{\bm\ell'} \rangle = (2\pi)^2\delta(\bm\ell+\bm\ell')C^{\gamma_\mathrm{E}\gamma_\mathrm{E}}_\ell\,.
    \end{equation}
Evaluating the correlation function
    \begin{equation}
    \langle \gamma^\mathrm{E}_{\bm\ell}\gamma^\mathrm{E}_{\bm\ell'} \rangle = \ell^4\int_{0}^{\bar r_z}d\bar r\,W(\bar r_z,\bar r)\int_{0}^{\bar r_z}d\bar r\,W(\bar r_z,\bar r) \langle \Psi_{\bm\ell}(\bar r)\Psi_{\bm\ell'}(\bar r') \rangle\,\,
    \end{equation}
under the Limber approximation in Eq.~(\ref{limber}), we conclude that
    \begin{equation}
    C^{\gamma_\mathrm{E}\gamma_\mathrm{E}}_\ell=\ell^4\int_{0}^{\bar r_z}d\bar r\,\frac{W^2(\bar r_z,\bar r)}{\bar r^2} P_\Psi\left( \frac{\ell}{\bar r} \right)\,.
    \end{equation}

An improvement over the flat-sky approximation can be achieved by adopting the
full-sky relation between the shear E-modes and the lensing scalar potential, see Appendix~\ref{ap:deflection} for details.
This leads to a modified Limber result in which the factor $\ell^4$ is replaced by $(\ell+2)!/(\ell-2)!$.

\subsection{Convergence spectrum}

Following the same procedure as in the previous subsection, we derive the flat-sky expression for the convergence from Eq.~(\ref{lin}),
    \begin{equation}
    \kappa=\left(2-\frac{1}{\HH\bar r_z} \right)\Psi-\frac{4}{\bar r_z}\int_0^{\bar r_z}d\bar r\,\Psi+\int_0^{\bar r_z}d\bar r\, W(\bar r_z,\bar r)\left( \Psi_{,xx}+\Psi_{,yy} \right)\,.
    \end{equation}
Here we have also substituted the results derived in Appendix~\ref{app:sources}, obtaining explicit expressions in terms of the Weyl potential. 
Throughout this subsection, we neglect time derivatives, peculiar velocities, and assume vanishing anisotropic stress. The motivation for these assumptions is discussed in the next section. We further discarded perturbations evaluated at the observer position as they contribute exclusively to the monopole $\ell=0$.

After Fourier transforming to harmonic space
    \bea
    \kappa_{\bm\ell}&=&\left(2-\frac{1}{\HH\bar r_z} \right)\Psi_{\bm\ell}-\frac{4}{\bar r_z}\int_0^{\bar r_z}d\bar r\,\Psi_{\bm\ell}-\int_0^{\bar r_z}d\bar r\, W(\bar r_z,\bar r)\ell^2\Psi_{\bm\ell}
    \nonumber\\
    &\equiv& \gamma^E_{\bm\ell}+\left(2-\frac{1}{\HH\bar r_z} \right)\Psi_{\bm\ell}(\bar r_z)-\frac{4}{\bar r_z}\int_0^{\bar r_z}d\bar r\,\Psi_{\bm\ell}(\bar r)\,,
    \eea
we find that in the Jacobi map formalism, the convergence does not coincide with the shear E-modes, even in the flat-sky limit, but receives additional relativistic corrections. As a consequence, the angular power spectrum of the convergence differs from that of the shear E-modes. Using Eqs.~(\ref{nolimber}) and~(\ref{limber}), we find the difference in the angular power spectra:
    \begin{multline}
    C^{\kappa\kappa}_\ell = C^{\gamma_\mathrm{E}\gamma_\mathrm{E}}_\ell-\frac{8}{\bar r^3_z}\left(2-\frac{1}{\HH\bar r_z} \right)P_\Psi\left( \frac{\ell}{\bar r_z}\right)+\frac{8}{\bar r_z}\int_0^{\bar r_z}\frac{d\bar r}{\bar r^2}\left[ W(\bar r_z,\bar r)\ell^2+\frac{2}{\bar r_z} \right]P_\Psi\left( \frac{\ell}{\bar r} \right)
    \\
    +\frac{1}{\bar r^2_z}\left(2-\frac{1}{\HH\bar r_z} \right)^2\int_{-\infty}^{\infty} \frac{d k_{\parallel }}{ 2\pi }   P_{\Psi}\left(\sqrt{k^2_\parallel+\ell^2/\bar r^2}\right)\,. \label{eq:true-kappa}
    \end{multline}

\section{Second-order angular power spectra}\label{ref:ps}

In this section, we compute the angular power spectrum of cosmic shear B-modes at second order in relativistic perturbation theory, and compare it with the power spectrum of the image rotation. We also compute the angular power spectrum of the ellipticity B-modes. We assume statistical homogeneity and isotropy of the underlying cosmological perturbations and adopt a number of simplifying approximations. In particular, we adopt the flat-sky approximation, apply the Limber approximation to simplify line-of-sight integrals, and use Wick's theorem to express four-point correlations in terms of products of two-point functions. Moreover, we assume vanishing anisotropic stress and neglect time derivatives of the gravitational potentials as these are typically subdominant compared to spatial gradients, especially on small scales. We also restrict the analysis to correlations evaluated at the same redshift, that is, we neglect evolution effects along the line of sight.

\begin{figure}[t!]
    \centering
    \includegraphics[width=0.45\textwidth]{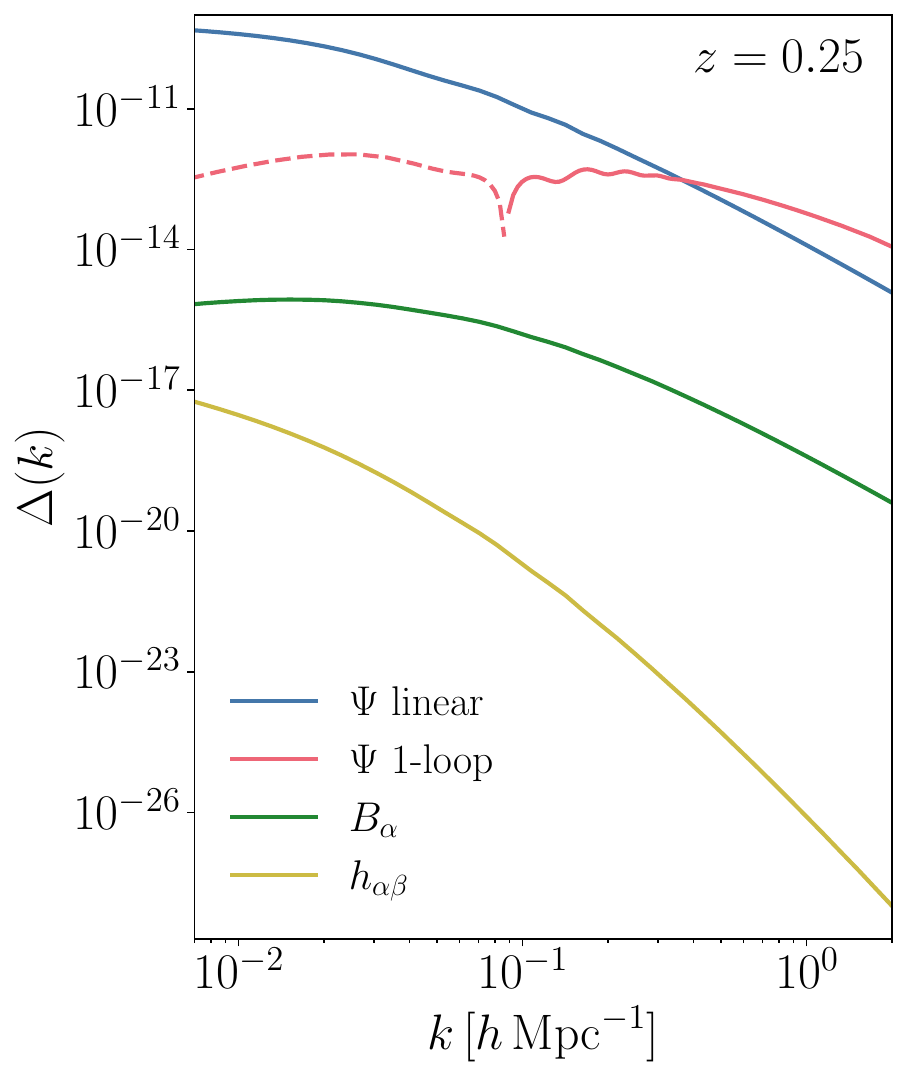}
    \caption{
Dimensionless power spectra for the metric perturbations $\Psi$, $B_\alpha$, and $h_{\alpha\beta}$ at redshift $z=0.25$ from perturbation theory up to 1-loop order. The 1-loop calculation for $\Psi$, which we do with \texttt{CLASS-PT} \cite{Chudaykin:2020aoj} here, also contains a contribution from the correlation between first and third order perturbations, sometimes denoted as $P_{13}$ in the literature. This contribution allows the 1-loop result to become negative, indicated with a dashed line style in the plot. The full power spectrum of $\Psi$ (not shown in the plot) is given by the sum of linear and 1-loop result. $B_\alpha$ and $h_{\alpha\beta}$ vanish at first order, hence their power spectra are given by their $P_{22}$ contribution only, i.e.\ the autocorrelation of the second-order perturbations. Their expressions can be found in App.~\ref{ap:numer-Four}.
    }
    \label{fig:1loopspectra}
\end{figure}

Concerning the relative importance of different contributions from second-order perturbation theory, we exploit the fact that scalar, vector and tensor perturbations are sourced differently in $\Lambda$CDM. The Weyl potential is sourced by the matter density contrast itself, which receives large second-order corrections on scales smaller than the nonlinear scale. However, this second-order correction, being itself a scalar, only affects the convergence and the E-modes of the shear. Second-order scalar-induced B-modes and rotation do not originate from this correction, but rather from quadratic couplings of the first-order Weyl potential, such as lens-lens coupling or post-Born corrections. The frame-dragging potential is sourced by the curl of the momentum density, which appears at second order from the product of first-order velocity and first-order density contrast. Although the first-order velocity is a pure gradient, this product contains gradient and curl components. However, we should keep in mind that the velocity of cold dark matter always remains small, of the order $10^{-3}$ even in the nonlinear regime. This renders the frame-dragging potential much smaller than the one-loop correction to the Weyl potential and therefore subdominant in the calculation of the convergence and the E-modes of the shear at second order. On the other hand, frame dragging sources the B-modes of the shear directly, and the hierarchy between its contribution and the one from the Weyl potential is less clear in this case. As we will demonstrate, frame dragging can become the dominant source of B-modes at low multipoles. Contrary to shear B-modes, rotation is not produced by frame dragging at this order. Finally, the tensor perturbations of the metric are sourced by the product of two first-order velocities. Because velocities are always small, this product is much smaller than the source term of frame dragging on the scales of interest. The contributions of tensor perturbations are therefore always subdominant to frame dragging and we neglect them from now on. The described hierarchy of one-loop perturbations is illustrated in Fig.~\ref{fig:1loopspectra}.

It is important to stress that the situation described here depends on the assumption, valid in $\Lambda$CDM, that the matter sector is non-relativistic, i.e.\ that rest-mass energy completely dominates the stress-energy tensor. The presence of relativistic contributions, for example from a dynamical dark-energy field, could potentially change the situation significantly and render vector- and tensor perturbations much more relevant.

\subsection{Wick contractions}\label{sec:wick}

To compute angular power spectra at second order in perturbation theory, we employ Wick’s theorem to decompose four-point correlation functions in harmonic space into products of two-point functions. In Sec.~\ref{subsec:flat}, we derived the two-point correlators of generic scalar perturbations entering the expressions for the lensing observables, working in the flat-sky and Limber approximations. We now build on those results to evaluate the four-point functions required at second order.
    
For the practical expressions encountered in this section, it will not always be possible to apply the Limber approximation. To illustrate this point, consider a generic contribution of the form $\langle X _{\bm\ell -\bm L}(\bar r)  X _{\bm L}(\bar r ') X _{\bm\ell '-\bm L'}(\bar x)  X _{\bm L'}(\bar x ') \rangle$. When all evaluation points along the line of sight are different, the Limber approximation in Eq.~($\ref{limber}$) can be applied, leading to
    \begin{multline}\label{wicklimb}
    \langle X _{\bm\ell -\bm L}(\bar r)  X _{\bm L}(\bar r ') X _{\bm\ell '-\bm L'}(\bar x)  X _{\bm L'}(\bar x ') \rangle  =  ( 2\pi )^{4} \delta _{D}( \bm\ell +\bm\ell ')\bigg[\delta _{D}( \bm L+\bm L') \frac{\delta _{D}(\bar r -\bar x)}{\bar r^{2}}\frac{\delta _{D}(\bar r '-\bar x ')}{\bar r ^{\prime 2}}
    \\
    {\scriptstyle \times} P_X\left(\frac{|\bm\ell -\bm L|}{\bar r}\right) P_X\left(\frac{L}{\bar r '}\right)
    +  \delta _{D}(\bm \ell -\bm L+\bm L') \frac{\delta _{D}(\bar r -\bar x ')}{\bar r^{2}}
    \frac{\delta _{D}(\bar r '-\bar x)}{\bar r ^{\prime 2}} P_X\left(\frac{|\bm\ell -\bm L|}{\bar r}\right) P_X\left(\frac{L}{\bar r '}\right) \bigg]\,.
    \end{multline}
However, when two evaluation points coincide, e.g., $\bar x\equiv\bar r=\bar r_z$, the Limber approximation cannot be applied, since it would lead to the product of two Dirac delta distributions evaluated at the same point, which is not well defined. In this case, we instead use Eq.~($\ref{nolimber}$) to obtain
    \begin{multline}\label{wicknolimb}
    \langle X _{\bm\ell -\bm L}(\bar r_{z})  X _{\bm L}(\bar r) X _{\bm\ell '-\bm L'}(\bar r_{z})  X _{\bm L'}(\bar x) \rangle =( 2\pi )^{4} \delta _{D}( \bm\ell +\bm\ell ') \bigg[ \delta _{D}( \bm L+\bm L') \frac{\delta _{D}(\bar r -\bar x)}{\bar r_{z}^{2}\bar r^{2}}P_X\left(\frac{L}{\bar r}\right)
    \\
    {\scriptstyle \times} \int _{-\infty }^{+\infty }\frac{dk_{\parallel }}{2\pi } P_X\left(\sqrt{k_{\parallel }^{2}+|\bm \ell -\bm L|^2/\bar r^2_{z} }\right)
    \\
    +\delta _{D}( \bm\ell -\bm L+\bm L') \frac{\delta _{D}(\bar r -\bar r_{z})}{\bar r_{z}^{2}}\frac{\delta _{D}(\bar x -\bar r_{z})}{\bar r_{z}^{2}} P_X\left(\frac{|\bm\ell -\bm L|}{\bar r_{z}}\right) P_X\left(\frac{L}{\bar r_{z}}\right)\bigg]\,.
    \end{multline}
Eqs.~($\ref{wicklimb}$) and ($\ref{wicknolimb}$) correspond to the sum of the Wick contractions of the pairings $(1,3)(2,4)$ and $(1,4)(2,3)$. Contributions at zero multipole ($\ell=0$) and disconnected terms have been omitted.

At second order in perturbation theory, metric perturbations include vector modes in addition to scalars (see Eq.~\eqref{SVT}). The definition of the three-dimensional Fourier power spectrum in Eq.~\eqref{ps3d} must therefore be generalized to account for the transversality of the vector perturbations.
    \bea\label{ps3dB}
    \langle B_\al(\bm k)B_\beta(\bm k') \rangle=(2\pi)^3\delta_D(\bm k+\bm k')P_{B}(k)\Pi_{\al\beta}(\bm k)\,, \qquad\qquad \Pi_{\al\beta}(\bm k)=\delta_{\al\beta}-\frac{k_\al k_\beta}{k^2}
    \eea
such that the projector $\Pi_{\al\beta}(\bm k)$ ensures the presence of only two independent components.

The angular power spectrum can be computed by performing a two-dimensional Fourier transform, as in Eq.~(\ref{flatfour}), for each transverse component of the vector field. The resulting angular correlators can then be related to the corresponding three-dimensional power spectrum in close analogy with the scalar case discussed in Sec.~\ref{subsec:flat}. Considering correlations evaluated at two distinct redshifts, we obtain:
    \bea\label{limberB}
    \langle B_{\al,\bm\ell}(\bar r)B_{\beta,\bm\ell'}(\bar r')\rangle=(2\pi)^{2}\frac{1}{\bar r^{2}} \delta_D\left(\bm\ell +\bm\ell '\frac{\bar r}{\bar r '}\right)\int_{-\infty}^{\infty} \frac{d k_{\parallel }}{ 2\pi } e^{\ii(\bar r -\bar r ') k_{\parallel }} \bigg[P_{B}(k)
    \Pi_{\al\beta}(\bm k)\bigg]_{\bm k_\perp=\tfrac{\bm\ell}{\bar r}}\,.
    \eea
Under the Limber approximation, we derive an expression similar to Eq.~\eqref{limber}, differing only by the presence of the additional transverse projector.

\subsection{Rotation spectrum}

To obtain the flat-sky expression for the image rotation, we start from the full-sky result in Eq.~\eqref{rot}, and follow the procedure outlined in Sec.~\ref{subsec:flat} to take the flat-sky limit.

We first replace angular derivatives with derivatives along the flat-sky directions
    \begin{equation}\label{rotation}
    \omega=2\int_0^{\bar r_z}d\bar r\,W(\bar r_z,\bar r)\int_0^{\bar r}d\bar r'\,W(\bar r,\bar r')\bigg\{\big[\Psi_{,xx}(\bar r)-\Psi_{,yy}(\bar r)\big]\Psi_{,xy}(\bar r')-\Psi_{,xy}(\bar r)\big[\Psi_{,xx}(\bar r')-\Psi_{,yy}(\bar r')\big]\bigg\}\,.
    \end{equation}
We then move to harmonic space by taking the Fourier transform of this expression.
    \begin{equation}\label{rotationl}
    \omega_{\bm\ell}=2\int_0^{\bar r_z}d\bar r\,W(\bar r_z,\bar r)\int_0^{\bar r}d\bar r'\,W(\bar r,\bar r')\int\frac{d^2\bm L}{(2\pi)^2}\,(\ell_xL_y -\ell_yL_x)(\bm\ell\cdot\bm L-L^2)\Psi_{\bm\ell-\bm L}(\bar r)\Psi_{\bm L}(\bar r')\,.
    \end{equation}
    
The angular power spectrum of the image rotation is defined as
    \begin{equation}
    \langle \omega_{\bm \ell}\omega_{\bm \ell'} \rangle=(2\pi)^2\delta(\bm \ell+\bm \ell')C^{\omega\omega}_\ell\,.
    \end{equation}
Due to statistical isotropy, the angular power spectrum~$C^{\omega\omega}_\ell$, is independent of the orientation of the Fourier modes that are correlated, therefore, we are free to choose $\bm \ell=(\ell,0)$ and $\bm \ell'=(\ell',0)$ to simplify the computations. Using the Wick contractions given in Eq.~($\ref{wicklimb}$), we obtain the final result for the rotation angular power spectrum:
    \begin{equation}\label{Celrot}
    C^{\omega\omega}_\ell=4\int_0^{\bar r_z}d\bar r\,\frac{W^2(\bar r_z,\bar r)}{\bar r^2}\int_0^{\bar r}d\bar r'\,\frac{W^2(\bar r,\bar r')}{\bar r'^2}\int\frac{d^2\bm L}{(2\pi)^2}\,\ell^2L^2_y (\ell L_x-L^2)^2 P_\Psi\left(\frac{L}{\bar r'} \right)P_\Psi\left( \frac{|\bm \ell-\bm L|}{\bar r}\right)\,.\qquad~
    \end{equation}
Notice that the contraction proportional to $\delta(\bar r-\bar x')\delta(\bar r-\bar x')$ does not give a contribution, due to the radial structure of the integrals.

\subsection{Shear B-modes spectrum}\label{subsec:B spectrum}

We now compute the angular power spectrum of the shear B-modes and compare it with the image rotation angular power spectrum obtained in the previous subsection. The shear B-mode angular power spectrum is defined as follows:
    \begin{equation}
    \langle \gamma^\mathrm{B}_{\bm \ell}\gamma^\mathrm{B}_{\bm \ell'} \rangle=(2\pi)^2\delta(\bm \ell+\bm \ell')C^{\gamma_\mathrm{B}\gamma_\mathrm{B}}_\ell\,.
    \end{equation}
As in the case of the image rotation, we take advantage of the statistical isotropy of the angular power spectrum to simplify the calculation. Since~$C^{\gamma_\mathrm{B}\gamma_\mathrm{B}}_\ell$ is independent of the orientation of the Fourier modes, we choose $\bm \ell=(\ell,0)$ and $\bm \ell'=(\ell',0)$ without loss of generality. Substituting our choice into the flat-sky definition of the shear B-modes in Eq.~($\ref{EBflat}$) shows that~$C^{\gamma_\mathrm{B}\gamma_\mathrm{B}}_\ell$ is equivalent to~$C^{\gamma_2\gamma_2}_\ell$, the angular power spectrum of~$\gamma_2$. This is particularly convenient, since~$\gamma_2$ is directly related to the image rotation as in Eq.~($\ref{shear2}$). We therefore focus on the computation of $\langle \gamma_{2,\bm \ell}\gamma_{2,\bm \ell'}\rangle$.

Following the prescription outlined in Sec.~\ref{subsec:flat}, the full-sky expression in Eq.~($\ref{Gamma2}$) reduces, in the flat-sky limit, to
\begin{multline}
(\gamma_2)_{\rm S}=\omega+2\int _{0}^{\bar{r}_{z}} d\bar{r} \ W(\bar{r}_{z} ,\bar{r} )( \Psi_{,xy}-2\Psi _{,x} \Psi _{,y} -2\Psi _{,xy} \Psi )-2\Psi_{\rm o}\int _{0}^{\bar{r}_{z}} d\bar{r} \left[\frac{1}{\bar{r}}\left(\frac{1}{\mathcal{H}_{z}\bar{r}_{z}} +1\right) -\frac{1}{\bar{r}_{z}} \right] \Psi _{,xy}
\\
-4\int _{0}^{\bar{r}_{z}} d\bar{r} \ W(\bar{r}_{z} ,\bar{r}) \ \Psi _{,xxy} \ \int _{0}^{\bar{r}} d\bar{r} '\ W(\bar{r} ,\bar{r} ')\ \Psi _{,x}
-4\int _{0}^{\bar{r}_{z}} d\bar{r} \ W(\bar{r}_{z} ,\bar{r}) \ \Psi _{,xyy} \ \int _{0}^{\bar{r}} d\bar{r} '\ W(\bar{r} ,\bar{r} ')\ \Psi _{,y}
\\
-4\int _{0}^{\bar{r}_{z}} d\bar{r} \ W(\bar{r}_{z} ,\bar{r} )\Psi _{,xx}\int _{0}^{\bar{r}} d\bar{r} '\ W(\bar{r} ,\bar{r} ')\Psi _{,xy} 
-4\int _{0}^{\bar{r}_{z}} d\bar{r} \ W(\bar{r}_{z} ,\bar{r} )\Psi _{,xy}\int _{0}^{\bar{r}} d\bar{r} '\ W(\bar{r} ,\bar{r} ')\Psi _{,yy}
\\
+2\Psi(\bar r_z)\int _{0}^{\bar{r}_{z}} d\bar{r} \left[\frac{1}{\bar{r}}\left(\frac{1}{\mathcal{H}_{z}\bar{r}_{z}} -2\right) +\frac{2}{\bar{r}_{z}} \right] \Psi _{,xy}
+\frac{8}{\bar{r}_{z}}\int _{0}^{\bar{r}_{z}} d\bar{r}' \ \Psi \int _{0}^{\bar{r}_{z}}\frac{d\bar{r}}{\bar{r}} \Psi _{,xy}
\\
+\frac{4}{\bar{r}_{z}}\int _{0}^{\bar{r}_{z}}\frac{d\bar{r}}{\bar{r}}   \bigg[ \Psi _{,y} \int _{0}^{\bar{r}} d\bar{r} '\  \Psi _{,x} +  \Psi _{,x}  \int _{0}^{\bar{r}} d\bar{r} '\  \Psi _{,y} - \Psi _{,xy}  \int _{0}^{\bar{r}} d\bar{r} '\  \Psi +\Psi \int_0^{\bar r_z}d\bar r  \left( \frac{\bar r_z}{\bar r}-\frac{\bar r}{\bar r'} \right)\Psi_{,xy}\bigg]\,. \label{eq:Clgamma}
\end{multline}
Here we neglect the contributions of the vector potential~$B_\alpha$ and retain only scalar perturbations, as indicated by the subscript~${}_\mathrm{S}$. The impact of~$B_\alpha$ is analysed separately in Sec.~\ref{subsec:framedragging}. Notice also that we have neglected the peculiar velocities at the source and the observer, since they are subdominant, as discussed at the beginning of this section.

Taking the Fourier transform and adopting the preferred orientation of the Fourier modes specified above, we obtain:
\begin{multline}\label{gamma2s}
(\gamma_{2,\bm\ell})_{\rm{S}}=\omega_{\bm\ell}-\frac{8}{\bar{r}_{z}}\int _{0}^{\bar{r}_{z}}\frac{d\bar{r}'}{\bar{r}'}\int _{0}^{\bar{r}_{z}} d\bar{r}   \int \frac{d^{2}\bm{L}}{( 2\pi )^{2}}  L_{x} L_{y}  \Psi _{\bm\ell-\bm L}(\bar{r}) \Psi _{\bm L}(\bar{r} ')
\\
-2\int _{0}^{\bar{r}_{z}} d\bar{r}  \left[\frac{1}{\bar{r}}\left(\frac{1}{\mathcal{H}_{z}\bar{r}_{z}} -2\right) +\frac{2}{\bar{r}_{z}} \right] \int \frac{d^{2}\bm{L}}{( 2\pi )^{2}} L_{x} L_{y} \Psi _{\bm\ell-\bm L}(\bar{r}_{z}) \Psi _{\bm L}(\bar{r})
\\
+\frac{4}{\bar{r}_{z}}\int _{0}^{\bar{r}_{z}}\frac{d\bar{r}}{\bar{r}}\int _{0}^{\bar{r}} d\bar{r} ' \int \frac{d^{2}\bm{L}}{( 2\pi )^{2}}\Bigl[ -2 \ell L_{y}  +\ \left( 3-\frac{\bar{r}_{z}}{\bar{r}}  +\frac{\bar{r}}{\bar{r} '}\right) L_{x} L_{y}\Bigr] \Psi _{\bm\ell-\bm L}(\bar{r}) \Psi _{\bm L}(\bar{r} ')\,.
\end{multline}
We remark that the linear term~$\Psi_{,xy}$ and the quadratic contributions involving perturbations of the Weyl potential at the observer position drop out of the computation. In Fourier space, such linear term is directly proportional to~$\ell_y$, which vanishes for our choice of orientation of the Fourier mode,~$\ell \equiv \ell_x$. The same holds for terms of the form~$\Psi_{\rm o}\Psi_{,xy}$, since~$\Psi_{\rm o}$ does not depend on angular coordinates.

The angular power spectrum of the shear B-modes, restricted to scalar perturbations, is obtained by evaluating the two-point correlator of the equation above and applying Wick’s theorem, as described in Sec.~\ref{sec:wick}.
\begin{multline}\label{CBmodes}
\left(C^{\gamma_\mathrm{B}\gamma_\mathrm{B}}_\ell\right)_{\rm S}-C^{\omega\omega}_\ell=\int \frac{d^{2}\bm{L}}{( 2\pi )^{2}}
\Bigg\{\Bigg[\frac{64}{\bar{r}_{z}^{2}}\int _{0}^{\bar{r}_{z}}\frac{d\bar{r} '}{\bar{r} ^{\prime 3}}\int _{0}^{\bar{r}_{z}}\frac{d\bar{r}}{\bar{r}^{3}}\left[ -\ell L_x L^2_y  + L_{x}^{2} L_{y}^{2}\left(1+\frac{\bar{r}}{\bar{r} '}\right)\right]
\\
+\frac{16}{\bar{r}_{z}^{2}}\int _{0}^{\bar{r}_{z}}\frac{d\bar{r}}{\bar{r}^{4}}\int _{0}^{\bar{r}}\frac{d\bar{r} '}{\bar{r}^{'2}}\Bigl[ L_{x}^{2} L_{y}^{2}  \left( 3-\frac{\bar{r}_{z}}{\bar{r}} +\frac{\bar{r}}{\bar{r} '}\right)\left( -\frac{\bar{r}_{z}}{\bar{r}} -1-3\frac{\bar{r}}{\bar{r} '}\right) +8\frac{\bar{r}}{\bar{r} '} \ell L_{x} L_{y}^{2}\Bigr]
\\
+\frac{16}{\bar{r}_{z}}\int _{0}^{\bar{r}_{z}}\frac{d\bar{r}}{\bar{r}^{3}}\int _{0}^{\bar{r}}\frac{d\bar{r} '}{\bar{r} ^{\prime 2}} \ W(\bar r_z,\bar r) W(\bar r,\bar r') \left[ L_{x}\left( 3-\frac{\bar{r}_{z}}{\bar{r}} -\frac{\bar{r}}{\bar{r} '}\right)  -\ell \right] \ell L_{y}^{2} (\ell L_{x} -L^{2} )\Bigg]
\\
{\scriptstyle \times} P_{\Psi}\left(\frac{|\bm\ell -\bm L|}{\bar{r}}\right) P_{\Psi}\left(\frac{L}{\bar{r} '}\right)
+\frac{4}{\bar{r}_{z}^{6}}\left(\frac{1}{\mathcal{H}_{z}\bar{r}_{z}}\right)^{2} L_{x} L_{y}^{2}( L_{x} -\ell ) P_{\Psi}\left(\frac{|\bm\ell -\bm L|}{\bar{r_{z}}}\right) P_{\Psi}\left(\frac{L}{\bar{r}_{z}}\right)
\\
+\frac{4}{\bar{r}_{z}^{2}}\int _{0}^{\bar{r}_{z}}\frac{d\bar{r}}{\bar{r}^{4}}\left[\frac{1}{\mathcal{H}_{z}\bar{r}_{z}}  +2\frac{\bar{r}}{\bar{r}_{z}}-2\right]^{2} L_{x}^{2} L_{y}^{2} \, P_{\Psi}\left(\frac{L}{\bar{r}}\right)\int _{-\infty }^{+\infty }\frac{dk_{\parallel }}{2\pi }\, P_{\Psi}\left(\sqrt{k_{\parallel }^{2}+{|\bm\ell -\bm L|}^2/\bar{r}^2_{z}} \right)
\\
+\frac{16}{\bar{r}_{z}^{4}}\int _{0}^{\bar{r}_{z}}\frac{d\bar{r}}{\bar{r}^{3}}\left(\frac{1}{\mathcal{H}_{z}\bar{r}_{z}} +2\frac{\bar{r}}{\bar{r}_{z}}-2\right)\left( -2+\frac{\bar{r}_{z}}{\bar{r}}\right) L_{x}^{2} L_{y}^{2}\, P_{\Psi}\left(\frac{|\bm\ell -\bm L|}{\bar{r}_{z}}\right) P_{\Psi}\left(\frac{L}{\bar{r}}\right)
\Bigg\}\,. 
\end{multline}
We can distinguish four types of contributions to the difference between the two angular power spectra, corresponding to the different combinations of the scalar power spectrum~$P_\Psi$. The first term contains all possible products of Eq.~($\ref{gamma2s}$) with itself, namely both cross terms and squares of the individual contributions, except for those involving the rotation. The second and third terms arise from the squares of individual contributions, again excluding the rotation. Finally, the last term is generated by cross correlations among contributions that do not involve the rotation.

The angular power spectrum of the rotation is given in Eq.~($\ref{Celrot}$), and in terms of angular multipoles it scales as~$\ell^4$. The contributions appearing in the right-hand-side of the equation above instead scale as~$\ell^2$, or as~$\ell^3$ in the case of cross terms involving the rotation. Naively, this suggests that the difference between the angular power spectra of shear B-modes and rotation should be small at high~$\ell$. At low~$\ell$, however, it is not possible to determine a priori which contribution dominates, since the integrals are involved and different terms can combine in a non-trivial way. A quantitative analysis is presented in Sec.~\ref{ref:num}.

\subsection{Ellipticity B-modes spectrum}

As discussed in Sec.~\ref{ellipt}, the quantity directly accessible to observations is not the cosmic shear extracted from the amplification matrix, but the ellipticity of galaxy images. Beyond linear order, these two quantities are no longer proportional, due to the mixing between first-order shear and convergence.

The ellipticity is a spin-2 quantity, just like the cosmic shear, and can therefore be decomposed into E- and B-modes in an analogous way. Combining Eqs.~($\ref{epspm}$) and~($\ref{EBflat}$) we obtain the expression for the ellipticity B-modes in the flat-sky limit,
    \begin{equation}
    -\frac14\epsilon^\mathrm{B}_{\bm\ell}=[\gamma_{2,\bm\ell}+(\kappa\ast\gamma_2)_{\bm\ell}]\cos(2\al_\ell)-[(\gamma_{1,\bm\ell}+(\kappa\ast\gamma_1)_{\bm\ell}]\sin(2\al_\ell)\,.
    \end{equation}
Once again, statistical isotropy allows us to simplify the computation of the angular power spectrum by setting $\al_\ell=0$, so that only the first square bracket contributes,
    \begin{equation}\label{epsB}
    \frac{1}{16} \langle \epsilon^\mathrm{B}_{\bm\ell}\epsilon^\mathrm{B}_{\bm\ell'} \rangle =\langle \gamma_{2,\bm\ell} \gamma_{2,\bm\ell'}\rangle +\langle \gamma_{2,\bm\ell} (\kappa \ast \gamma _{2} )_{\bm\ell '} \rangle +\langle \gamma_{2,\bm\ell'} (\kappa \ast \gamma _{2} )_{\bm\ell } \rangle +\langle (\kappa \ast \gamma _{2} )_{\bm\ell } (\kappa \ast \gamma _{2} )_{\bm\ell '} \rangle \,.
    \end{equation}
We have already obtained the expression for the second-order shear component~$\gamma_{2,\bm\ell}$ and its angular power spectrum in the previous subsection. Therefore, in order to compute the ellipticity B-modes angular power spectrum, it is sufficient to derive the contribution arising from the convolution of the first-order convergence with~$\gamma_{2,\bm\ell}$. From Eq.~($\ref{lin}$), we find
    \bea
    (\kappa \ast \gamma _{2} )_{\bm\ell } =2\int _{0}^{\bar r_{z}} d\bar r\, W(\bar r_{z} ,\bar r )\int _{0}^{\bar r_{z}} d\bar r ' \, W(\bar r_{z} ,\bar r ')\int \frac{d^{2}\bm L}{( 2\pi )^{2}} |\bm \ell -\bm L|^{2} L_{x} L_{y} \Psi _{\bm L} (\bar r )\Psi _{\bm\ell -\bm L} (\bar r ')\,,
    \eea
where we approximated the convergence~$\kappa_{\bm\ell}$ with the shear E-modes~$\gamma^\mathrm{E}_{\bm\ell}$. Using Eq.~($\ref{wicklimb}$) for the Wick contractions, we can compute the corresponding correlation function:
    \begin{multline} \label{eq:eps-main}
    \langle (\kappa \ast \gamma _{2} )_{\bm\ell } (\kappa \ast \gamma _{2} )_{\bm\ell '} \rangle = 4\int _{0}^{\bar r_{z}}d\bar r\,\frac{W^{2}(\bar r_{z} ,\bar r )}{\bar r^{2}}\int _{0}^{\bar r_{z}}d\bar r '\,\frac{W^{2}(\bar r_{z} ,\bar r ')}{\bar r ^{\prime 2}}  \int \frac{d^{2}\bm L}{( 2\pi )^{2}} |\bm\ell -\bm L|^{2} L_{y}^{2} L_{x}\left[ L_{x} |\bm \ell -\bm L|^{2}\right.
    \\
    \left.+ L^{2}( L_{x} -\ell )\right] ( 2\pi )^{2} \delta _{D}( \bm\ell +\bm\ell ') P_{\Psi}\left(\frac{|\bm\ell -\bm L|}{\bar r}\right) P_{\Psi}\left(\frac{L}{\bar r '}\right)\,. 
    \end{multline}
    
From Eq.~(\ref{epsB}), combining the results obtained above with those derived in the previous subsection, we compute the angular power spectrum of the ellipticity B-modes:
    \begin{multline} \label{Celeps}
\left(C^{\epsilon_\mathrm{B}\epsilon_\mathrm{B}}_\ell\right)_{\rm S}=\left(C^{\gamma_\mathrm{B}\gamma_\mathrm{B}}_\ell\right)_{\rm S}
    + \int \frac{d^{2}\bm L}{( 2\pi )^{2}}\Bigg\{4\int _{0}^{\bar r_{z}}\frac{d\bar r}{\bar r^{2}}\int _{0}^{\bar r_{z}}\frac{d\bar r '}{\bar r ^{\prime 2}}\,W^2(\bar r_z,\bar r)W^2(\bar r_z,\bar r')   |\bm\ell -\bm L|^{2} L_{y}^{2} L_{x}\left[ L_{x} |\bm \ell -\bm L|^{2}\right.
    \\
    \left.+ L^{2}( L_{x} -\ell )\right]+4\int _{0}^{\bar{r}_{z}}\frac{d\bar{r}}{\bar{r}^{2}}  \int _{0}^{\bar{r}}\frac{d\bar{r} '}{\bar{r} ^{\prime 2}} \ W (\bar{r}_{z} ,\bar{r} )^{2}W(\bar{r} ,\bar{r} ')W(\bar{r}_{z} ,\bar{r} ') \ell L_{y}^{2} (\ell L_{x} -L^{2} )\left[ L_{x} |\bm\ell -\bm L|^{2} +( L_{x} -\ell ) L^{2}\right]
    \\
    +\frac{8}{\bar{r}_{z}}\int _{0}^{\bar{r}_{z}}\frac{d\bar{r}}{\bar{r}^{3}} \int _{0}^{\bar{r}}\frac{d\bar{r} '}{\bar{r} ^{\prime 2}} \ W(\bar{r}_{z} ,\bar{r} ) W(\bar{r}_{z} ,\bar{r} ')\Bigl[ -2\ell L_{y}^{2}\left[ L_{x} |\bm\ell -\bm L|^{2} +( L_{x} -\ell ) L^{2}\right] +\left( 3-\frac{\bar{r}_{z}}{\bar{r}} +\frac{\bar{r}}{\bar{r} '}\right)
    \\
    {\scriptstyle \times}\left[( 3L_{x} -\ell ) |\bm\ell -\bm L|^{2} +( L_{x} -\ell ) L^{2}\right] L_{x} L_{y}^{2}\Bigr]
    -\frac{16}{\bar{r}_{z}}\int _{0}^{\bar{r}_{z}}\frac{d\bar{r} '}{\bar{r} ^{\prime 3}} W(\bar{r}_{z} ,\bar{r} ')\int _{0}^{\bar{r}_{z}}\frac{d\bar{r}}{\bar{r}^{2}} \ W(\bar{r}_{z} ,\bar{r} ) L_{x} L_{y}^{2}
    \\
    {\scriptstyle \times}\left[ 2L_{x} |\bm\ell -\bm L|^{2} +( L_{x} -\ell )\left( L^{2} +|\bm\ell -\bm L|^{2}\right)\right]
    \Bigg\} P_{\Psi}\left(\frac{|\bm\ell -\bm L|}{\bar r}\right)P_{\Psi}\left(\frac{L}{\bar r '}\right)\,.
    \end{multline}

\subsection{Frame-dragging contributions}
\label{subsec:framedragging}

In this subsection, we analyse the contribution of the second-order vector potential to the angular power spectrum of the shear B-modes, or in other words the frame-dragging contribution. This was neglected in the previous part of this section, where only scalar fluctuations were included. Following the same strategy adopted in the computation of the spectrum of the shear B-modes sourced by scalar perturbations, we focus exclusively on the shear component $\gamma_2$ in Eq.~($\ref{Gamma2}$):
    \begin{equation}\label{gamma2v}
    (\gamma_2)_{\mathrm{V}}=
    - \frac1{\sin\theta}\int_0^{\bar r_z}d\bar r\,W(\bar r_z,\bar r)   (B_{\parallel,\theta\phi}-\cot\theta B_{\parallel,\phi})+  \frac1{2\sin\theta}\int_0^{\bar r_z}\frac{d\bar r}{\bar r}(B_{\theta,\phi}-\cos\theta B_\phi+\sin\theta B_{\phi,\theta})\,,\qquad
    \end{equation}
where the subscript~$_{\rm V}$ denotes the contribution from vector perturbations.

To derive the flat-sky limit of this expression, it is not sufficient to directly apply the arguments of Sec.~\ref{subsec:flat}, where we discussed the flat-sky limit of full-sky expressions. That discussion implicitly assumed scalar quantities. For vector perturbations, the components projected onto the sky differ from those defined on the flat-sky plane. To illustrate this point, consider that the vector perturbation~$\bm B$ can be expressed in the following two bases:
    \begin{equation}
    \bm B=B_\parallel \bm n + B_\theta \bm\theta + B_\phi \bm\phi
    \equiv
     B_z \bm e_z+B_x \bm e_x+ B_y \bm e_y\,.
    \end{equation}
We expand the spherical-coordinate unit vectors defining the full-sky components in the small-$\theta$ limit, since the flat sky is defined as the tangent plane at the north pole. We do not expand in~$\phi$, as this coordinate is degenerate at the pole. This yields at lowest order in~$\theta$:
    \begin{equation}
    \bm n\approx (0,0,1)\equiv\bm e_z\,,\qquad \bm \theta\approx (\cos\phi,\sin\phi,0) \,,\qquad \bm \phi\approx (-\sin\phi,\cos\phi,0)\,.
    \end{equation}
This implies that the line-of-sight projection of~$\bm B$ aligns with the flat-sky reference direction~$\bm e_z$, while the angular components do not align with the~$x$ and~$y$ axes of the tangent plane but are instead rotated:
    \bea
    B_\parallel\approx B_z\,,\qquad B_\theta\approx B_x\cos\phi+B_y\sin\phi\,,\qquad B_\phi\approx -B_x\sin\phi+B_y\cos\phi\,.
    \eea
Using the relations between trigonometric functions of~$\phi$ and the flat-sky coordinates~$x$ and~$y$ derived in Sec.~\ref{subsec:flat}, we obtain the flat-sky limit of Eq.~(\ref{gamma2v}):
    \begin{equation}
    (\gamma_2)_{\rm V}=
    - \int_0^{\bar r_z}d\bar r\,W(\bar r_z,\bar r)   B_{z,xy}+  \frac1{2}\int_0^{\bar r_z}\frac{d\bar r}{\bar r}(B_{x,y}+ B_{y,x})\,.
    \end{equation}
In harmonic space, given the preferred choice of Fourier mode orientation $\bm\ell=(\ell,0)$, we derive
    \begin{equation}
    (\gamma_{2,\bm\ell})_{\rm V}=
      \ii  \frac{\ell}{2}\int_0^{\bar r_z}\frac{d\bar r}{\bar r} B_{y,{\bm\ell}}(\bar r)\,.
    \end{equation}
The corresponding two-point correlation function, evaluated using Eq.~(\ref{limberB}) and adopting the Limber approximation, reads
    \bea
    \langle(\gamma_{2,\bm\ell})_{\rm V}(\gamma_{2,\bm\ell'})_{\rm V}\rangle&=&
    -  \frac{\ell\ell'}{4}\int_0^{\bar r_z}\frac{d\bar r}{\bar r}\int_0^{\bar r_z}\frac{d\bar r'}{\bar r'} \langle B_{y,{\bm\ell}}(\bar r)B_{y,{\bm\ell'}}(\bar r')\rangle
    \nonumber\\&=&
    (2\pi)^2\delta_D(\bm\ell+\bm\ell')\frac{\ell^2}{4}\int_0^{\bar r_z}\frac{d\bar r}{\bar r^4}P_B\left(\frac{\ell}{\bar r} \right)\,.
    \eea

Due to statistical isotropy, the frame-dragging contribution to the shear B-modes angular power spectrum is given by
    \begin{equation}
    (C^{\gamma_\mathrm{B}\gamma_\mathrm{B}}_\ell)_{\rm V}=\frac{\ell^2}{4}\int_0^{\bar r_z}\frac{d\bar r}{\bar r^4}P_B\left(\frac{\ell}{\bar r} \right)\,.
    \end{equation}
The total angular power spectrum of the shear B-modes is given by the sum of the scalar and vector contributions. In principle, cross terms involving correlations between the scalar potential~$\Psi$ and the vector perturbation~$\bm B$ could also contribute. However, we find numerically that these mixed terms are subdominant and can be safely neglected (see discussion in Sec.~\ref{sec:numcel}).

For completeness, we compute the vector contribution to the convergence and the shear E-modes. We start with the convergence, which, from Eq.~\eqref{lensingobs}, is given by the trace of the Jacobi map. Using the results derived in Appendix~\ref{app:sources}, we obtain the full-sky expression
    \begin{multline}
    (\kappa)_{\rm V} = -\frac12\int_0^{\bar r_z}d\bar r\,W(\bar r_z,\bar r) \left(\cot\theta B_{\parallel,\theta}+B_{\parallel,\theta\theta}+\frac{1}{\sin^2\theta}B_{\parallel,\phi\phi}+2\bar r B'_\parallel\right) 
    \\
    +\frac1{2\sin\theta}\int_0^{\bar r_z}\frac{d\bar r}{\bar r}(B_{\phi,\phi}+\cos\theta B_\theta+\sin\theta B_{\theta,\theta})\,.
    \end{multline}
Neglecting temporal derivatives, the flat-sky limit leads to
    \begin{equation}\label{kappaV}
    (\kappa)_{\rm V}=-\frac12\int_0^{\bar r_z}d\bar r\,W(\bar r_z,\bar r) \left(B_{z,xx}+B_{z,yy}\right)+\frac1{2}\int_0^{\bar r_z}\frac{d\bar r}{\bar r}\,(B_{y,y}+B_{x,x})\,,
    \end{equation}
which in harmonic space, given the choice $\ell\equiv\ell_x$, becomes
    \begin{equation}
    (\kappa_{\bm\ell})_{\rm V}=\frac{\ell^2}2\int_0^{\bar r_z}d\bar r\,W(\bar r_z,\bar r) B_{z,\bm\ell}-\ii\,\frac{\ell}{2}\int_0^{\bar r_z}\frac{d\bar r}{\bar r}B_{x,{\bm\ell}}\,.
    \end{equation}
Finally, we obtain the frame-dragging contribution to the angular power spectrum:
    \begin{equation}
    (C^{\kappa\kappa}_\ell)_{\rm V}=\frac{\ell^4}4\int_0^{\bar r_z}\frac{d\bar r}{\bar r^2}\,W^2(\bar r_z,\bar r)P_B\left(\frac{\ell}{\bar r}\right)\,.
    \end{equation}
This result follows from the fact that~$\bm B$ is transverse, which implies that, under the Limber approximation, the projector~$\Pi_{\al\beta}$ in Eq.~\eqref{limberB} selects only the component along~$\bm e_z$. This is in agreement with the expression previously derived in the literature~\cite{Andrianomena:2014sya}.

In the flat-sky approximation, the angular power spectrum of the shear E-mode is given by that of~$\gamma_1$ under our choice of orientation of the Fourier modes (see Eq.~\eqref{EBflat}). We can therefore work directly with~$\gamma_1$ to simplify the problem. Taking the flat-sky limit of Eq.~\eqref{shear1}, we obtain an expression similar to that of the convergence:
    \begin{equation}
    (\gamma_{1})_{\rm V}=\frac12\int_0^{\bar r_z}d\bar r\,W(\bar r_z,\bar r) \left(B_{z,yy}-B_{z,xx}\right)-\frac1{2}\int_0^{\bar r_z}\frac{d\bar r}{\bar r}\,(B_{y,y}-B_{x,x})\,,
    \end{equation}
consistently with Eq.~\eqref{lensingobs}.

Given the similarity with Eq.~\eqref{kappaV} and the transversality of the vector potential, we conclude that the frame-dragging contribution to the shear E-modes and convergence is identical in the flat-sky Limber approximation: $(C^{\gamma_E\gamma_E}_\ell)_{\rm V}=(C^{\kappa\kappa}_\ell)_{\rm V}$.

\section{Numerical results}
\label{ref:num}
\subsection{$N$-body simulation and ray tracing}
\label{ref:sim}

To test our analytic predictions, we generate full-sky synthetic weak lensing maps from a high-resolution relativistic $N$-body simulation that provides us with a non-perturbative realisation of the gravitational Weyl potential $\Psi$ and the frame-dragging potential $B_\alpha$. The $N$-body simulation is run with the code \texttt{gevolution} \cite{Adamek:2015eda,Adamek:2016zes} using a mesh of $1920^3$ cells and the same number of particles. We simulate a cosmological volume with box size $1440\,h^{-1}\,\mathrm{Mpc}$, so our resolution is $750\,h^{-1}\,\mathrm{kpc}$. The cosmological parameters are based on the \textit{Euclid} reference cosmology \cite{Euclid:2024yrr}, with $\Omega_\mathrm{b} = 0.049$, $\Omega_\mathrm{c} = 0.2686$, $h = 0.67$, and a minimal-mass neutrino content. Initial conditions are set at $z_\mathrm{ini} = 31$ using the linear prediction from \texttt{CLASS} \cite{Blas:2011rf} for a primordial power spectrum with amplitude $A_\mathrm{s} = 2.215{\scriptstyle \times}10^{-9}$ and spectral index $n_\mathrm{s} = 0.9619$ at the pivot scale $k_\mathrm{pivot} = 0.05\,\mathrm{Mpc}^{-1}$.

To compute the convergence, shear and rotation non-perturbatively, we use the methodology of Ref.~\cite{Lepori:2020ifz}, solving the null geodesic equations and the geodesic deviation equations backwards in time from the observation point. This method is formally exact and provides the components of the amplification matrix for each line of sight directly. Whereas the analysis of convergence and rotation is then straightforward, the decomposition of the shear into E and B modes is non-local on the sky, and due to its small amplitude, the B-mode signal is notoriously sensitive to numerical artifacts. To overcome this issue, we use full-sky maps (to avoid leakage from masking) at 3.2 Gigapixel resolution, $N_\mathrm{side} = 16384$ in the \texttt{HEALPix} map~\cite{Gorski:2004by}, corresponding to an angular resolution of just under 13 arcseconds. Furthermore, we perform the ray tracing using a fourth-order Runge-Kutta solver for the ordinary differential equations, and we interpolate the gravitational fields from the Cartesian mesh of a single snapshot using tricubic interpolation to ensure that second derivatives remain continuous everywhere. Ray tracing is embarrassingly parallel (each geodesic is independent), and we make use of hardware acceleration to solve the numerical integrals efficiently. Our code can be found at \url{https://github.com/JulianAdamek/snapshot-raytracer}.

For the purpose of our numerical comparison, we use a snapshot of the Weyl potential $\Psi$ and the frame-dragging potential $B_\alpha$ at $z = 0.25$ and run the ray tracer in three different settings. In one setting, we use the full metric that includes both types of perturbations. For the other two settings, we set either $\Psi$ or $B_\alpha$ to zero; this allows us to compute the contribution of each type of perturbation separately, ignoring cross-correlation terms that appear at higher order in perturbation theory. Note that our ray tracer integrates each ray backwards in time for a fixed look-back coordinate time that is uniform across the sky. We therefore do not take into account redshift perturbations due to peculiar motions and gravitational redshift variations in the source plane. Compared to setting boundary conditions at fixed observed redshift, we are therefore missing some boundary terms that contribute to the convergence at first order and to the E-modes of the ellipticity at second order. The B-modes of ellipticity, as well as image rotation, are not affected by these terms at second order.

\subsection{Angular power spectra}
\label{sec:numcel}
From the simulated maps of the area distance, ellipticity, and image rotation, we extract the angular power spectra of the convergence, the shear E- and B-modes, and the image rotation. We make extensive use of \texttt{healpy}, the \texttt{python} implementation of the \texttt{HEALPix} library~\cite{Gorski:2004by}. Convergence and shear maps are constructed from the area distance and ellipticity following Ref.~\cite{Lepori:2020ifz}.

The convergence map is defined as the fluctuations of the area distance with respect to its background value. The real and imaginary components of the shear can be reconstructed from the ellipticity, convergence, and rotation, as reported in Eqs.~\eqref{eq:ellip1} and~\eqref{eq:ellip2}. We use the fully nonlinear shear--ellipticity relation and rescale our ellipticity so that the ellipticity and shear are identical in linear perturbation theory.

For comparison, we also extract convergence, shear, and rotation from maps of the deflection angle. The relation between the deflection angle and the weak lensing fields in the standard weak lensing formalism is described in Appendix~\ref{ap:deflection}.

We downgrade the resolution of all maps from the original $N_\mathrm{side} = 16384$ to $N_\mathrm{side} = 8192$, and extract the full-sky angular power spectra of the fields using the \texttt{anafast} estimator. The maps are initially produced at high resolution to reduce numerical noise in the spin-2 quantities; however, on the angular scales of interest, our results are insensitive to the final pixel resolution.

\begin{figure}[t]
    \centering
    \includegraphics[width=\textwidth]{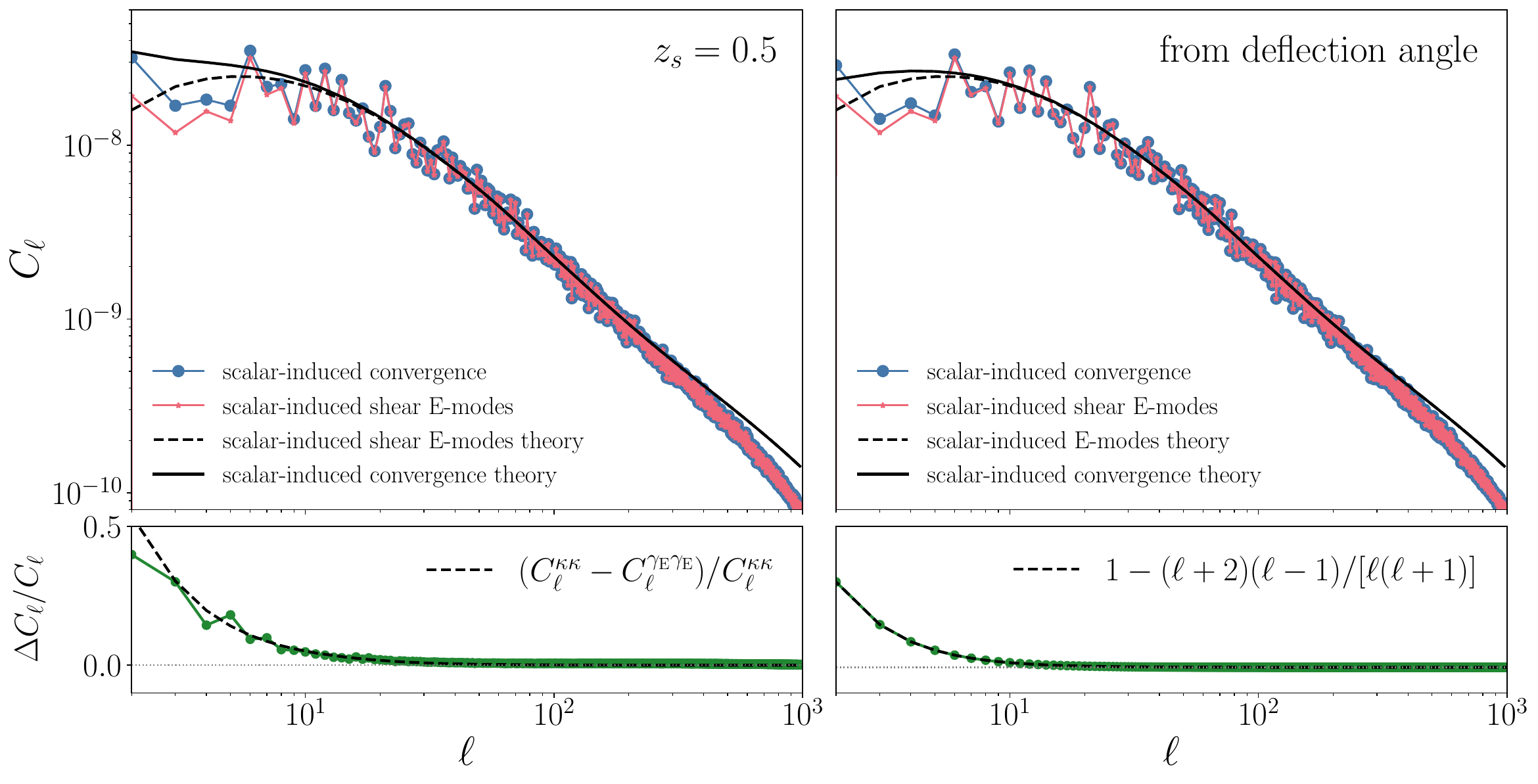}
    \caption{
        Angular power spectra of scalar-induced convergence and shear E-modes (top panels) and their relative difference $\Delta C_\ell/C_\ell \equiv (C_\ell^{\kappa\kappa} - C_\ell^{\gamma_{\rm E}\gamma_{\rm E}})/C_\ell^{\kappa\kappa}$ (bottom panels).
        \textbf{Left:} Simulated spectra are computed from the fully relativistic Jacobi map. 
        The theoretical prediction for the convergence includes the relativistic corrections from Eq.~\eqref{eq:true-kappa}.
        \textbf{Right:} Simulated spectra are extracted from maps of the deflection angle. Theoretical predictions are estimated using the standard weak lensing formalism and do not include relativistic corrections to the convergence.
        The source redshift is $z_{\rm s} = 0.5$. 
    }
    \label{fig:scalar_convergence_shearE}
\end{figure}

In this section, we compare the simulated angular power spectra with their theoretical predictions. For consistency with our simulation setup, all theoretical predictions are evaluated using the Fourier-space spectra of the Weyl potential $\Psi$ and the frame-dragging potential $B_\alpha$ at a fixed redshift, $z = 0.25$, corresponding to the simulation snapshot used for the ray-tracing.
We have verified that this approximation has a negligible impact on the results.

First, we focus on the scalar-induced power spectra, extracted from maps in which the vector potential $B_\alpha$ has been set to zero. In Fig.~\ref{fig:scalar_convergence_shearE}, we show the angular power spectra of the scalar-induced convergence and shear E-modes at a source redshift $z_{\rm s} = 0.5$, which receive non-zero contributions from linear scalar perturbations.
In the left panel, we show the spectra computed by solving the evolution equations for the full Jacobi matrix, whereas the right panel shows the same quantities extracted from the deflection angle maps.
The theoretical prediction is computed assuming the Limber approximation and the full-sky relation between the weak lensing fields and the scalar potential. 

As shown in Appendix~\ref{ap:deflection}, in the standard weak lensing formalism the convergence and shear E-modes share the same angular power spectrum in the flat-sky approximation, \emph{i.e.}, for $\ell \gg 1$. In full generality, the relative difference between the two is given by
\begin{equation}
    \left(\frac{\Delta C_\ell}{C_\ell}\right)_{\rm def}
    = 1 - \frac{(\ell + 2)(\ell - 1)}{\ell(\ell + 1)}  \,. \label{eq:ell-fact}
\end{equation}

This relation is exactly satisfied by construction for the spectra computed from the deflection angle, as shown in the bottom-right panel of Fig.~\ref{fig:scalar_convergence_shearE}, which displays the relative difference between convergence and shear E-modes. In Sec.~\ref{ref:psLin}, we showed that in the fully relativistic case this relation is not exact. We have numerically evaluated the correction to this relation, given by Eq.~\eqref{eq:true-kappa}. Comparing the left and right panels of Fig.~\ref{fig:scalar_convergence_shearE}, we see that the standard weak lensing formalism underestimates the convergence power spectrum and the difference between convergence and shear E-modes on large scales. We find excellent agreement between the relativistic corrections to the convergence power spectrum in our simulations and the theoretical expectations.

\begin{figure}[t]
    \centering
    \includegraphics[width=\textwidth]{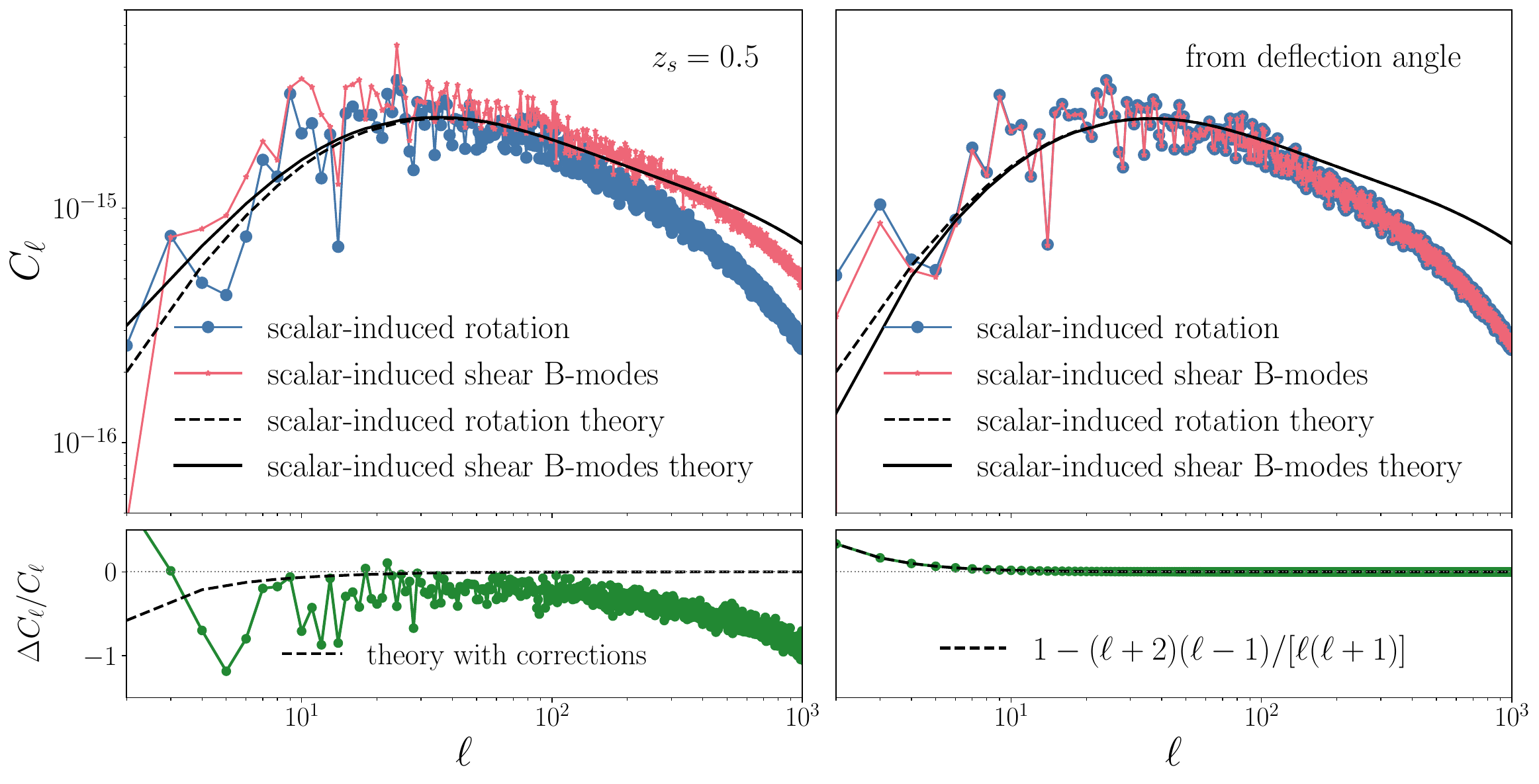}
    \caption{
        Angular power spectra of scalar rotation and shear B-modes (top panels) and their relative difference $\Delta C_\ell/C_\ell \equiv (C_\ell^{\omega\omega} - C_\ell^{\gamma_{\rm B}\gamma_{\rm B}})/C_\ell^{\omega\omega}$ (bottom panels).
        \textbf{Left:} Simulated spectra are computed from the fully relativistic Jacobi map. 
        \textbf{Right:} Simulated spectra are extracted from maps of the deflection angle. 
        The source redshift is $z_{\rm s} = 0.5$. 
    }
    \label{fig:scalar_rotation_shearB}
\end{figure}

In Fig.~\ref{fig:scalar_rotation_shearB} we consider the rotation and shear B-modes sourced by scalar perturbations. These quantities vanish identically in linear perturbation theory, and therefore a second-order computation is required to estimate their spectra. As discussed in Sec.~\ref{ref:ps}, the standard weak lensing formalism predicts that the rotation and shear B-modes share the same power spectrum in the limit $\ell \gg 1$, with their relative difference determined solely by the $\ell$-dependent factor in~\eqref{eq:ell-fact}.
In Sec.~\ref{ref:ps}, we show that this equivalence does not hold in the full Jacobi map formalism, and we provide for the first time the analytical expressions for all relativistic corrections that contribute to the angular power spectrum of the shear B-modes but not to the rotation.
We numerically evaluate the theoretical prediction for the shear B-mode power spectrum, including these corrections.
These corrections take a form similar to the convolution integral required to estimate the rotation, and we present the details of the numerical implementation in Appendix~\ref{ap:numer-th}.
 Relativistic corrections become significant on large angular scales and are of order $\sim 5\%$ on scales $\ell \approx 5$. The analytical calculations qualitatively reproduce the simulated power spectra on scales $\ell < 100$. In particular, while the standard weak lensing formalism predicts that the rotation angular power spectrum has a systematically larger amplitude than the shear B-modes, in the fully relativistic estimate the opposite is true.
Nevertheless, we stress that the discrepancy between the standard formalism and the correct estimate arises only on very large angular scales, and it will be extremely challenging to detect. On the scales relevant for current weak lensing surveys, such as \textit{Euclid} and LSST, the equivalence between the shear B-modes and the rotation angular power spectra is an excellent approximation. On small angular scales, there is a large discrepancy between the simulated data and the theoretical model due to resolution effects.

\begin{figure}[t]
    \centering
    \includegraphics[width=\textwidth]{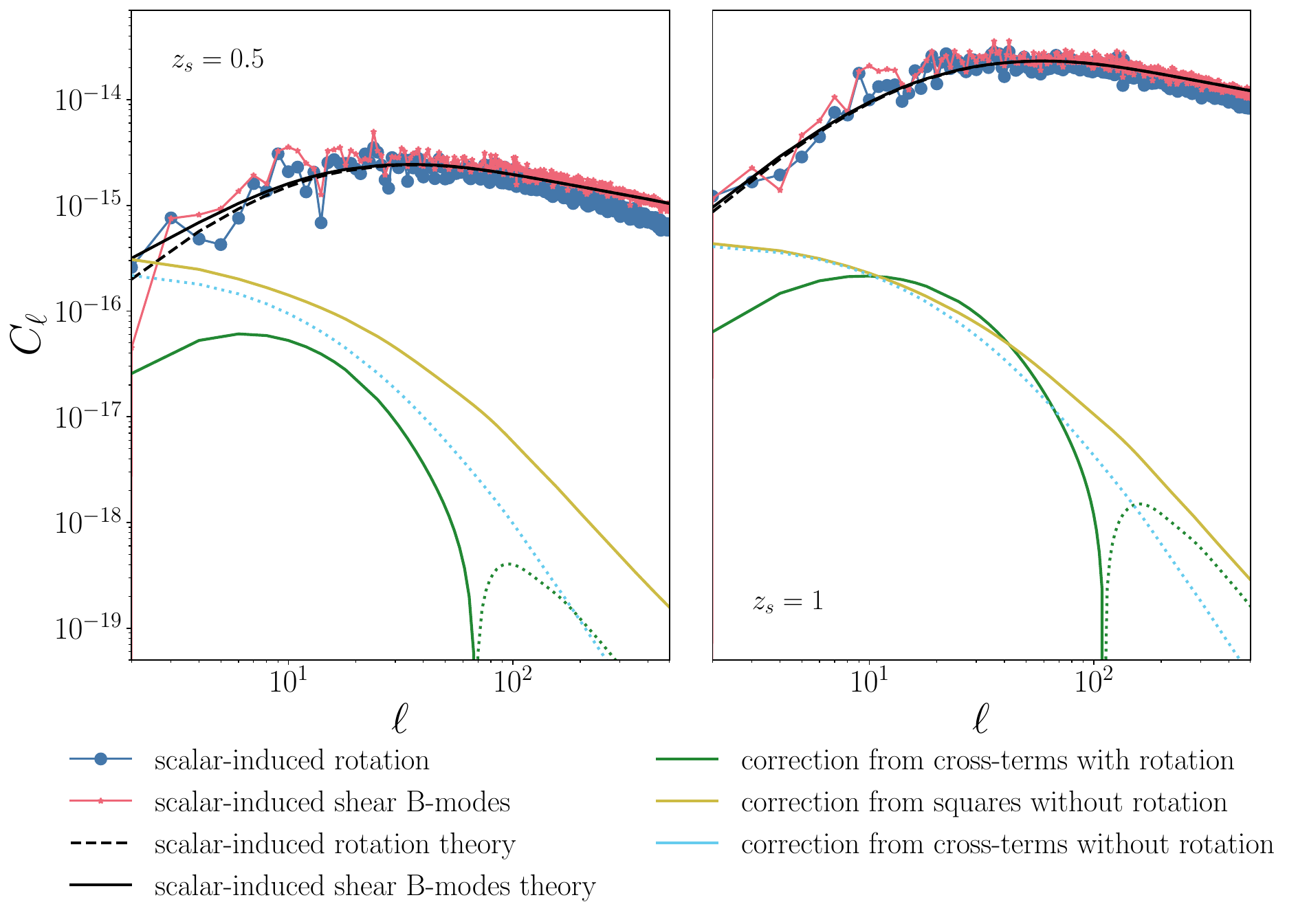}
    \caption{
        Contributions to the angular power spectrum of the scalar shear B-modes. 
        Dotted lines indicate negative values of the relativistic corrections.
        \textbf{Left:} Source redshift is $z_{\rm s} = 0.5$. 
        \textbf{Right:} Source redshift is $z_{\rm s} = 1$.  
    }
    \label{fig:scalar_rotation_shearB_contr}
\end{figure}

In Fig.~\ref{fig:scalar_rotation_shearB_contr}, we compare the relevance of the relativistic corrections at $z_{\rm s} = 0.5$ and $z_{\rm s} = 1$.
The relativistic contributions can be grouped into three types: corrections in which the rotation-like term is cross-correlated with the additional shear contributions (green line), auto-correlations of the additional shear contributions (squares, yellow line), and cross-correlations among the different relativistic corrections (light blue).
At low redshift, the overall relativistic contribution is dominated by the auto-correlations of the additional shear contributions.
Indeed, all of these corrections are positive, whereas the cross-correlation terms can be either positive or negative and partially cancel each other out.
We also find that the relative importance of the relativistic corrections to the total power spectrum is larger at lower redshift, where the overall lensing signal is smaller.

\begin{figure}[t]
    \centering
    \includegraphics[width=\textwidth]{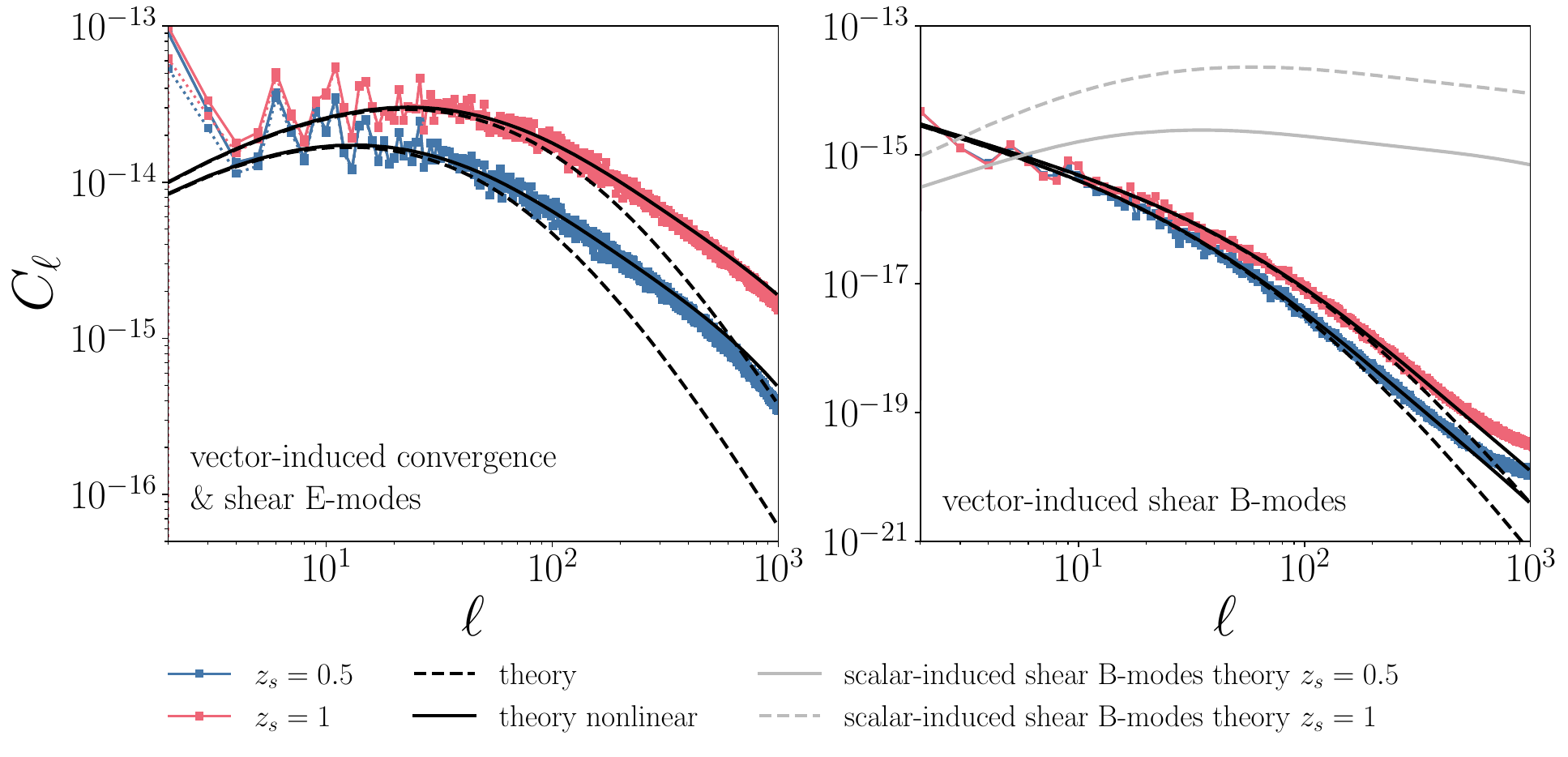}
    \caption{
Frame-dragging contributions to the weak lensing observables at source redshift $z_{\rm s} = 0.5$ and $z_{\rm s} = 1$.
\textbf{Left:} Lensing convergence (solid lines) and shear E-modes (dotted lines).
\textbf{Right:} Shear B-modes. Frame dragging does not contribute at leading order to the rotation, and its contribution is approximately ten orders of magnitude smaller than the frame-dragging contribution to the shear B-modes. For this reason, it is not shown in this plot.
    }
    \label{fig:vector_kappa_shear}
\end{figure}

In general relativity, the metric contains scalar-, vector-, and tensor-type degrees of freedom. In the remainder of this section, we quantify the contributions of vector-type perturbations to weak lensing observables by comparing their analytical predictions with our simulation data. This work presents the first simulations of frame-dragging contributions to weak lensing observables.
In Fig.~\ref{fig:vector_kappa_shear}, we show the frame-dragging contributions to the lensing convergence and to the cosmic shear E and B-modes.
Vector perturbations do not contribute to the rotation at leading order, and their subleading contribution is very subdominant, being approximately ten orders of magnitude smaller than the contribution to the shear B-modes.
Therefore, in the fully relativistic treatment, vector-induced shear B-modes and image rotation do not share the same angular power spectrum, as is also the case for scalar perturbations.
The frame-dragging contribution to convergence and shear E-modes, similar to the scalar-induced spectra, is the same on $\ell \gg 1$, and, in agreement with the theoretical expectation.  
The vector-induced convergence and shear E-modes are approximately seven orders of magnitude smaller than their scalar-induced counterparts, making them a highly subdominant contribution for current weak lensing surveys on all scales. In the right panel of Fig.~\ref{fig:vector_kappa_shear}, we see that the contribution of frame dragging to shear B-modes becomes comparable to the second-order scalar contributions on large angular scales, $\ell \lesssim 10$. Furthermore, the relative importance of the signal is greater at low redshift, as the scalar-induced contributions grow with redshift faster than the vector-induced ones on these scales.

\begin{figure}[t]
    \centering
    \includegraphics[width=0.7\textwidth]{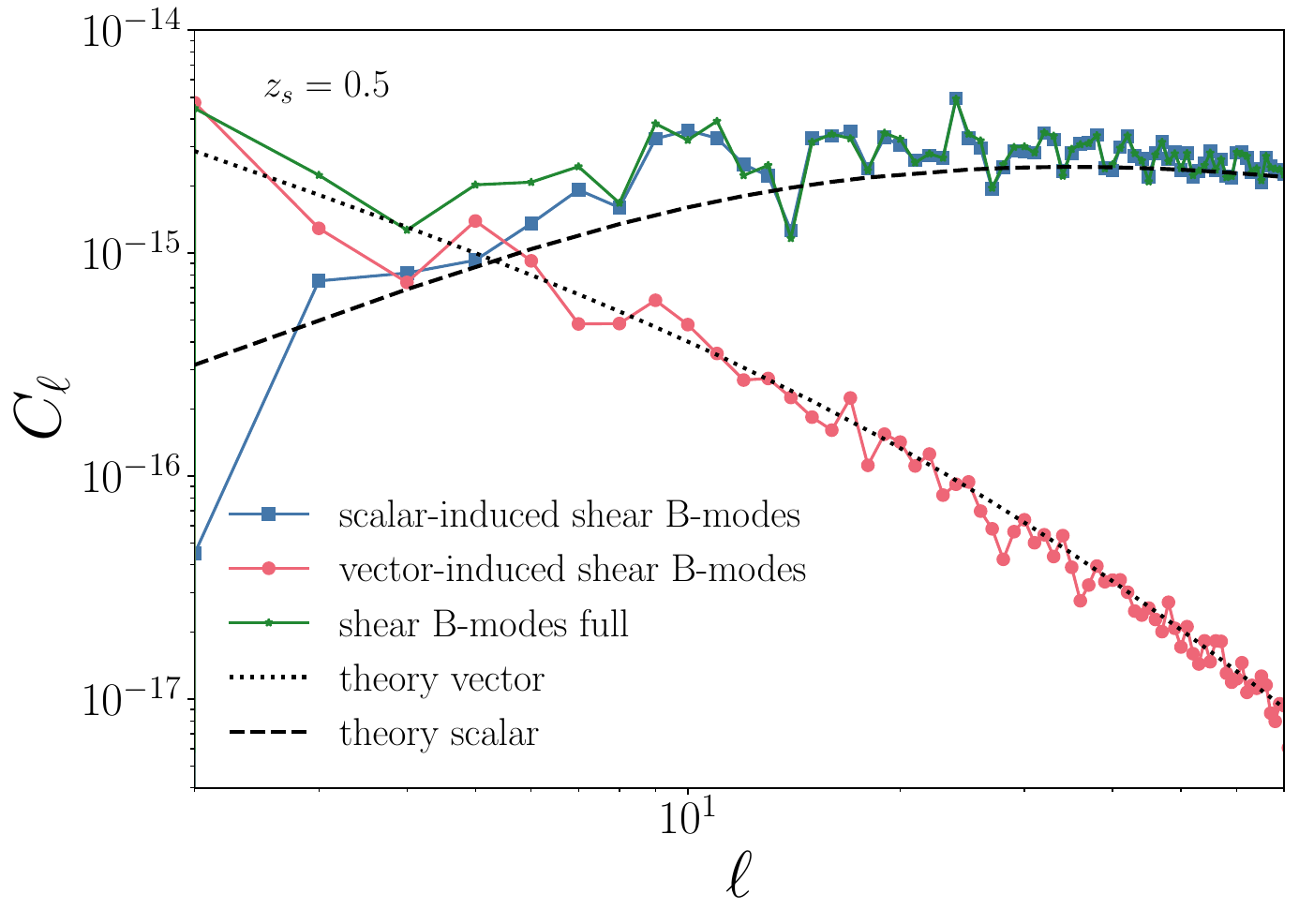}
    \caption{
        Shear B-modes induced by scalar perturbations (blue data), vector perturbations (red data), and the full observable including both contributions and their cross-correlations (green data).
        The source redshift is $z_{\rm s} = 0.5$. 
    }
    \label{fig:vector_shearB_z05}
\end{figure}

So far, we have discussed the scalar- and vector-induced contributions to the power spectra separately. However, both affect weak lensing observations, and therefore their cross-correlations also contribute to the total power spectra. In Fig.~\ref{fig:vector_shearB_z05}, we compare the scalar-induced (blue data) and vector-induced (red data) shear B-modes with the simulated spectra that include both contributions and their cross-correlation (green data). We find that the full power spectrum is well approximated by the sum of the individual scalar- and vector-induced contributions, suggesting that their cross-correlation is small.

\begin{figure}[t]
    \centering
    \includegraphics[width=\textwidth]{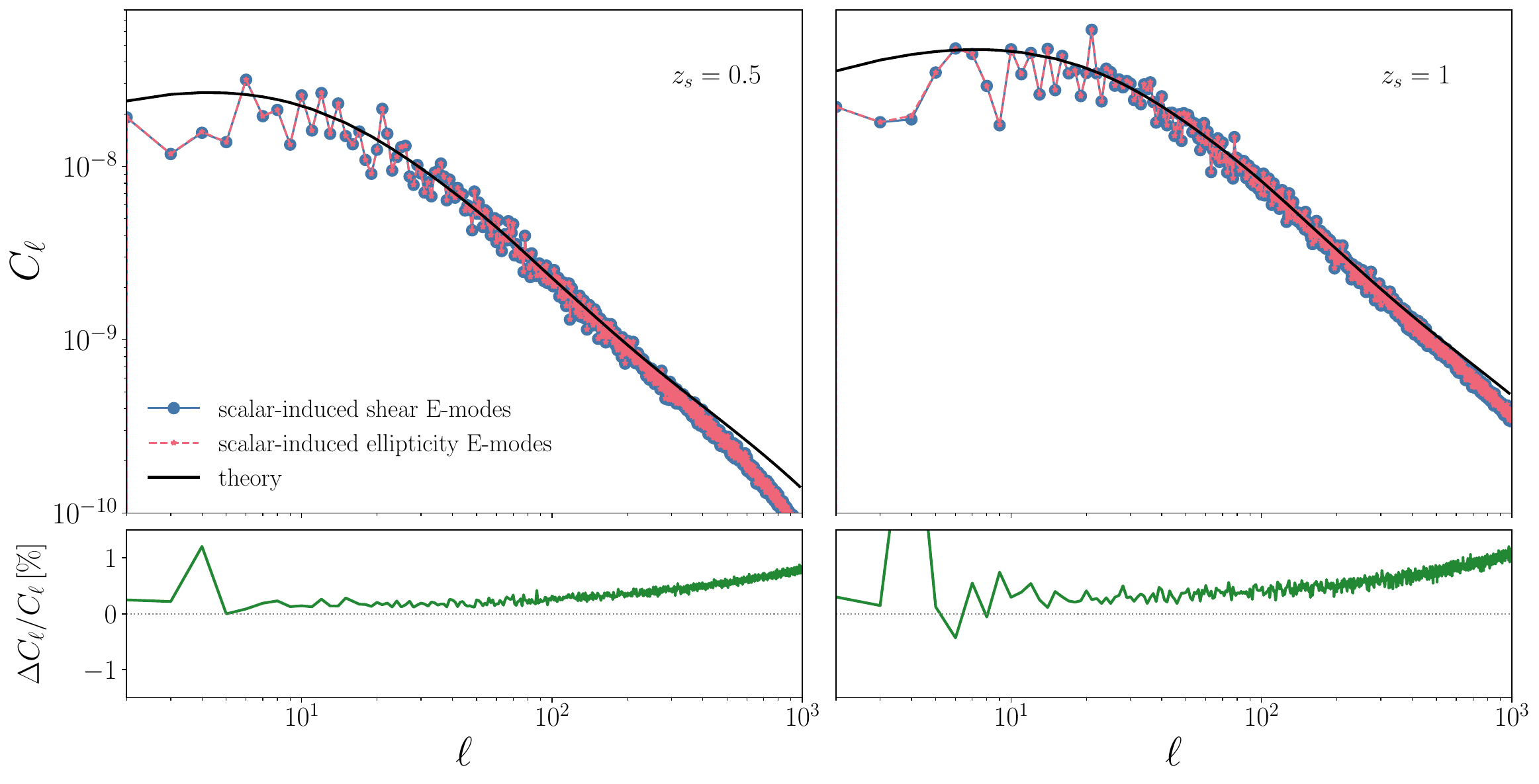}
    \caption{
    Impact of the `reduced shear' correction on the ellipticity E-modes at source redshift $z_{\rm s} = 0.5$ (left panel) and $z_{\rm s} = 1$ (right panel).
{\bf Top}: Angular power spectra of the scalar-induced E-modes for the shear and the ellipticity.
A black line indicates our theory prediction, which includes only the linear contribution estimated from the nonlinear power spectrum modelled with the \textit{Halofit} prescription.
{\bf Bottom}: Relative difference $\Delta C_\ell/C_\ell \equiv (C_\ell^{\epsilon_{\rm E}\epsilon_{\rm E}} - C_\ell^{\gamma_{\rm E}\gamma_{\rm E}})/C_\ell^{\epsilon_{\rm E}\epsilon_{\rm E}}$ (in percentage) of the two simulated power spectra.
    }
    \label{fig:ellipticity_shearE}
\end{figure}

\begin{figure}[t]
    \centering
    \includegraphics[width=\textwidth]{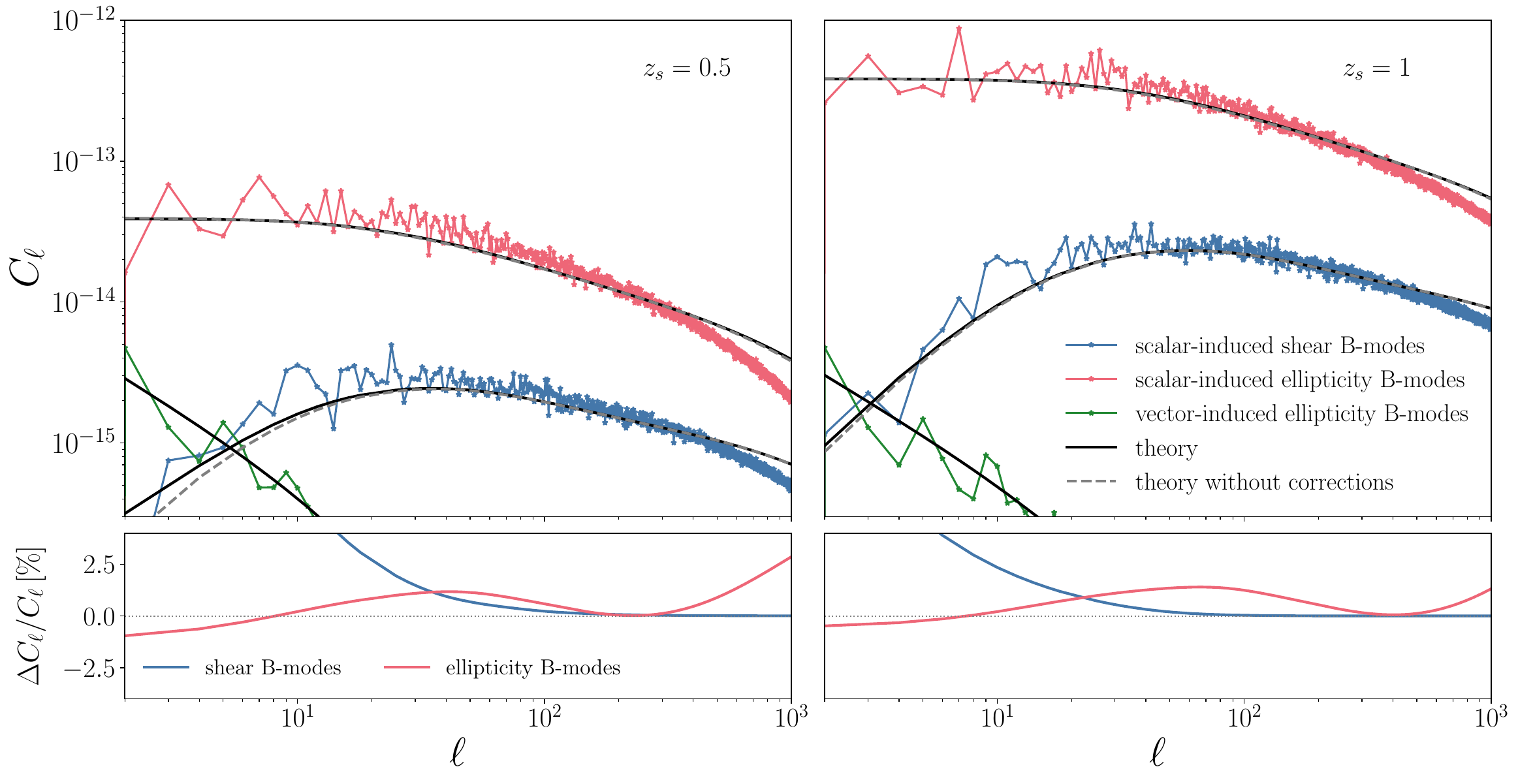}
    \caption{
    Impact of the `reduced shear' correction on the ellipticity B-modes at source redshift $z_{\rm s} = 0.5$ (left panel) and $z_{\rm s} = 1$ (right panel).
{\bf Top}: Angular power spectra of the scalar-induced B-modes for the shear and the ellipticity.
We also show, for comparison, the vector-induced ellipticity B-modes, which are only visible on large angular scales. Black lines indicate our theory predictions for these three quantities, including the nonlinear relativistic corrections discussed in Sec.~\ref{ref:ps}, while the grey lines represent the theoretical estimates from the standard weak lensing formalism.
{\bf Bottom}: Contribution of the scalar-induced nonlinear relativistic correction to the shear and ellipticity B-modes from our analytical estimate. 
For the shear B-modes, this relative contribution is given by
$\Delta C_\ell/C_\ell \equiv \left[\left(C^{\gamma_\mathrm{B}\gamma_\mathrm{B}}_\ell\right)_{\rm S}-C^{\omega\omega}_\ell\right]/\left(C^{\gamma_\mathrm{B}\gamma_\mathrm{B}}_\ell\right)_{\rm S}$.  For the ellipticity B-modes, it is defined as
$\Delta C_\ell/C_\ell \equiv \left[\left(C^{\epsilon_\mathrm{B}\epsilon_\mathrm{B}}_\ell\right)_{\rm S}-C^{\omega\omega}_\ell-
C^{\epsilon_\mathrm{B}\epsilon_\mathrm{B}}_{(\kappa \ast \gamma _{2} )^2}
\right]/\left(C^{\epsilon_\mathrm{B}\epsilon_\mathrm{B}}_\ell\right)_{\rm S}$, where $C^{\epsilon_\mathrm{B}\epsilon_\mathrm{B}}_{(\kappa \ast \gamma _{2} )^2}$
is the main contribution to the ellipticity B-modes power spectrum, given by Eq.~\eqref{eq:eps-main}.
    }
    \label{fig:ellipticity_shearB}
\end{figure}

As discussed in Sec.~\ref{ellipt}, weak lensing surveys do not measure the shear of galaxy images directly, but rather the galaxy ellipticity. 
In linear theory, these two quantities are equivalent (up to a constant factor that depends on the convention used to define the ellipticity\footnote{As previously noted, we have rescaled the ellipticity in such a way that ellipticity and shear are identical in linear perturbation theory.}).
At second order in perturbation theory, however, these two quantities differ and the correction to the observed ellipticity due to this difference is called `reduced shear'~\cite{Shapiro:2008yk, Krause:2009yr}.
Using our simulation framework, we have extracted ellipticity E- and B-modes and quantified the impact of the `reduced shear' non-perturbatively.

In Fig.~\ref{fig:ellipticity_shearE} we compare the shear and ellipticity E-modes. Since the E-modes are dominated by the linear contribution, the `reduced shear' correction to the power spectra is subdominant, reaching the percent level on scales $\ell \gtrsim 1000$. This does not necessarily imply that the effect can be neglected, since upcoming surveys such as \textit{Euclid} and LSST will measure E-modes with very high precision down to small angular scales~\cite{Deshpande:2019sdl}. Note that on these scales our simulations are significantly affected by resolution effects and cannot be fully trusted.
In Fig.~\ref{fig:ellipticity_shearB} we compare the shear and ellipticity B-modes.
Since the shear B-modes are identically zero in linear perturbation theory, the `reduced shear' contributes to the ellipticity B-modes at leading order and represents the dominant contribution to the signal, roughly one to two orders of magnitude larger than the shear B-modes.
In the bottom panel, we compare the contribution of the nonlinear (scalar-induced) relativistic corrections to the shear and ellipticity B-modes. While the relative impact of these relativistic corrections becomes substantial on large angular scales for the shear B-modes, they affect the ellipticity only at the percent level on all scales.
The `reduced shear' contribution arises from a convolution of the linear convergence and shear, and therefore acts as a contaminant for the relativistic signals in the ellipticity power spectrum. Detecting these small relativistic effects will require precise modelling and removal of this dominant contribution. Strategies analogous to CMB B-mode delensing~\cite{Belkner:2023duz} could be employed: high-precision measurements of shear E-modes, potentially combined with independent constraints from lensing magnification, can be used to reconstruct and subtract the `reduced shear' contribution. Developing such techniques to accurately remove this signal will be necessary to pursue the detection of the subtle nonlinear relativistic effects studied in this work.

\section{Conclusions}

\label{ref:conc}

In this paper we computed the angular power spectra of the weak lensing observables, namely convergence, image rotation, and cosmic shear, in a fully relativistic framework up to second order in perturbation theory. For convergence and shear E-modes, we evaluated the leading-order frame-dragging contribution from vector modes sourced by scalar perturbations in $\Lambda$CDM. For shear B-modes, we performed a complete second-order calculation. We derived for the first time the full relativistic expression for the angular power spectrum including both scalar and vector perturbations, and evaluated these contributions numerically. Importantly, our approach captures relativistic effects that are absent in the standard weak lensing formalism.

Indeed, the standard formalism constructs lensing observables solely from light propagation in a perturbed spacetime. In doing so, it lacks a physical definition of intrinsic sizes and shapes in the local rest frame and does not provide a consistent way to compare separations at different spacetime positions. These limitations are resolved in the Jacobi map approach adopted here, where rest-frame quantities are defined using tetrads and geometric properties are compared through the parallel transport of the Sachs basis along the photon trajectory.

The additional terms arising in the Jacobi map approach compared to the standard formalism restore the gauge invariance of the lensing observables, as manifest in Eq.~($\ref{ampliJacob}$). Moreover, they encode physical effects that go beyond the lens–lens couplings and post-Born corrections already included in the standard formalism at second order in perturbation theory~\cite{Dodelson:2005zj}. Purely general-relativistic effects arise from the expansion of the Riemann tensor beyond linear order, as well as from couplings between the lens and perturbations at the observer and source positions.

Since the relativistic contributions enter the different observables in a distinct manner, the shear B-mode statistics are no longer identical to those of image rotation, in contrast with the prediction of the standard formalism. Our main analytical result is presented in Eq.~($\ref{CBmodes}$), where we compute the difference between the shear B-mode angular power spectrum and that of image rotation.
The calculation is performed in the Poisson gauge, retaining the dominant contributions while neglecting subdominant effects such as peculiar velocities, time evolution of gravitational potentials and anisotropic stress. Perturbations at the observer position are also neglected, as they do not affect higher multipoles. Furthermore, we worked in the flat-sky limit and under the Limber approximation. Despite these approximations, the analytical results are in good agreement with the full numerical computation, even at relatively low multipoles (see Fig.~\ref{fig:scalar_rotation_shearB} and Fig.~\ref{fig:scalar_rotation_shearB_contr}).

We find that these scalar-induced relativistic effects become relevant at large angular scales and low redshift; for example, at source redshift $z_{\rm s} = 0.5$, their contribution is approximately $5\%$ of the shear B-mode signal on scales $\ell \approx 5$.

Moreover, we have carried out an analytical and numerical study of the frame-dragging effect in weak lensing. While the contribution of frame dragging to the convergence and rotation fields is very subdominant, its contribution to the shear is comparable to the scalar-induced B-modes on large angular scales. Whereas in our analytical calculations we do not attempt to model the cross-correlation of scalar-induced and vector-induced B-modes, our numerical simulations consistently include them, and a direct comparison then indicates that these terms can be neglected. In general, we find that the relative importance of frame dragging is larger at low redshift, where the overall lensing signal is smaller and may be harder to measure.

When discussing the relative impact of frame dragging, it is important to keep in mind that our numerical results pertain to the $\Lambda$CDM model where frame dragging is naturally suppressed due to the non-relativistic nature of the matter sector. Relativistic sources of stress-energy, for example a dynamical dark energy, could potentially give rise to much larger vector perturbations. Our analytic formulas can be used to make predictions for the weak lensing observables in such scenarios.

We have also presented the computation of the ellipticity B-modes from $N$-body simulations. This is the quantity that cosmic shear surveys actually measure. We have shown that the main contribution to the ellipticity B-modes comes from a second-order effect known in the literature as `reduced shear'~\cite{Krause:2009yr}, which is the convolution of the linear convergence and shear fields. Since the linear convergence and shear fields can be accurately measured from the ellipticity E-modes, there is hope that adapted techniques will be able to model and remove this signal in order to reconstruct the shear field and attempt to detect frame dragging on cosmological scales.

This work can be extended in several directions. In our analytical and numerical framework, we have not included the effect of source redshift clustering, which can also source shear B-modes~\cite{Schneider:2001af}, and we have neglected astrophysical effects that can act as contaminants to our signal, such as intrinsic alignment~\cite{Georgiou:2023xgo}. Comparing these effects with the relativistic contributions studied here will be important.\footnote{An alternative approach to mitigating intrinsic alignment systematics has been proposed in~\cite{Francfort:2022laa}, where the authors introduce an estimator for cosmic shear that exploits polarization information from radio galaxies and therefore does not require the assumption that galaxy ellipticities are uncorrelated.} Finally, while this work presents a first analytical and numerical estimation of the frame-dragging contribution to shear B-modes, a quantitative study of the detectability of this signal is left to future work. Detecting frame dragging on cosmological scales would provide a unique test of general relativity in the weak-field, large-scale regime, complementing existing local and astrophysical constraints~\cite{Everitt:2011hp}.

\acknowledgments

FL thanks Giulio Fabbian for useful discussions. FL and JA acknowledge financial support from the Swiss National Science Foundation. MM was supported by IBS under the project code IBS-R018-D3. FL acknowledges additional support from UZH Candoc/Postdoc
Grant FK-23-125. JA acknowledges additional support from the Dr.\ Tomalla Foundation for Gravity Research. This work was supported by a grant from the Swiss National Supercomputing Centre (CSCS) under project ID sm97 on \textit{Alps}.

\appendix

\section{Standard weak lensing formalism}
\label{ap:deflection}

In the standard weak lensing formalism, the lensing map is fully determined by the deflection angle~$\bm{\alpha}(\bn)$, which relates the observed (lensed) angular position $\bn$ to the true (unlensed) position $\bn'$ via
\be
\bn' = \bn + \bm{\alpha}(\bn)\,.
\ee

The deflection angle is a vector field on the sphere and can be generically decomposed into a gradient and a curl component:
\be
\bm{\alpha}(\bn) = \bnabla \psi(\bn) - \bnabla \times \Omega(\bn)\,,
\ee
where $\psi$ is a scalar potential and $\Omega$ is a pseudo-scalar potential. The gradient component is responsible for lensing convergence and shear, while the curl component accounts for lensing rotation, which vanishes in linear scalar perturbation theory but may arise from vector and tensor modes, or from nonlinear effects.

Since $\bm{\alpha}(\bn)$ is a spin-1 field, it can be expressed in terms of the spin-raising and spin-lowering operators $\spl$ and $\smin$ (as defined, e.g., in~\cite{Durrer_2020}) as
\be
\bm{\alpha}(\bn) = - \spl \left[\psi(\bn) + \ii\, \Omega(\bn)\right]\,. \label{eq:def-spins}
\ee

The Jacobian of the lensing map defines the so-called amplification matrix $\mathcal{A}$, which encodes the local distortions of the image:
\be\label{stdobs}
\mathcal{A} \equiv \left(\!\begin{array}{cc} 1-\ka-\gga_1 & \omega -\gga_2 \\ -\omega -\gga_2 & 1-\ka+ \gga_1\end{array}\!\right) \equiv \frac{\partial \bn'}{\partial \bn} = \mathbb{I}_2 + \frac{\partial \bm{\alpha}(\bn)}{\partial \bn}\,,
\ee
where $\mathbb{I}_2$ is the $2{\scriptstyle \times} 2$ identity matrix, $\kappa$ denotes the convergence, $\gamma_1$ and $\gamma_2$ the two components of the shear, and $\omega$ represents the rotation. 

These quantities can be obtained from the deflection angle using the spin-raising and spin-lowering operators~\cite{Carron:2025qxj}:
\begin{align}
&\kappa(\bn) + \ii\,\omega(\bn) = \frac{1}{2} \smin \bm{\alpha}(\bn)\, , \\
&\gga_1(\bn) + \ii\,\gga_2(\bn) =  \frac{1}{2} \spl \bm{\alpha}(\bn)\,.
\end{align}
Substituting Eq.~\eqref{eq:def-spins}, we obtain:
\begin{align}
\kappa(\bn) &= -\frac{1}{2} \smin \spl \psi(\bn)\,, \\
\omega(\bn) &= -\frac{1}{2} \smin \spl \Omega(\bn)\,, \\
\gga_1(\bn) + \ii \gga_2(\bn) &= \left(-\frac{1}{2} \spl \spl \psi(\bn) \right) + \ii  \left(-\frac{1}{2} \spl \spl \Omega(\bn) \right)\,, \\
\gga_1(\bn) - \ii \gga_2(\bn) &=  \left(-\frac{1}{2} \smin \smin \psi(\bn) \right) + \ii  \left(\frac{1}{2} \smin \smin \Omega(\bn) \right)\,.
\label{eq:shear-pot}
\end{align}
Each of the relevant lensing quantities can be decomposed in terms of spin-weighted spherical harmonics ${}_s Y_{\ell m}$:
\begin{align}
& \bm{\alpha}(\bn) = -\sum_{\ell, m} \left(G_{\ell m} + \ii C_{\ell m}\right)\,{}_1Y_{\ell m}(\bn)\,, \\
& \psi(\bn) = \sum_{\ell, m} \psi_{\ell m}\, {}_0Y_{\ell m}(\bn)\,, \\
& \ka(\bn) = \sum_{\ell, m} \ka_{\ell m}\, {}_0Y_{\ell m}(\bn)\,, \\
& \omega(\bn) = \sum_{\ell, m} \omega_{\ell m}\, {}_0Y_{\ell m}(\bn)\,, \\
& \gga_1(\bn) \pm \ii \gga_2(\bn) = \sum_{\ell, m} A_{\pm 2, \ell m}\, {}_{\pm 2}Y_{\ell m}(\bn) = 
\sum_{\ell, m} \left(\gga^{\rm E}_{\ell m} \pm \ii \gga^{\rm B}_{\ell m}\right) \, {}_{\pm 2}Y_{\ell m}(\bn)\,.
\end{align}
To relate these harmonic coefficients, we make use of the following identities for the spin-raising and lowering operators~\cite{Durrer_2020}:
\begin{align}
 & \spl \left( {}_s Y_{\ell m}\right) = \sqrt{(\ell - s) (\ell + s + 1)}\,{}_{s+1} Y_{\ell m}\,, \\
 & \smin \left({}_s Y_{\ell m}\right) = - \sqrt{(\ell + s) (\ell - s + 1)}\,{}_{s-1} Y_{\ell m}\,. \label{eq:spin-rel}
\end{align}
Applying these relations yields:
\begin{align}
& \ka_{\ell m} = \frac{1}{2} \ell\,(\ell + 1) \psi_{\ell m} = \frac{1}{2} \sqrt{\ell\,(\ell + 1)}\,G_{\ell m}\,, \label{eq:ka_lm}\\
& \omega_{\ell m} = \frac{1}{2} \ell\,(\ell + 1) \Omega_{\ell m} = \frac{1}{2} \sqrt{\ell\,(\ell + 1)}\,C_{\ell m}\,, \label{eq:om_lm} \\
 & \gga^{\rm E}_{\ell m} = - \frac{1}{2} (A_{2, \ell m} + A_{-2, \ell m}) = \frac{1}{2} \sqrt{\frac{(\ell + 2)!}{(\ell - 2)!}} \psi_{\ell m} = \frac{1}{2} \sqrt{(\ell+2)(\ell -1)} G_{\ell m}\,, \label{eq:gaE_lm}\\
& \gga^{\rm B}_{\ell m} = - \frac{1}{2\ii} (A_{2, \ell m} - A_{-2, \ell m}) = \frac{1}{2} \sqrt{\frac{(\ell + 2)!}{(\ell - 2)!}} \Omega_{\ell m} = \frac{1}{2} \sqrt{(\ell+2)(\ell -1)} C_{\ell m}\,.
\label{eq:gaB_lm}
\end{align}
Thus, up to an $\ell$-dependent normalisation that tends to unity at high multipoles ($\ell \gg 1$), the rotation and shear B-mode components share the same power spectrum, and similarly, the convergence and shear E-mode spectra are equivalent.

    \section{Parallel transport of the Sachs basis}
\label{app:paralleltransport}
To solve Eq.~\eqref{JacobConf}, we require an expression for the conformal screen basis~$\hat e^\mu_I(\lambda)$ at a generic point along the null congruence. This can be derived by solving the evolution equation for $D\hat e^\mu_I/D\lambda=\hat k^\nu\hat{\nabla}_\nu\hat e^\mu_I$, which follows from the fact that the Sachs basis satisfies the parallel-transport condition:~$D e^\mu_I/D\Lambda=0$. A direct computation yields
    \begin{equation}
    \frac{D}{D\lambda}\hat e^\mu_I = \frac{D}{D\lambda}\left( \Omega e^\mu_I \right) =\hat e^\mu_I \hat k^\nu \partial_\nu \ln \Omega + \Omega\, \hat k^\nu \hat \nabla_\nu e^\mu_I\,.
    \end{equation}
The last term can be simplified by expressing the covariant derivative compatible with~$\hat g_{\mu\nu}$ in terms of the covariant derivative compatible with~$g_{\mu\nu}$, making the parallel transport condition manifest. The relationship between covariant derivatives of vector fields for conformally related metrics is given by~\cite{Wald:1984rg}
    \begin{equation}
    \hat \nabla_\nu e_I^\mu = \nabla_\nu e_I^\mu + C^\mu_{\nu\rho} e^\rho_I\,,
    \qquad\quad
    C^\mu_{\nu\rho} =  - \delta^\mu_\nu \partial_\rho \ln \Omega -\delta^\mu_\rho \partial_\nu \ln \Omega
    +g_{\nu\rho}  g^{\mu\sigma} \partial_\sigma \ln \Omega\,,
    \end{equation}
where~$C^\mu_{\nu\rho}$ is the difference between the Christoffel symbols $C^\mu_{\nu\rho}=\hat \Gamma^{\mu}_{\nu\rho}-\Gamma^\mu_{\nu\rho}$ based on~$\hat g_{\mu\nu}$ and~$g_{\mu\nu}$, respectively. By direct substitution, we obtain
    \begin{equation}
    \frac{D}{D\lambda}\hat e^\mu_I = (\hat k_\rho \hat e_I^\rho) \hat g^{\mu\nu} \partial_\nu \ln \Omega - \hat k^\mu \hat e^\nu_I \partial_\nu \ln \Omega\,.
    \end{equation}
The term in round brackets is proportional to~$k_\rho e^\rho_I$, which is zero for our choice of screen basis (see the discussion in Sec.~\ref{sub:jacobimap}). Thus, our task reduces to solving
    \begin{equation}\label{notparallel}
    \frac{D}{D\lambda}\hat e^\mu_I + \hat k^\mu \hat e^\nu_I \partial_\nu \ln \Omega = 0\,.
    \end{equation}
To proceed, we first introduce a convenient parametrisation of~$\hat e^\mu_I$ that satisfies the orthogonality condition~$\hat k_\rho \hat e^\rho_I = 0$. Since~$\hat k^\mu$ is a null vector, one may consider a decomposition of the form $\hat e^\mu_I \supset A_i \Phi^i_I \hat k^\mu$. However, this alone is not sufficient, since the normalisation condition for the screen basis $\hat g_{\mu\nu}\hat e^\mu_i \hat e^\nu_j = \delta_{ij}$, must also be satisfied. To ensure both orthogonality and normalisation, we introduce~$E^\alpha_I = E^\alpha_i \Phi^i_I$, a set of orthonormal spacelike vectors defined by
    \begin{equation}
    \hat g_{\alpha\beta}E^\alpha_I E^\beta_J = \delta_{IJ}\,, \qquad\qquad\quad \hat k_\alpha E^\alpha_I = 0\,,
    \end{equation}
such that the conformally projected screen basis takes the form
    \begin{equation}\label{param}
    \hat e^\mu_I = \hat e^\mu_i \Phi^i_I = \left(A_i \hat k^0\,,\, E^\alpha_i + A_i \hat k^\alpha \right)^\mu \Phi^i_I\,.
    \end{equation}

Although Eqs.~\eqref{notparallel}–\eqref{param} are exact, our interest is in first-order perturbative solutions around an FLRW background, with the spacetime metric specified in Eqs.~\eqref{metric}-\eqref{SVT}. In perturbation theory, the conformal wave vector~$\hat k^\mu$ is decomposed into a background contribution plus perturbations and can be parametrised as follows:
    \begin{equation}\label{hatk}
    \hat k^\mu = (1 + \delta \nu, -n^\alpha - \delta n^\alpha)^\mu\,,
    \end{equation}
where $\delta\nu$ represents the frequency perturbation and $\delta n^\al$ the perturbation in the spatial propagation direction, both determined by integrating the null geodesic equation. The quantities $A_i$ and $E^\al_i$ can also be expanded perturbatively. Since $\hat e_i$ form a spacelike triad, only~$E^\al_i$ has a non-vanishing background contribution, while~$A_i$ is purely perturbative.

Substituting the parametrisation of the wave vector given in Eq.~\eqref{hatk} into Eq.~\eqref{param}, we obtain the first-order expression for the screen basis at a generic point along the null congruence
    \begin{equation}\label{paramcc}
    \hat e^\mu_I = \big[A_i\,,\, \delta^\alpha_i - n^\alpha (\delta n_i + A_i) \big]^\mu \Phi^i_I~,
    \end{equation}
where~$E^\al_i$ has been written in terms of the perturbation in the photon propagation direction, left implicit for the moment.

To uniquely solve Eq.~\eqref{notparallel}, we must specify an initial condition, which we take to be the value of $\hat e^\mu_I$ at the observer position $O$. At this point, we impose that the timelike tetrad vector coincides with the observer four-velocity, $\hat e^\mu_0 \equiv (1 - 2\Psi_{\rm o},\, \mathcal{U}_{\rm o}^\alpha)^\mu$, where the temporal component is fixed by the normalisation condition $u_\mu u^\mu=-1$. The spatial triad vectors $\hat e^\mu_i$ are, by construction, orthogonal to~$\hat e^\mu_0$. This implies that at the observer position, the coefficient $A_i$ is determined by the observer peculiar velocity:    \begin{equation}\label{sachsO}
    \hat e^\mu_I{}_{\rm o} = \big[(\mathcal{U}_{i})_{\rm o}\,, \delta^\alpha_i - n^\alpha (\delta n_i + \mathcal{U}_i)_{\rm o} \big]^\mu \Phi^i_I\,.
    \end{equation}
To express the initial condition explicitly in terms of metric and matter perturbations, we derive the light wave vector $\hat k^\mu_{\rm o}$ at the observer position.
The relationship between the conformal wave vector $\hat k^\mu_{\rm o}$ and $k^\mu_{\rm o}$ is:
    \bea
    \hat k_{\rm o}^\mu =\mathbb{C} \Omega_{\rm o}^2 k_{\rm o}^\mu= \mathbb{C} \Omega_{\rm o}^2 e_a^\mu k_{\rm o}^a = \mathbb{C}(\Omega_{\rm o} \nu_{\rm o})\Omega_{\rm o} e_a^\mu (1,-n^i)^a\,.
    \eea
Our choice for the conformal constant is such that $\mathbb{C}(\Omega \nu)_{\rm o} \equiv 1$, therefore,
    \begin{equation}
    \hat k^\mu_{\rm o} = (\hat e^\mu_0 - n^i \hat e^\mu_i)_{\rm o}\,.
    \end{equation}
Substituting previous expressions yields the temporal and spatial fluctuations in the conformal photon wave vector at the observer position,
    \begin{equation}
    \delta \nu_{\rm o} =  -2\Psi_{\rm o} - n_i\, \mathcal{U}^i_{\rm o} ~, \qquad\qquad
    \delta n_{\rm o}^\alpha = -\mathcal{U}_{\rm o}^\alpha -n_i n^\alpha (\delta n^i +  \mathcal{U}^i)_{\rm o}\,.
    \end{equation}
The null condition $(\hat k_\mu \hat k^\mu)_{\rm o} = 0$ allows us to replace $n_i \delta n^i_{\rm o} = - n_i\, \mathcal{U}_{\rm o}^i$, leading to the final expression:
    \begin{equation}
    \delta \nu_{\rm o} = -2\Psi_{\rm o} - n_i\, \mathcal{U}^i_{\rm o}\,, \qquad\qquad
    \delta n_{\rm o}^\alpha = -\mathcal{U}_{\rm o}^\alpha\,.
    \end{equation}
Substituting these into Eq.~\eqref{sachsO}, we obtain the initial condition for Eq.~\eqref{notparallel}
    \begin{equation}\label{einit}
    \hat e^\mu_I{}_{\rm o} = \big[(\mathcal{U}_i)_{\rm o}\,,\, \delta^\alpha_i \big]^\mu \Phi^i_I~.
    \end{equation}
Note that we implicitly oriented the spacelike triad to be aligned with the FLRW coordinates at zeroth order.

The last ingredient to solve Eq.~\eqref{notparallel} is the conformal wave vector at an arbitrary point of the congruence $\hat k^\mu(\lambda)$. Solving the geodesic equation $\hat k^\nu \hat{\nabla}_\nu \hat k^\mu = 0$ we derive
    \begin{equation}\label{deltan}
    \delta \nu = \left(2 \Psi - n^\alpha\, \mathcal{U}_\alpha \right)_{\rm o} - 4 \Psi - 2 \int_0^{\bar r} dr \, \Psi'\,, \qquad\qquad
    \delta n^\alpha = -\mathcal{U}^\alpha_{\rm o} - 2 \int_0^{\bar r} dr \, \partial^\alpha \Psi\,,
    \end{equation}
where a prime stands for partial derivative with respect to conformal time.

In summary, we need to solve Eq.~\eqref{notparallel} at first order in perturbation theory, with conformal factor $\Omega=a\,e^{\varphi}$,
    \begin{equation}
    \frac{d}{d\lambda} \hat e^\mu_I + \hat \Gamma^\mu_{\nu\rho} \bar{\hat k}^\nu \bar{\hat e}^\rho_I + \mathcal{H} \bar{\hat k}^\mu \hat e^0_I + \bar{\hat k}^\mu \bar{\hat e}^\nu_I \partial_\nu \varphi = 0\,,
    \end{equation}
where the initial condition is given by Eq.~\eqref{einit}, and~$\hat e^\mu_I(\lambda)$ is parametrised by Eq.~\eqref{paramcc}.
A direct substitution yields the differential equations
    \begin{equation}
    \frac{d}{d\lambda} A_i +\mathcal{H} A_i + \partial_i \psi=0\,, \qquad\qquad
    \frac{d}{d\lambda} \delta n_i  -2  \partial_i \Psi=0\,,
    \end{equation}
which have the following solutions
    \begin{equation}\label{solutions}
    A_i =\frac1a (\mathcal{U}_i)_{\rm o} + \frac{1}{a} \int_0^{\bar r} dr \, a \, \partial_i \psi\,, \qquad\qquad
    \delta n_i = - (\mathcal{U}_i)_{\rm o} - 2  \int_0^{\bar r} dr \, \partial_i \Psi\,.
    \end{equation}
Note that the solution for~$\delta n_i$ is consistent with Eq.~\eqref{deltan}.

\section{Source terms for the evolution of the Jacobi map}
\label{app:sources}
Combining the results of Appendix~\ref{app:paralleltransport} with the direct computation of the Riemann tensor constructed from the conformal metric in Eq.~\eqref{metric}, we obtain the following expression up to second order:
    \begin{multline}\label{source}
    \hat{R}_{\mu\nu\rho\sigma} \, \hat{k}^\nu \hat{k}^\rho \, \hat{e}_i^\mu \hat{e}_j^\sigma =
    - 2 \partial_{j}\partial_{i}\Psi +  \partial_{(i}B'_{j)} 
    - n^{\alpha} \partial_{\alpha}\partial_{(i}B_{j)}
      + n^{\al}\partial_{j}\partial_{i}B_{\al} + \frac{1}{2} n^{\alpha} n^{\beta} \partial_{j}\partial_{i}h_{\alpha\beta} \\ 
     +  n^{\al} \partial_{(i}h'_{j)\al}+\frac{1}{2} h''_{ij} -  n^{\alpha} \partial_{\alpha}h'_{ij}   -   n^{\alpha} n^{\beta} 
    \partial_{\beta}\partial_{(i}h_{j)\al}  + \frac{1}{2} n^{\alpha} n^{\beta} \partial_{\alpha}\partial_{\beta}h_{ij}
    \\
    -4 \partial_{i}\Psi \partial_{j}\Psi + 8\,\partial_{j}\partial_{i}\Psi \int_0^{\bar{r}}dr\,{\Psi'}  - 8 {\Psi}_{\rm o} \partial_{j}\partial_{i}\Psi + 4 n_\alpha \,\mathcal{U}^\alpha_{\rm o} \partial_{j}\partial_{i}\Psi + 8 \Psi \partial_{j}\partial_{i}\Psi \\
    - 4 n^\alpha\bigg[  (\mathcal{U}_{(j})_{\rm o} + 2 \int_0^{\bar r} dr \, \partial_{(j} \Psi \bigg] \partial_{\al}\partial_{i)}\Psi\,.
    \end{multline}
The quantities in the first two lines are linear in the perturbations, while those in the last two lines are quadratic. The linear terms have the same structure as the first-order expression, but here they contribute up to the second order in perturbation theory.

Notice that the tetrad component $A_i$ in Eq.~\eqref{paramcc}, with solution in Eq.~\eqref{solutions}, does not appear in the final result. This can be understood as follows: the basis vector $\hat{e}^\mu_i$ contains~$A_i$ in the form $\hat{e}^\mu_i \ni A_i \bar{\hat{k}}^\mu$, but any contribution from $A_i$ must enter through terms like $\hat{R}_{\mu\nu\rho\sigma} \,  \bar{\hat{k}}^\nu \bar{\hat{k}}^\rho \bar{\hat{e}}_j^\sigma A_i$, or with $\mu \leftrightarrow \sigma$ and $i \leftrightarrow j$. That is because the conformal Riemann tensor is at least first order in perturbations, and $A_i$ is also of first order. However, these contributions vanish due to the antisymmetry of the Riemann tensor in the indices $\mu$ and~$\nu$. As a result, $A_i$ drops out of the second-order computation.

When Eq.~\eqref{source} is integrated along the line of sight, so-called post-Born corrections arise from expanding the integrand around the background light path,
    \begin{equation}\label{pb}
    \hat{R}_{\mu\nu\rho\sigma} \, \hat{k}^\nu \hat{k}^\rho \, \hat{e}_i^\mu \hat{e}_j^\sigma = \hat{\RR}_{ij}- 2\de x_{\bar r}^\mu \partial_{\mu}\partial_i\partial_{j}\Psi\,,
    \end{equation}
where the first term on the right-hand side corresponds to the conformal Riemann tensor evaluated along the background light path, $\hat{R}_{\mu\nu\rho\si}(\bar x_{\bar r})$, and retains the same functional form as in Eq.~\eqref{source}. The second term instead accounts for the deviation~$\delta x^\mu_{\bar r}$ of the photon trajectory from straight coordinate lines in the perturbed spacetime.

Such distortion can be computed by formally integrating the conformal wave vector along the photon trajectory,
    \bea
    \delta\eta_{\bar r}&=&\delta\eta_{\rm o}-\int_0^{\bar r}dr\,\delta \nu=\de\eta_{\rm o}
    -\bar r\left(2 \Psi - \mathcal{U}_\pa \right)_{\rm o} +4\int_0^{\bar r}dr\,\Psi+2\int_0^{\bar r}dr\,(\bar r-r)\Psi'\,,
   \nonumber\\
    \delta x^\al_{\bar r}&=&\delta x^\al_{\rm o}+\int_0^{\bar r}dr\,\delta n^\al=\delta x^\al_{\rm o}-\bar r\,\mathcal{U}^\alpha_{\rm o} - 2 \int_0^{\bar r} dr \,(\bar r-r) \partial^\alpha \Psi
    \,,
    \eea
where the coordinate lapse $\de\eta_{\rm o}$ and shift $\de x^\al_{\rm o}$ at the observer position depend on the motion of the observer \cite{Yoo:2018qba}. Projecting the spatial distortion in spherical coordinates, we obtain
    \bea
    \de r&=&n_\al \de x^\al_{\bar r}=(\de x_\pa)_{\rm o}+\bar r(2\Psi-\mathcal{U}_\pa)_{\rm o}-2\int_0^{\bar r}dr\,\Psi+2\int_0^{\bar r}dr\,(\bar r-r)\Psi'\,,
    \nonumber\\
    \de\theta&=&\theta_\al \frac{\de x^\al_{\bar r}}{\bar r}=\frac1{\bar r}(\delta x_\theta)_{\rm o}-(\mathcal{U}_\theta)_{\rm o} - 2 \int_0^{\bar r} dr \,W(\bar r,r)\Psi_{,\theta}
    \,,
    \nonumber\\
    \de\phi&=&\phi_\al \frac{\de x^\al_{\bar r}}{\bar r\sin\theta}=\frac1{\bar r\sin\theta}(\delta x_\phi)_{\rm o}-\frac{(\mathcal{U}_\phi)_{\rm o}}{\sin\theta} - 2 \int_0^{\bar r} dr \,W(\bar r,r)\frac{\Psi_{,\phi}}{\sin{\theta}}\,.
    \eea

To derive the projected components~$\hat{\mathcal{R}}_{IJ}$ from Eqs.~\eqref{source} and \eqref{pb}, we first express derivatives with respect to Cartesian coordinates in terms of derivatives with respect to spherical coordinates,
    \begin{equation}
    \partial_i=n_i\partial_r+\theta_i\frac{\partial_\theta}{r}+\phi_i\frac{\partial_\phi}{r\sin\theta}\,.
    \end{equation}
Using standard identities for unit vectors on the sphere, we can express second derivatives with respect to Cartesian coordinates as:
    \begin{multline}
    \partial_j\partial_i = n_j n_i \partial^2_r-(n_j\theta_i+n_i\theta_j)\frac{\partial_\theta}{r^2}+(n_j\theta_i+n_i\theta_j)\frac{\partial_r\partial_\theta}{r}-(n_j\phi_i+n_i\phi_j)\frac{\partial_\phi}{r^2\sin\theta}\\+(n_j\phi_i+n_i\phi_j)\frac{\partial_r\partial_\phi}{r\sin\theta}
    +\theta_j\theta_i\frac{\partial^2_\theta}{r^2}+\theta_j\theta_i\frac{\partial_r}{r}-(\theta_j\phi_i+\theta_i\phi_j)\frac{\cot\theta}{\sin\theta}\frac{\partial_\phi}{r^2}
    \\+(\theta_j\phi_i+\theta_i\phi_j)\frac{\partial_\theta\partial_\phi}{r^2\sin\theta}
    +\phi_j\phi_i\frac{\partial^2_\phi}{r^2\sin^2\theta}+\phi_j\phi_i\frac{\partial_r}{r}+\phi_j\phi_i\cot\theta\frac{\partial_\theta}{r^2}\,.
    \end{multline}
Taking into account the last two equations, and contracting Eq.~($\ref{source}$) with $\theta^i\theta^j$, we obtain
    \begin{multline}
    \bar r^2\hat{\mathcal{R}}_{11} = -2 \Psi_{,\theta\theta}-2\bar r\Psi_{,\bar r}+ B_{\parallel,\theta\theta}+\bar r B'_{\parallel}-\frac{d}{d\bar r} ( \bar r B_{\theta,\theta})
    \\+\frac{\bar r^2}{2}\left[\frac{d^2}{d\bar r^2}h_{\theta\theta}- n^\al\theta^i\theta^j\frac{d}{d\bar r}\left( \partial_{(j}h_{i)\al} \right)+n^\al n^\beta \theta^i\theta^j\partial_i\partial_j h_{\al\beta}\right]
    -4 (\Psi_{,\theta})^2\\+4( \Psi_{,\theta\theta}+\bar r\Psi_{,\bar r})\left[ 2\Psi-2\Psi_{\rm o}+(\mathcal{U}_{\parallel})_{\rm o}+2 \int_0^{\bar r}dr\,\Psi'\right]
    +4(\Psi_{,\theta}-\bar r\Psi_{,\bar r\theta})\left[ (\mathcal{U}_\theta )_{\rm o} +2\int_0^{\bar r}\frac{dr}r\,\Psi_{,\theta}\right]\,,
    \end{multline}
where $d/d\bar r=n^\al\partial_\al-\partial_\eta$ is the derivative along the background light path.

Similarly, contracting with $\phi^i\phi^j$, yields
    \begin{multline}
    \bar r^2\hat{\mathcal{R}}_{22} = -\frac{2}{\sin^2\theta}\Psi_{,\phi\phi}-2\bar r\Psi_{,\bar r}-2\cot\theta\Psi_{,\theta}+\frac{B_{\parallel,\phi\phi}}{\sin^2\theta}+\cot\theta B_{\parallel,\theta}+\bar r B'_{\parallel}
    \\-\frac{1}{\sin\theta}\frac{d}{d\bar r} \left[\bar r(B_{\phi,\phi}+B_{\theta}\cos\theta)\right]
    +\frac{\bar r^2}{2}\left[\frac{d^2}{d\bar r^2}h_{\phi\phi}- n^\al \phi^i\phi^j\frac{d}{d\bar r}\left( \partial_{(j}h_{i)\al} \right)+n^\al n^\beta\phi^i\phi^j\partial_i\partial_j h_{\al\beta}\right]
    \\
    +\frac4{\sin\theta}(\Psi_{,\phi}-\bar r\Psi_{,\bar r\phi})\left[ (\mathcal{U}_\phi )_{\rm o} +2\int_0^{\bar r}\frac{dr}{r\sin\theta}\,\Psi_{,\phi}\right]
    -4 \left(\frac{\Psi_{,\phi}}{\sin\theta}\right)^2
    \\+4 \left(\frac{\Psi_{,\phi\phi}}{\sin^2\theta}+\bar r\Psi_{,\bar r}+\cot\theta\Psi_{,\theta}\right)\left[ 2\Psi-2\Psi_{\rm o}+(\mathcal{U}_\parallel)_{\rm o}+2 \int_0^{\bar r}dr\,\Psi'\right]\,.
    \end{multline}
Finally, contracting with $\theta^i\phi^j$ gives
    \begin{multline}
    \bar r^2\hat{\mathcal{R}}_{12} = -\frac{2}{\sin\theta}\Psi_{,\theta\phi}+2\frac{\cot\theta}{\sin\theta}\Psi_{,\phi}+ \frac1{\sin\theta}(B_{\parallel,\theta\phi}-\cot\theta B_{\parallel,\phi}) \\- \frac{1}{2\sin\theta}\frac{d}{d\bar r}\left[\bar r(B_{\theta,\phi}-\cos\theta B_\phi+\sin\theta B_{\phi,\theta})\right]
    +\frac{\bar r^2}{2}\left[\frac{d^2}{d\bar r^2}h_{\theta\phi}- n^\al \theta^i\phi^j\frac{d}{d\bar r}\left( \partial_{(j}h_{i)\al} \right)+n^\al n^\beta\theta^i\phi^j\partial_i\partial_j h_{\al\beta}\right]
    \\-\frac4{\sin\theta} \Psi_{,\theta}\Psi_{,\phi}
    +\frac4{\sin\theta}\left( \Psi_{,\theta\phi}-\cot\theta\Psi_{,\phi} \right)\left[ 2\Psi-2\Psi_{\rm o}+(\mathcal{U}_\parallel)_{\rm o}+2 \int_0^{\bar r}dr\,\Psi'\right]+\frac2{\sin\theta}(\Psi_{,\phi}-\bar r\Psi_{,\bar r\phi})
    \\
    {\scriptstyle \times}\left[ (\mathcal{U}_\theta )_{\rm o} +2\int_0^{\bar r}\frac{dr}{r}\,\Psi_{,\theta}\right]+2(\Psi_{,\theta}-\bar r\Psi_{,\bar r\theta})\left[ (\mathcal{U}_\phi )_{\rm o} +2\int_0^{\bar r}\frac{dr}{r\sin\theta}\,\Psi_{,\phi}\right]
    \,.
\end{multline}

In addition to the optical tidal matrix, the Jacobi map is also sourced by fluctuations in the observed redshift $\delta z$ defined as
    \begin{equation}
    1+z=\frac{\nu_{\rm s}}{\nu_{\rm o}}=\frac{1+\delta z}{a_{\rm s}}\,.
    \end{equation}
Given our choice $\mathbb{C}=(\Omega_{\rm o}\,\nu_{\rm o})^{-1}$, we can rewrite the frequency ratio as
    \begin{equation}
    \frac{\nu_{\rm s}}{\nu_{\rm o}}=\mathbb{C} \Omega_{\rm o} \nu_{\rm s}=\mathbb{C}\Omega_{\rm o}(-g_{\mu\nu}u^\mu k^\nu)_{\rm s}=-\frac{\Omega_{\rm o}}{\Omega_{\rm s}}(\hat g_{\mu\nu}\hat u^\mu \hat k^\nu)_{\rm s}\,.
    \end{equation}
Therefore,
    \begin{equation}
    \delta z=-a_{\rm o} (\hat g_{\mu\nu}\hat u^\mu \hat k^\nu)_{\rm s} e^{\varphi_{\rm o}-\varphi_{\rm s}}-1\,.
    \end{equation}
At first order in perturbation theory, we derive
    \begin{equation}
    \de z=(\varphi+\mathcal{H}\delta\eta+2\Psi-\mathcal{U}_\pa)_{\rm o}-(2\Psi+\varphi-\mathcal{U}_\pa)_{\rm s}-2\int_0^{\bar r_z} d\bar r\,\Psi'\,,
    \end{equation}
where we expanded the scale factor at the observer position $a_{\rm o}=1+\mathcal{H}_{\rm o}\de\eta_{\rm o}$.

Finally, we need to evaluate the fluctuation in the affine parameter, $\delta\lambda_{\rm s} = \lambda_{\rm s} - \lambda_z$, arising when the observed redshift is used as a physical parametrisation. Specifically, the observed redshift $z$ can be used to define a unique (conformal) time coordinate $\bar{\eta}_z$, given by the relation $1 + z = \frac{1}{a(\bar{\eta}_z)}$, where $a(\bar{\eta}_z)$ is the scale factor at time $\bar{\eta}_z$. This time coordinate is associated with the affine parameter~$\lambda_z$.

The temporal fluctuation~$\Delta\eta_{\rm s}$ around this reference time~$\bar{\eta}_z$ differs from the fluctuation~$\delta\eta_{\rm s}$ expanded around a generic gauge-dependent background. The distinction between the two fluctuations is precisely quantified by~$\delta\lambda_{\rm s}$. To compute this temporal distortion, we integrate over the light path:
    \begin{equation}
    \delta \eta_{\rm s}=\delta\eta_{\rm o}-\int_0^{\bar r_z}d\bar r\,\delta\nu=\de\eta_{\rm o}
    -\bar r_z\left(2 \Psi - n^\alpha\, \mathcal{U}_\alpha \right)_{\rm o} +4\int_0^{\bar r_z}d\bar r\,\Psi+2\int_0^{\bar r_z}d\bar r\,(\bar r_z-\bar r)\Psi'\,,
    \end{equation}
then, perturbing the relation $1+z=\frac1{a(\bar\eta_z)}$ we obtain
    \begin{equation}
    \Delta\eta_{\rm s}=\frac{\delta z}{\mathcal{H}_z}=\frac{1}{\mathcal{H}_z}\left[  (\varphi+\mathcal{H}\delta\eta+2\Psi-n^\alpha\,\mathcal{U}_\al)_{\rm o}-(2\Psi+\varphi-n^\alpha\,\mathcal{U}_\alpha)_{\rm s}-2\int_0^{\bar r_z} d\bar r\,\Psi'
 \right]\,.
    \end{equation}
Subtracting the two results, we obtain
    \begin{multline}
    \delta\lambda_{\rm s} = -\delta\eta_{\rm o}+\bar r_z(2 \Psi -  \mathcal{U}_\pa)_{\rm o}+\frac{1}{\mathcal{H}}(\varphi+\mathcal{H}\delta\eta+2\Psi-\mathcal{U}_\pa)_{\rm o}-\frac{1}{\mathcal{H}}(2\Psi+\varphi-\mathcal{U}_\pa)_{\rm s}
    \\
    -4\int_0^{\bar r_z}d\bar r\,\Psi-\frac{2}{\mathcal{H}}\int_0^{\bar r_z} d\bar r\,\Psi'-2\int_0^{\bar r_z}d\bar r\,(\bar r_z-\bar r)\Psi'\,.
    \end{multline}

\section{Notes on numerical integration}
\subsection{Fourier space power spectra for vector and tensor modes}
\label{ap:numer-Four}
For numerical computations, we employ the dimensionless power spectrum in Fourier space~$\Delta$ rather than the dimensionful power spectrum~$P$ introduced in the main text. For scalar perturbations like the Weyl potential $\Psi$, our convention for the relation between the two spectra is given by:
\bea
P_\Psi(k) \doteq\frac{2\pi ^{2}}{k^{3}} \Delta_\Psi(k)\,.
\eea
For the transverse vector potential~$B_\alpha$, in the main text we defined the dimensionful power spectrum per single polarisation state, as in Eq.~\eqref{ps3dB}. However, it is customary to introduce the dimensionless power spectrum as the total one, obtained by summing over the two independent polarizations:
\bea
2P_B(k)\doteq\frac{2\pi ^{2}}{k^{3}} \Delta_B(k)\,.
\eea
An analogous definition applies to the transverse-traceless tensor~$h_{\alpha\beta}$, which also possesses two independent polarization states.

The second-order dimensionless Fourier-space power spectra for the vector~$\Delta^{(2)}_B$ and tensor modes~$\Delta^{(2)}_h$ are reported in Appendix~B of Ref.~\cite{Adamek:2017uiq}. These spectra are expressed as convolution integrals over the wave vector $\mathbf{q}$ and can be cast into a form suitable for numerical evaluation by applying the following change of variables~\cite{Lu:2008ju, Adamek:2017grt}:
\be
w = \frac{q}{k}, \qquad
u = \sqrt{1 - 2 w \mu + w^2}, \qquad
\mu \equiv \frac{\mathbf{q}\cdot\mathbf{k}}{qk},
\ee
so that $q = k w$ and $\lvert \mathbf{k} - \mathbf{q} \rvert = k u$.
The resulting expressions for the vector and tensor modes are given by
\begin{multline}
\Delta^{(2)}_B(k) ={}
\frac{18 \mathcal{H}^4}{k^6}
\int_0^\infty \frac{dw}{w^2}
\int_{|w-1|}^{w+1} \frac{du}{u^2}\, \left\{\sum_i \Omega_i
\frac{T^{\theta}_i(k w)}{w^2} \, T^{\delta}_i(k u)\right\} {\scriptstyle \times}
\\
\left\{ \sum_j \Omega_j
\left[
\frac{T^{\theta}_j(k u)}{u^2} \, T^{\delta}_j(k w)
-
\frac{T^{\theta}_j(k w)}{w^2} \, T^{\delta}_j(k u)
\right]\right\}
\Delta_\zeta(k w)\, \Delta_\zeta(k u)
\left[
\frac{1}{4}\left(1 + w^2 - u^2\right)^2 - w^2
\right]\,,
\end{multline}
and 
\begin{multline}
\Delta^{(2)}_h(k) ={}
8 \int_0^\infty \frac{dw}{w^2}
\int_{|w-1|}^{w+1} \frac{du}{u^2}\,
\Biggl[
T^{\Psi}(k w)\, T^{\Psi}(k u)
+ \frac{3}{2}\,\mathcal{H}^2
\sum_{i}
\Omega_i
\frac{T^{\theta}_i(k w)}{k^2 w^2}
\frac{T^{\theta}_i(k u)}{k^2 u^2}
\Biggr]^2
\\[1ex]
{\scriptstyle \times}
\Delta_\zeta(k w)\, \Delta_\zeta(k u)\,
\left[
\frac{1}{4}\left(1 + w^2 - u^2\right)^2 - w^2
\right]^2\,,
\end{multline}
where $\Delta_\zeta$ is the power spectrum of the primordial curvature perturbation,
$T^\Psi$ is the linear transfer function for the scalar potential,
$T^\delta_i$ and $T^\theta_i$ are the linear transfer functions of the density contrast and velocity divergence for the matter species $i$, $\Omega_i$ is the redshift-dependent density parameter of the $i$-th matter species, and the sums over $i$ and $j$ run over all matter species.
The linear transfer functions are computed using the \texttt{CLASS} code~\cite{Blas:2011rf, Lesgourgues:2011re}.

\subsection{Angular power spectra}
\label{ap:numer-th}
The numerical estimation of the scalar-induced second-order angular power spectra requires to compute convolution integrals of the scalar gravitational potential $P_\Psi$. In this section, we cast 
the expression of the integrals in 
Eq.~\eqref{Celrot} (rotation), Eq~\eqref{CBmodes} (shear) and Eq.~\eqref{Celeps} (ellipticity) in a form suitable for numerical integration.

We consider as an example the rotation power spectrum.
The integral in $\mathbf{L}$ in Eq.~\eqref{Celrot} can be expressed in polar coordinates,
\begin{multline}
C^{\omega\omega}_\ell(\bar r_z)=\frac{4}{(2\pi)^2}\int_0^{\infty} dL \int_0^{2\pi} d\phi\, L^3 (\ell \sin{\phi})^2 \left(\ell L  \cos{\phi} - L^2\right)^2  \\
 {\scriptstyle \times}\int_0^{\bar r_z}d\bar r\ \frac{W^2(\bar r_z,\bar r)}{\bar r^2} \ P_\Psi\left[\frac{|\bm\ell-\bm L|}{\bar r}, z(\bar r)\right] \int_0^{\bar r}d\bar r'\ \frac{W^2(\bar r,\bar r')}{\bar r'^2}\ P_\Psi\left[\frac{L}{{\bar r}'}, z(\bar r')\right]\,.
\end{multline}
We can define

\be
C^{\Psi}_{L}[z(\bar r)] \equiv  \int_0^{\bar r}d\bar r'\ \frac{W^2(\bar r,\bar r')}{\bar r'^2}\,  P_\Psi\left[\frac{L}{{\bar r}'}, z(\bar r')\right], \label{eq:Cell_psi}
\ee
and write
\begin{multline}
C^{\omega\omega}_\ell (\bar r_z)=\frac{4}{(2\pi)^2}\int_0^{\infty} dL \int_0^{2\pi} d\phi\, L^3 (\ell \sin{\phi})^2 \left(\ell L  \cos{\phi} - L^2\right)^2  \\
{\scriptstyle \times}\int_0^{\bar r_z}d\bar r\ \frac{W^2(\bar r_z,\bar r)}{\bar r^2} \ P_\Psi\left[\frac{|\bm\ell -\bm L|}{\bar r}, z(\bar r)\right] C^{\Psi}_{L}[z(\bar r)] \,.
\end{multline}

Following the approach of~\cite{Pratten:2016dsm}, we introduce 
\be
M(L, |\bm\ell -\bm L|, \bar r_z) =  \int_0^{\bar r_z} d\bar r\ \frac{W^2(\bar r_z,\bar r)}{\bar r^2}\ P_\Psi\left[\frac{|\bm\ell -\bm L|}{\bar r}, z(\bar r)\right] C^{\Psi}_{L}[z(\bar r)]\,, \label{eq:mat}
\ee
and write the integral as 
\bea
C^{\omega\omega}_\ell (\bar r_z)&=&\frac{4}{(2\pi)^2}\int_0^{\infty} dL \int_0^{2\pi} d\phi\, L^3 (\ell \sin{\phi})^2 \left(\ell L  \cos{\phi} - L^2\right)^2 M(L, |\bm\ell -\bm L|, \bar r_z)  \,. \label{eq:fin}
\eea

Numerically, we can first estimate Eq.~\eqref{eq:Cell_psi} for $N_z$ comoving distances between $\bar{r} = 0$
and $\bar{r} = \bar{r}_z$, and $N_\ell$ values of $\bar{L}' = |\bm\ell -\bm L|$. 
We then estimate the matrix in~\eqref{eq:mat} at fixed comoving distance $
\bar r_z$, and finally we perform the integral over $L$ and $\phi$ in~\eqref{eq:fin}. 
This approach is the one adopted in the \texttt{CAMB} cosmology code.\footnote{\url{https://github.com/cmbant/CAMB}}

The contributions to the B-modes of the shear power spectrum can be written in a similar form. In~\eqref{CBmodes}, we can identify a few types of contributions. The integrals that we identify as `type~1' have the same formal expression as the rotation angular power spectrum. We can write these terms as
\bea
 [C^{\gamma_B}_\ell]^{\rm type1}_i (\bar r_z) &=&\frac{K_i}{(2\pi)^2}\int_0^{\infty} L \, dL \int_0^{2\pi} d\phi\, f_i(\phi, L) M_i(L, |\bm\ell -\bm L|, \bar r_z)  \,, \label{eq:int-type1}\\
 M_i(L, |\bm\ell -\bm L|, \bar r_z) &=& \int_0^{\bar r_z} d\bar r  \int_0^{\bar r} d\bar r' \, \tilde{W}^2_{A,i}(\bar r_z,\bar r)\,  \tilde{W}^2_{B, i}(\bar r,\bar r')\, P_\Psi\left[\frac{|\bm\ell-\bm L|}{\bar r}, z(\bar r)\right]\,  P_\Psi\left[\frac{L}{{\bar r}'}, z(\bar r')\right]\,. \qquad\qquad
\eea

In Table~\ref{tab:type1} we write the expression of $K$, $f(L,\phi)$, $(\tilde{W}_A)$, $(\tilde{W}_B)$ for all contributions in the shear power spectrum that can be written in this form.
These corrections include all the terms obtained by cross-correlating the relativistic corrections to the shear with the rotation term (`rot cross'), some squared relativistic corrections (`square'), and some cross-correlations among relativistic corrections (`cross no-rot').

\begin{table}[h]
    \centering
    \resizebox{\textwidth}{!}{
    \begin{tabular}{ccccc}
        \toprule
        Type-1 integrals\\
        \toprule
        Index & $K$ & $f(L,\phi)$ & $\tilde{W}^2_A$ & $\tilde{W}^2_B$ \\
        \midrule
        rotation & $4$ & $\ell^2 L^2 \sin^2{\phi}(L\ell\cos{\phi}-L^2)^2$ & $W^2(\bar r_z, \bar r)/{\bar r^2}$  &  $W^2(\bar r, \bar r')/{\bar r'^2}$ \\
        rot cross 1 & $-32/{\bar r_z}$ & $\ell L^3 \cos{\phi}\sin^2{\phi} (\ell L \cos{\phi}-L^2)$ &$W(\bar r_z, \bar r)/{\bar r^2}$  &  $W(\bar r, \bar r')/{\bar r'^3}$\\
        rot cross 2 & $-16/{\bar r_z}$ & $\ell^2 L^2 \sin^2{\phi} (\ell L \cos{\phi}-L^2)$ &$W(\bar r_z, \bar r)/{\bar r^3}$  &  $W(\bar r, \bar r')/{\bar r'^2}$\\
        rot cross 3 & $16/{\bar r_z}$ & $\ell L^3  \cos{\phi}\sin^2{\phi} (\ell L \cos{\phi}-L^2)$ &$W(\bar r_z, \bar r)/{\bar r^3}$  &  $\frac{W(\bar r, \bar r')}{{\bar r'^2}}\left( 3-\frac{\bar{r}_{z}}{\bar{r}} +\frac{\bar{r}}{\bar{r} '}\right)$\\
        square c & $64/{\bar r^2_z}$ & $\ell^2 L^2 \sin^2{\phi}$ &$\frac{1}{\bar r^4}$  &  $\frac{1}{\bar r'^2}$\\
        square d & $-64/{\bar r^2_z}$ & $\ell L^3 \cos{\phi}\sin^2{\phi}$ &$\frac{1}{\bar r^4}$  &  $\frac{1}{\bar r'^2} \left(3-\frac{r_z}{\bar r} + \frac{\bar r}{\bar{r}'}\right)$\\
        square e & $16/\bar{r}^2_z$ & $L^4 \cos^2{\phi}\sin^2{\phi} $ &$\frac{1}{\bar r^4}$  &  $\frac{1}{\bar r'^2} \left(3-\frac{\bar r_z}{\bar r} + \frac{\bar r}{\bar{r}'}\right)^2$\\
        cross no-rot 1 & $128/{\bar r^2_z}$ & $\ell L^3 \cos{\phi} \sin^2{\phi}$ &$\frac{1}{\bar r^3}$  &  $\frac{1}{\bar r'^3}$\\
        cross no-rot 2 & $-64/{\bar r^2_z}$ & $L^4 \cos^2{\phi}\sin^2{\phi}$ &$\frac{1}{\bar r^3}$  &  $\frac{1}{\bar r'^3}\left(3-\frac{\bar r_z}{\bar r} + \frac{\bar r}{\bar{r}'}\right)$\\
        cross no-rot 3 & $-64/{\bar r^2_z}$ & $\ell^2 L^2 \sin^2{\phi}$ &$\frac{1}{\bar r^4}$  &  $\frac{1}{\bar r'^2}$\\
        cross no-rot 4 & $-64/{\bar r^2_z}$ & $L^3\cos\phi \sin^2{\phi} (L \cos{\phi}-\ell)$ &$\frac{1}{\bar r^4}$  &  $\frac{1}{\bar r'^2} \left(3-\frac{\bar r_z}{\bar r} + \frac{\bar r}{\bar{r}'}\right)$\\
        eps cross-rot & $4$ & 
\begin{tabular}{c}
$\ell L^2 \sin^2{\phi}(\ell L \cos{\phi} - L^2)[L\cos{\phi} \cdot$ \\
$\cdot|\bm\ell-\bm L|^2+L^2(L\cos{\phi}-\ell)]$ 
\end{tabular}
&$W^2(\bar r_z, \bar r)/r^2$ &  $W(\bar r_z, \bar r') W(\bar r, \bar r')/r'^2$\\
        eps corr 2 & $-16/{\bar r_z}$ & 
\begin{tabular}{c}
$\ell L^2 \sin^2{\phi}[L \cos{\phi} \cdot$ \\
$\cdot|\bm\ell-\bm L|^2+L^2(L\cos{\phi}-\ell)]$ 
\end{tabular}
&$W(\bar r_z, \bar r)/r^3$ &  $W(\bar r_z, \bar r')/r'^2$\\
        eps corr 3 & $8/{\bar r_z}$ & 
\begin{tabular}{c}
$L^3 \cos{\phi} \sin^2{\phi}[(L \cos{\phi} - \ell)\cdot$ \\
$(L^2+|\bm\ell-\bm L|^2)+2L\cos{\phi}|\bm\ell-\bm L|^2]$ 
\end{tabular}
&$W(\bar r_z, \bar r)/r^3$ &  $\frac{W(\bar r_z, \bar r')}{r'^2} \left(3-\frac{\bar r_z}{\bar r} + \frac{\bar r}{\bar{r}'}\right)$\\
        \bottomrule
    \end{tabular}
    }
    \caption{Table for the integrals that take the form in Eq.~\eqref{eq:int-type1}.}
    \label{tab:type1}
\end{table}

Two integrals coming from the squares of relativistic effects can be written in the following form:
\bea
 [C^{\gamma_B}_\ell]^{\rm type2}_i (\bar r_z) &=&\frac{K_i}{(2\pi)^2}\int_0^{\infty} L \, dL \int_0^{2\pi} d\phi\, f_i(\phi, L) C^X_{|\bm\ell-\bm L|}(\bar r_z) C^Y_{L}(\bar r_z)\,, \label{eq:int-type2a}
\\
 C^{X/Y}_{L'}(\bar r_z) &=& \int_0^{\bar r_z} d\bar r \, \tilde{W}^2_{X/Y}(\bar r_z,\bar r)\,  P_\Psi\left[\frac{L'}{\bar r}, z(\bar r) \right]\,.
\eea

Two cross-correlations terms that do not involve the rotation, can be written as
\bea
 [C^{\gamma_B}_\ell]^{\rm type2}_i (\bar r_z) &=&\frac{K_i}{(2\pi)^2}\int_0^{\infty} L \, dL \int_0^{2\pi} d\phi\, f_i(\phi, L) 
 P_\Psi\left[\frac{|\bm\ell-\bm L|}{{\bar r_z}}, z(\bar r_z)\right] C^X_{L}(\bar r_z)\,, \label{eq:int-type2b}
\\
 C^{X}_{L}(\bar r_z) &=& \int_0^{\bar r_z} d\bar r \,  \tilde{W}^2_{X}(\bar r_z,\bar r)\,  P_\Psi\left[\frac{L}{\bar r}, z(\bar r) \right]\,.
\eea

We group both these types of corrections together as `type 2' integrals, and report their expression for $K$, $f(L,\phi)$, $\tilde{W}^2_X$, $\tilde{W}^2_Y$ in Table~\ref{tab:type2}.

\begin{table}[h]
    \centering
    \resizebox{\textwidth}{!}{
    \begin{tabular}{ccccc}
        \toprule
        Type-2 integrals\\
        \toprule
        Index & $K$ & $f(L,\phi)$ & $\tilde{W}^2_X$ & $\tilde{W}^2_Y$ \\
        \midrule
        square a & $64/{\bar r^2_z}$ & $L^4 \cos^2{\phi} \sin^2{\phi}$ &$\frac{1}{\bar r^2}$  &  $\frac{1}{\bar r^4}$\\
        square b & $64/{\bar r^2_z}$ & $L^3  \cos{\phi}\sin^2{\phi} (L \cos{\phi} - \ell)$ &$\frac{1}{\bar r^3}$  &  $\frac{1}{\bar r^3}$\\
        cross no-rot 5 & $-16/{\bar r^4_z}$ & $L^4 \cos^2{\phi}\sin^2{\phi}$ &$\frac{1}{\bar r^3} \left(\frac{1}{\bar{r}_z \mathcal{H}_z} -2 + 2 \frac{\bar r}{\bar r_z}\right)\left( 2+\frac{\bar{r}_{z}}{\bar{r}}\right)$  & - \\
        cross no-rot 6 & $32/{\bar r^3_z}$ & $L^4 \cos^2{\phi}\sin^2{\phi}$ &$\frac{1}{\bar r^4} \left(\frac{1}{\bar{r}_z \mathcal{H}_z} - 2 + 2 \frac{\bar r}{\bar r_z}\right)$  & - \\
        eps main & $4$ & 
\begin{tabular}{c}
$L^3 \cos{\phi} \sin^2{\phi} |\bm\ell-\bm L|^2 [L\cos{\phi}\cdot$ \\
$\cdot|\bm\ell-\bm L|^2 + L^2(L\cos{\phi}-\ell)]$ 
\end{tabular}
&$W^2(\bar r_z, \bar r)/r^2$ &  $W^2(\bar r_z, \bar r)/r^2$\\
        eps corr 1 & $-16/{\bar r_z}$ & 
\begin{tabular}{c}
$L^3 \cos{\phi} \sin^2{\phi} [2L\cos{\phi}|\bm\ell-\bm L|^2+$ \\
$+(L\cos{\phi}-\ell)(L^2+|\bm\ell-\bm L|^2)]$ 
\end{tabular}
&$W(\bar r_z, \bar r)/r^2$ &  $W(\bar r_z, \bar r)/r^3$\\
        \toprule
        square 1 & $4/{\bar r^2_z}$& $L^4 \cos^2{\phi}\sin^2{\phi} $ &$\frac{1}{\bar r^4} \left(\frac{1}{\bar{r}_z \mathcal{H}_z} -2 + 2 \frac{\bar r}{\bar r_z}\right)^2$ & - \\
        square 2 & $4/{\bar r^6_z}\left(\frac{1}{\bar{r}_z \mathcal{H}_z}\right)^2$& $L^3 \cos{\phi}\sin^2{\phi}  (L \cos{\phi} -\ell)$ & - & - \\
        \bottomrule
    \end{tabular}
    }
    \caption{Table for the integrals that take the form in Eq.~\eqref{eq:int-type2a} (`square a/b', `eps main', and `eps corr 1'), Eq.~\eqref{eq:int-type2b} (`cross no-rot 5/6'), Eq.~\eqref{sq1} (`square 1'), and Eq.~\eqref{sq2} (`square 2').}
    \label{tab:type2}
\end{table}

The remaining two contributions to the shear, coming from the squares of relativistic corrections (`square 1' and `square 2') can be written as
\begin{equation}
[C^{\gamma_B}_\ell]^{\rm square1} (\bar r_z) =\frac{K_i}{(2\pi)^2}\int_0^{\infty} L \, dL \int_0^{2\pi} d\phi\, f_i(\phi, L)  C^X_{L}(\bar r_z) \int_{-\infty}^{\infty} \frac{d k_\parallel}{2\pi} P_\Psi\left[ \sqrt{\left(\frac{|\bm\ell-\bm L|}{{\bar r_z}}\right)^2 + k_\parallel^2}, z(\bar r_z)\right]\,, \label{sq1}
\end{equation}
and
\begin{equation}
[C^{\gamma_B}_\ell]^{\rm square2} (\bar r_z) =\frac{K_i}{(2\pi)^2}\int_0^{\infty} L \, dL \int_0^{2\pi} d\phi\, f_i(\phi, L) 
 P_\Psi\left[\frac{|\bm\ell-\bm L|}{{\bar r_z}}, z(\bar r_z)\right] P_\Psi\left[\frac{L}{{\bar r_z}}, z(\bar r_z)\right]\,. \label{sq2}
\end{equation}
We report the corresponding expression of $K$, $f(L,\phi)$, and $\tilde{W}^2_X$ in Table~\ref{tab:type2}.

The ellipticity B-modes are given by the shear B-mode power spectrum plus the extra relativistic terms presented in Eq.~\eqref{Celeps}. 
The main correction, arising from the square of the convolution of the linear convergence and the linear shear, is given by Eq.~\eqref{eq:eps-main}, and it can be cast in the same form as the shear B-mode integrals of type~2. 
The remaining terms include a correction coming from the cross-correlation of the rotation with the convolution of the linear convergence and shear (`eps cross rot'), and three additional terms. 
All these four contributions can be cast in the form of `type~1' or `type~2' integrals.  
We report their expressions in Table~\ref{tab:type1} and Table~\ref{tab:type2}.

The code used to estimate these theory predictions can be found at:
\url{https://github.com/leporif/relativistic_effects_weak_lensing.git}.

\bibliographystyle{JHEP}
\bibliography{refs}

\end{document}